\documentclass[12pt]{article}
\usepackage{graphicx}
\usepackage{amscd,amsmath}
\usepackage{amssymb}
\usepackage{braket}

\usepackage{tikz-cd}

\usepackage{wrapfig}
\usepackage{xcolor}
\usepackage{indentfirst}
\usepackage{amsbsy}
\usepackage{wasysym}
\usepackage{hyperref}
\usepackage{cancel}
\usepackage{verbatim}
\usepackage{empheq}
\usepackage{adjustbox}

\def\hybrid{\topmargin 0pt      \oddsidemargin 0pt
        \headheight 0pt \headsep 0pt
        \voffset=-0.5cm
        \hoffset=-0.25in
        \textwidth 6.75in
        \textheight 9.5in       % A4 paper
        \marginparwidth 0.0in
        \parskip 5pt plus 1pt   \jot = 1.5ex}
\catcode`\@=11
\def\marginnote#1{}

\newcount\hour
\newcount\minute
\newtoks\amorpm
\hour=\time\divide\hour by60 \minute=\time{\multiply\hour by60
\global\advance\minute by-\hour}
\edef\standardtime{{\ifnum\hour<12 \global\amorpm={am}%
        \else\global\amorpm={pm}\advance\hour by-12 \fi
        \ifnum\hour=0 \hour=12 \fi
        \number\hour:\ifnum\minute<10 0\fi\number\minute\the\amorpm}}
\edef\militarytime{\number\hour:\ifnum\minute<10 0\fi\number\minute}

\def\draftlabel#1{{\@bsphack\if@filesw {\let\thepage\relax
   \xdef\@gtempa{\write\@auxout{\string
      \newlabel{#1}{{\@currentlabel}{\thepage}}}}}\@gtempa
   \if@nobreak \ifvmode\nobreak\fi\fi\fi\@esphack}
        \gdef\@eqnlabel{#1}}
\def\@eqnlabel{}
\def\@vacuum{}
\def\draftmarginnote#1{\marginpar{\raggedright\scriptsize\tt#1}}
\def\draftlabel#1{{\@bsphack\if@filesw {\let\thepage\relax
   \xdef\@gtempa{\write\@auxout{\string
      \newlabel{#1}{{\@currentlabel}{\thepage}}}}}\@gtempa
   \if@nobreak \ifvmode\nobreak\fi\fi\fi\@esphack}
        \gdef\@eqnlabel{#1}}
\def\@eqnlabel{}
\def\@vacuum{}
\def\draftmarginnote#1{\marginpar{\raggedright\scriptsize\tt#1}}

\def\draft{\oddsidemargin -.5truein
        \def\@oddfoot{\sl preliminary draft \hfil
        \rm\thepage\hfil\sl\today\quad\militarytime}
        \let\@evenfoot\@oddfoot \overfullrule 3pt
        \let\label=\draftlabel
        \let\marginnote=\draftmarginnote
   \def\@eqnnum{(\theequation)\rlap{\kern\marginparsep\tt\@eqnlabel}%
\global\let\@eqnlabel\@vacuum}  }

\def\numberbysection{\@addtoreset{equation}{section}
        \def\theequation{\thesection.\arabic{equation}}}

\def\underline#1{\relax\ifmmode\@@underline#1\else
        $\@@underline{\hbox{#1}}$\relax\fi}

\def\titlepage{\@restonecolfalse\if@twocolumn\@restonecoltrue\onecolumn
     \else \newpage \fi \thispagestyle{empty}\c@page\z@
        \def\thefootnote{\fnsymbol{footnote}} }

\def\endtitlepage{\if@restonecol\twocolumn \else  \fi
        \def\thefootnote{\arabic{footnote}}
        \setcounter{footnote}{0}}  %\c@footnote\z@ }
\numberbysection \hybrid

\newcounter{mo}

\newcommand{\tr}{{\rm tr}}
\newcommand{\ti}[1]{\tilde{#1}}

\newcommand{\la}{\lambda}

\newcommand{\al}{\alpha}
\newcommand{\be}{\beta}
\newcommand{\ga}{\gamma}
\newcommand{\om}{\omega}
\newcommand{\vth}{\vartheta}

\newcommand{\Mat}{ {\rm Mat}(N,\mathbb C) }

\newcommand{\mC}{\mathbb C}

\newcommand{\z}{{\zeta}}

\newtheorem{theor}{Theorem}%[section]
\newtheorem{predl}{Proposition}%[section]
\newtheorem{example}{Example}%[section]
\newtheorem{lemma}{Lemma}%[section]

\def\beq{\begin{equation}}
\def\eq{\end{equation}}
\def\p{\partial}

\def\res{\mathop{\hbox{Res}}\limits}

\begin{document}

\setcounter{page}{1}

\

\vspace{-5mm}

%\begin{flushright}
% ITEP-TH-??/21\\
%\end{flushright}
%\vspace{0mm}

\begin{center}
%\vspace{10mm}
%
 {\Large{ Dualities in quantum integrable many-body systems   }}
 \\ \ \\
  {\Large{ and integrable probabilities - I}}
%  {\Large{ versus QQ dualities in integrable systems}}
\vspace{20mm}

{\large  {A. Gorsky}}\,\footnote{Institute for Information Transmission Problems RAS, 127051 Moscow, Russia;
Moscow Institute for Physics and Technology,Dolgoprudny 141700, Russia;
E-mail: shuragor@mail.ru
}
 \quad\quad\quad
{\large  {M. Vasilyev}}\,\footnote{
Skolkovo Institute of Science and Technology,
143026, Moscow, Russia;
Steklov Mathematical Institute of
Russian Academy of Sciences, 8 Gubkina St., Moscow 119991, Russia;
Department of Mathematics, NRU HSE, Moscow, Russia;
E-mail: mikhail.vasilyev@phystech.edu
}
 \quad\quad\quad
{\large   {A. Zotov}}\,\footnote{Steklov Mathematical Institute of Russian Academy of Sciences, 8 Gubkina St., Moscow 119991, Russia; National Research University Higher School of Economics, Usacheva str. 6,  Moscow, 119048, Russia;
 E-mail: zotov@mi-ras.ru
 }
\end{center}

\vspace{10mm}

\begin{abstract}
In this study we map the dualities observed in the framework of  integrable probabilities
into the dualities familiar in a realm of integrable many-body systems.
The dualities between  the pairs of
stochastic processes involve one representative from Macdonald-Schur  family, while the second representative is  from  stochastic higher spin six-vertex model of TASEP  family. We argue that these dualities are counterparts and  generalizations of the familiar quantum-quantum
(QQ) dualities between  pairs of integrable systems. One integrable system from QQ dual pair  belongs to the family
of inhomogeneous XXZ spin chains, while the second to the Calogero-Moser-Ruijsenaars-Schneider (CM-RS) family. The wave functions of the Hamiltonian
system from CM-RS family are known to be related to solutions to (q)KZ equations
at the inhomogeneous spin chain side. When the wave function gets substituted
by the measure, bilinear in wave functions,  a similar correspondence holds true. As an example, we have elaborated in some details a new
duality between the discrete-time  inhomogeneous multispecies TASEP model on the circle and the quantum Goldfish model from the RS family. We present the precise map of the inhomogeneous multispecies
TASEP  and 5-vertex model to the trigonometric and  rational Goldfish models respectively, where
the  TASEP local jump rates get identified as the coordinates in the Goldfish
model. Some  comments
concerning the relation of dualities in the stochastic processes with the
dualities in SUSY gauge models with surface operators included are made.
\end{abstract}

\bigskip

\newpage

{\small
\tableofcontents
}

%\newpage

%%%%%%%%%%%%%%%%%%%%%%%%%%%%%%%%%%%%%%%%%%%%%%%%%%%%%%%%%%%%%%%%%%%%%%%%%%%%%%%%%%%%%%%%%%%%%%%%%%%%%%
%%%%%%%%%%%%%%%%%%%%%%%%%%%%%%%%%%%%%%%%%%%%%%%%%%%%%%%%%%%%%%%%%%%%%%%%%%%%%%%%%%%%%%%%%%%%%%%%%%%%%%

%\section{Introduction: brief review and summary}
\section{Introduction}
\setcounter{equation}{0}
Integrable probabilities is the rapidly developing research area at the border
of a non-equilibrium statistical physics and representation theory of
affine groups.  The probabilities in stochastic process  are treated within
the conventional tools familiar in a realm of  integrable models
like transfer matrices and Bethe ansatz equations. The most developed
stochastic models belong to the ASEP-SEP-TASEP (1+1)-dimensional family with some possible
modifications. The stochastic higher spin six-vertex model
is at the top of this family if we restrict ourselves by the trigonometric
level. More recently, another family of the stochastic processes
has been introduced  involving the Schur process \cite{okoun1} with the Macdonald process at the top \cite{borodin11}.
The degenerations of the Macdonald process also involve the Whittaker, Hall-Littlewood and Jack processes. The term Macdonald process indicates that  the probability involves the product of  the
ordinary and skew Macdonald polynomials. The
Macdonald process can be considered as the stochastic process in (2+1) dimensions.
The reviews on this subject from the statistical physics perspective
can be found in \cite{spohn2,asep,rag}, while from a representation theory
perspective in \cite{cor12,bprev}.

The Macdonald polynomials $P_{\vec{\lambda}}(\vec{x}|q,t)$ are  two-parametric wave  functions of the trigonometric
RS model, hence we can expect a natural relation between the Macdonald stochastic process and
the RS model. On the other hand, the dynamics in the ASEP family is Markovian and the operators of Markovian evolution are related to the Hamiltonians of the
XXZ and XXX spin chains. Hence, we have a natural relation of the stochastic dynamics
in the ASEP family with the spin chains from XXZ family. From a world of
integrable systems we know that they enjoy at least two types of dualities:
the so-called quantum-classical (QC) duality which can be lifted to the
quantum-quantum(QQ) duality and the spectral duality. At the level of many-body systems the latter turns into Ruijsenaars duality also known as p-q
duality or  action-coordinate duality or bispectral problem. The QC and QQ dualities map the
data from representative in Macdonald family to the data from representative in
XXZ family while the spectral duality
maps the data within the representatives inside the families
at the spin chain  \cite{adams, mmzz,mukhin}  and CM-RS sides \cite{p-q,fgnr,gr}.

According to QC duality at the level of Hamiltonian systems  the Bethe ansatz
equations in spin chains get mapped into the equations for intersection
of two Lagrangian submanifolds at the
CM-RS side. The inhomogeneities at the spin chain side become the coordinates (positions of particles),
non-local spin chain Hamiltonians become momenta, while
twists get mapped into the eigenvalues of the Lax operators,
which provide the classical conservation laws in CM-RS model. Two
intersecting Lagrangian manifolds in the phase space  correspond to submanifold
of fixed coordinates and submanifold of fixed Hamiltonians. The
solutions to Bethe ansatz
equations yield the momenta provided the coordinates and Hamiltonians
to be fixed. The Yang-Yang
function at the spin side becomes the generating function for the canonical
transformations in the dual system. The coupling constant at RS side gets identified with the Planck
constant $\hbar$ at the spin side. At the QQ level  the Planck
constant at CM-RS side $\hbar_{RS}$ gets mapped into the coefficient in front of derivative
term in KZ and qKZ equations at the spin chain side.

The QQ duality generalizing QC duality  relates the wave functions from the CM-RS family of integrable many-body systems
with the solutions to the KZ and qKZ equations for the family of inhomogeneous
spin chains, like inhomogeneous XXX, XXZ spin chains,
supersymmetric generalizations and their limits known as Gaudin models \cite{M1,Ch1,Ch2,ZZ1}.
The relation between wave function and solutions to KZ and qKZ involves
the so-called Matsuo-Cherednik projection, which selects some particular
symmetric component from the solutions to KZ and qKZ equations.
Such type of relations
between integrable systems was studied for a long time from different
perspectives \cite{TsuboiZZ,AKLTZ,Z,GZZ,KV,GK}.

The algebraic counterpart of the QQ duality goes back to the works of Givental
with collaborators, where it was shown that  open  quantum Toda chain
wave functions are related to the peculiar spin chain-like model attributed
to  cohomologies of the flag manifolds \cite{givental,GK}. Similarly the
quantum relativistic open Toda chain is related to the K-theory on
the flag manifolds \cite{gl}. This algebraic correspondence has been  recently generalized
to the relation between the quantum CM and RS models  and
cohomologies and K-theory on the
cotangent bundles to the flag manifolds \cite{Koroteev:2018azn,koroteev17} respectively.
Since the cohomologies of the flag
manifolds are closely related to  vacua of the SUSY YM theory with defects
the QQ duality between two families of the integrable models has been
identified as the particular transformation of the brane configurations
representing the corresponding SUSY YM theories
on their worldvolumes \cite{GaK,BG}.

The nontrivial relations between two
families of integrable discrete statistical stochastic models have been found recently.
They were recognized at the level of "wave functions", measures and
the processes. At the level of wave function the non-symmetric
Macdonald polynomials were identified as the partition function
of the statistical inhomogeneous higher spin model of lattice  paths
on the cylinder \cite{BW19} with nontrivial boundary conditions.
Such partition function obeys the qKZ equation as a function of
inhomogeneities, which are arguments (specialization) of the
Macdonald polynomial.

The duality relations have been identified for the measure
as well. In this case
it was found that expectation values
of  particular observables evaluated with Hall-Littlewood
measure from Macdonald family emerging in $q=0$ limit
of Macdonald measure coincide with the expectation
values of the height functions in inhomogeneous higher spin six-vertex stochastic model from another
family  \cite{borodin}. At both sides of the correspondence the expectation values are evaluated
with the corresponding measures on partitions. The inhomogeneities
in six-vertex model get identified with the arguments of the
Hall-Littlewood polynomials as expected.

There are also two examples
of dualities  for  corresponding observables
evaluated for  processes, where the measure  is defined on the arrays of partitions. The first example involves the observables
evaluated for Hall-Littlewood
process at one side and
observables evaluated with the measure on degeneration of the inhomogeneous higher spin
six-vertex model at another side \cite{BBW}.
Another dual pair concerns
the inhomogeneous q-TASEP models with varying jump rates and the
q-Whittaker process from Macdonald family \cite{orr}.
The local jump rates in q-TASEP turn
out to be the arguments of the q-Whittaker functions \cite{orr} which are
the wave functions of the relativistic Toda chain. However the
origin of these observed dualities has not been clarified.

According to QQ duality we have the correspondence
$$ \Psi_{\lambda}(x) \Longleftrightarrow solution\ to\  KZ(qKZ)$$
between the wave functions $\Psi_{\lambda}(x)$ from the CM-RS family and solution
to the KZ(qKZ) equation supplemented with the Matsuo-Cherednik projection
at the inhomogeneous spin chain side. We will argue that the representation
of the Macdonald polynomial in terms of partition function of
stochastic model on a cylinder is the counterpart of QQ duality  in the
world of integrable probabilities.  The realization involves the partition
function of the colored $SU(N)$ paths on the cylinder with
nontrivial boundary conditions \cite{cantini,garbali,BW19}. It is
this realization which plays the important role in our study.

Within this realization the solution to qKZ equations emerges
and yields  the matrix product representations for the steady
state probabilities in the inhomogeneous ASEP model. The solutions
to qKZ yield the non-symmetric Macdonald polynomials and the
additional averaging of the non-symmetric polynomials
amounts to the symmetric Macdonald polynomials. This is exact
counterpart of the Matsuo-Cherednik projection \cite{M1,Ch2},
which yields the Macdonald polynomials and their degenerations
from solutions to qKZ.

We shall comment how the generalization of QQ  correspondence for
the wave function is extended to
expectation values of particular observables evaluated
with bilinear measure
$$ \Psi^{+}_{\lambda}(x)\Psi_{\lambda}(y)\,.$$
This measure on the spectrum at the Macdonald side and the corresponding measure at the inhomogeneous six-vertex higher spin  model
can be interpreted in the conventional quantum mechanical
setting.
We postpone the interpretation of duality for the processes
which involves more complicated measure
$$\Psi_{\lambda_1}(x_1)\Psi_{{\lambda_2/\lambda_1}} (x_2) \dots
\Psi_{{\lambda_N /\lambda_{N-1}}}(x_{N-1})\Psi_{\lambda_N}(x_N)
$$
in terms of integrable many-body system for a separate publication.
Note that "skew wave functions" $\Psi_{{\lambda_N /\lambda_{N-1}}}$
correspond to the matrix elements
of some operator $<\lambda_N|A(x_i)|\lambda_{N-1}>$.

The relation between the inhomogeneous multispecies TASEP model and some
peculiar limit of Macdonald process, which we identify
as the Goldfish process will be our main example.
The wave function of  dual
quantum Goldfish model defines the probability
of the corresponding stochastic process. These two processes can
be considered as  limit of trigonometric RS model
which corresponds to the Macdonald process and limit from the inhomogeneous
XXZ spin chain which is it's QQ dual. This limit is interesting by itself
since the QQ dual of Goldfish model at the spin chain side was unknown
so far even without the reference to the integrable probabilities.
As usual the inhomogeneous jump rates in TASEP model will
be identified with the coordinates in the Goldfish models.
The positions of the TASEP particles $x_i(t)$
get mapped into the spectral parameters of the Goldfish
wave functions $\Psi_{\vec{\lambda}}(\vec{z})$ $x_i \Longleftrightarrow \lambda_i$
which is natural since $x_i$ variables correspond to the boundary of Gelfand-Tsetlin
pattern providing the geometry of Macdonald process.

Our key findings are as follows
\begin{itemize}

\item
We identify at the level of wave functions the QQ duality between the representatives from the inhomogeneous higher rank spin chain
family and from  the CM-RS family in the framework of  integrable probabilities.
The analogue of QC duality for the stochastic processes
is formulated.
\item
The new duality between the inhomogeneous periodic multi-species TASEP model
and Goldfish process has been formulated and investigated in details.
It can be considered as the special interesting limit of duality between the Macdonald process and inhomogeneous periodic multi-species ASEP model.
%The wave functions of inhomogeneous TASEP model are related to the Grothendieck %polynomials
%$G^{\beta}_{\vec{\lambda}}(\vec{x})$ evaluated at $\beta=-1$ which are the %K-theoretic(relativistic)
%generalization of Schur polynomials. On the other hand there is some  %connection \cite{knizel} between
%inhomogeneous TASEP and the Schur process which involves the Schur polynomials
%that is $\beta=0$ limit of Grothendieck(?) polynomials. How the argument from %\cite{knizel} fits with our finding?

\end{itemize}

One could wonder if there is any deep reason behind a representation of a
quantum mechanical system in terms of an auxiliary statistical stochastic model.
We shall conjecture that such representation can be considered as dual Feynman
path integral in the following sense. Recall that a wave function $\Psi_E(x)$
has the Feynman representation as the sum over the paths in the coordinate space ending at point $x$, where
the energy $E$ enters the measure in summation over  paths. Let us look
at a wave function through the different glasses and consider $\Psi_x(E)$.
Our conjecture is that such viewpoint corresponds to the consideration
of the ensemble of paths in the Hilbert space which end up at state with energy $E$
while the coordinate $x$ enters the measure of the dual Feynman path integral. Not much is
known concerning  such representation of a wave function as
a partition function of some statistical model. Note however
the representation of the oscillator wave function in terms of ensemble of paths
on supertrees \cite{gnv}, representation of the wave function via matrix models \cite{krefl}
or the QDE/(statistical model) correspondence when the solution to the Baxter
equation in the statistical model is related to the spectral determinant of
the particular quantum mechanical model \cite{dorey}. We shall try to argue that representation
of the Macdonald polynomial in terms of the ensemble of colored lattice paths on the cylinder can be considered as the dual Feynman path integral.

The paper is organized as follows. In Section 2 we briefly review
dualities between different integrable systems.
In Section 3
the definitions of the two families of stochastic processes
and the relation of Markovian evolution at the ASEP-TASEP
side with the XXZ spin chain and its degenerations are
presented.  In Section 4 we present the identification
of QQ  dualities in integrable many-body systems as
dualities in integrable probabilities. Classical Goldfish
model is reviewed in Section 5 while general facts concerning
5-vertex model are presented in Section 6. Section 7 and
Section 8 are devoted to the QC and QQ dualities between
Goldfish and multi-species inhomogeneous TASEP model.
The Inozemtsev limit at the level of KZ equation for open Toda system
which is spectrally dual to Goldfish model
is formulated in Section 9
as the limit from affine Hecke algebra  to affine Nil-Hecke
algebra. The results of the study are summarized and some open questions are
formulated in Conclusion. In the Appendix A we present the complete
classification of $R$-matrices for the 5-vertex models, and
in the Appendix B we present  the relations between
the wave functions of many-body systems and correlators of
height functions in  polymer representations.
Appendix C is devoted to the toy example
of a representation of the oscillator wave function
in terms of the "dual Feynman path integral" in the
Hilbert space. In Appendix D we
consider the relation of our findings with  some aspects
of SUSY YM theories supplemented with the surface operators
in different dimensions.

%%%%%%%%%%%%%%%%%%%%%%%%%%%%%%%%%%%%%%%%%%%%%%%%%%%%%%%%%%%%%%%%%%%%%%%%%%%%%%%%%%%%%%%%%%%%%%%%%%
%%%%%%%%%%%%%%%%%%%%%%%%%%%%%%%%%%%%%%%%%%%%%%%%%%%%%%%%%%%%%%%%%%%%%%%%%%%%%%%%%%%%%%%%%%%%%%%%%%
%%%%%%%%%%%%%%%%%%%%%%%%%%%%%%%%%%%%%%%%%%%%%%%%%%%%%%%%%%%%%%%%%%%%%%%%%%%%%%%%%%%%%%%%%%%%%%%%%%
\section{Dualities in integrable systems}

Here we briefly review a number of dualities in integrable systems and interrelations between them. This review
of course does not cover all possible dualities, but only the most known, which are related to our study. Two main phenomena are the spectral duality and the Matsuo-Cherednik projection in QQ duality. The first one can be lifted to the level of Gaudin models and/or spin chains as well as (q)KZ equations. Different approaches and formulations give rise to interrelations, which we describe below. The Matsuo-Cherednik projection (from (q)KZ equations to quantum-many body problems) is well defined in semi-classical limit. Being formulated through Lax representation it provides the QC duality, which is also important in our study.

%%%%%%%%%%%%%%%%%%%%%%%%%%%%%%%%%%%%%%%%%%%%%%%%%%%%%%%%%%%%%%%%%%%%%%%%%%%%%%%%%%%%%%%%%%%%%%%%%%
\subsection{Many-body systems and Ruijsenaars dualities}
Here we discuss integrable many-body systems of the Calogero-Moser-Sutherland \cite{Calogero0} and the Ruijsenaars-Schneider \cite{RS} families. These are the Liouville integrable models of classical Hamiltonian mechanics. For $N$-body
system the phase space coordinates are given by canonical momenta $p_1,...,p_N$ and positions of particles $q_1,...,q_N$ on a complex plane
(we do not discuss real reductions here). Although these models can be constructed for an arbitrary root system, in this paper
we deal with $A_{N-1}$ type models only. That is, the Hamiltonians depend on the differences of particles positions.

\subsubsection*{\underline{Classical many-body systems}}
For the Calogero type models the first nontrivial Hamiltonian is  of the form:
\beq\label{y1}
\displaystyle{
 H=\sum\limits_{k=1}^N\frac{p_k^2}{2}-\nu^2\sum\limits^N_{i<j}U(q_i-q_j)\,,
 }
\eq
where $\nu\in\mC$ is a coupling constant. The potential of the pairwise interaction has the form $U(x)=1/x^2$ for the rational models, and $U(x)=1/\sinh^2(x)$ for trigonometric
 models\footnote{Trigonometric $U(x)=1/\sin^2(x)$ and hyperbolic $U(x)=1/\sinh^2(x)$ cases are indistinguishable in complex variables. This becomes important in the case of real variables.}.

 In contrast to (\ref{y1}) the Hamiltonians  of the Ruijsenaars-Schneider family have trigonometric (exponential) type dependence on momenta. The first nontrivial Hamiltonian is defined as
\beq\label{y2}
\displaystyle{
 H=\sum\limits_{i=1}^N e^{p_i/c}\prod\limits_{j:j\neq i}^N\frac{h(q_i-q_j+\eta)}{h(q_i-q_j)}\,,\quad \eta,c\in\mC\,,
 }
\eq
where $h(x)=\sinh(x)$ for the trigonometric model and $h(x)=x$ for the rational model.
 This Hamiltonian generalizes (\ref{y1}) in the sense that (\ref{y1}) is reproduced from (\ref{y2}) in the limit
 $c\rightarrow \infty$ together with the substitution $\eta=\nu/c$. In this paper we will also consider a different limit, when $\eta\rightarrow \infty$ with $c=1$. This limit provides the so-called goldfish model.

 Together with the canonical Poisson brackets
\beq\label{y3}
\displaystyle{
 \{p_i,q_j\}=\delta_{ij}\,,\quad \{p_i,p_j\}=0\,,\quad \{q_i,q_j\}=0
 }
\eq
a given Hamiltonian of type (\ref{y1}) or (\ref{y2}) provides equations of motion, which are equivalently written in the Lax form , i.e. they are represented in the matrix
form
\beq\label{y401}
\displaystyle{
 {\dot L}=\{H,L\}=[L,M]\,,\quad L,M\in\Mat\,.
  }
\eq
For example, for the rational Calogero-Moser models we have the following explicit expression for the Lax matrix (without spectral parameter):
\beq\label{y4}
\displaystyle{
 L_{ij}=\delta_{ij}p_i+\nu\frac{1-\delta_{ij}}{q_i-q_j}\,,
  }
\eq
For the rational Ruijsenaars-Schneider it is as follows:
\beq\label{y5}
\displaystyle{
 L_{ij}=\frac{\eta{\dot q}_j}{q_i-q_j+\eta}\,,\quad {\dot q}_j=e^{p_j/c}\frac{1}{c}\prod\limits_{k:k\neq j}^N\frac{q_j-q_k+\eta}{q_j-q_k}\,.
  }
\eq
The accompany $M$-matrices are known as well. These matrices depend on a choice of a flow (Hamiltonian).

It follows from the Lax equations that $H_k=(1/k)\tr(L^k)$ are conserved quantities for any $k$. For example,
 the Hamiltonian (\ref{y1}) is $H_2$ for the Lax matrix (\ref{y4}), and the Hamiltonian (\ref{y2}) is
 $H_1$ for the Lax matrix (\ref{y5}). Commutativity of Hamiltonians $\{H_k,H_m\}=0$ for each of the models
 with respect to the
  Poisson structure (\ref{y3}) follows from
 existence of the classical $r$-matrix structure.

%In trigonometric and rational cases

\subsubsection*{\underline{Dualities in classical many-body systems}}
Consider the rational Calogero-Moser model (\ref{y1}) and the eigenvalue problem for its Lax matrix (\ref{y4}):
\beq\label{y6}
\displaystyle{
 L\Psi=\Psi\Lambda\,,\quad \Lambda={\rm diag}(\lambda_1,...,\lambda_N)\,,
  }
\eq
where $\Psi\in{\rm Mat}_N$ is the matrix of eigenvectors.
The eigenvalues $\lambda_1,...,\lambda_N$ are the action variables. Indeed, the Hamiltonians $H_k$ are power sums
$H_k=(1/k)\sum_{i=1}^N\la_i^k$, and $\{\la_i,\la_j\}=0$ due to  $\{H_k,H_m\}=0$.
It is easy to see from its explicit form (\ref{y4}) that the Lax matrix satisfies equation
\beq\label{y7}
\displaystyle{
 [Q,L]={\mathcal O}\,,\quad {\mathcal O}_{ij}=\nu(1-\delta_{ij})\,.
  }
\eq
%
% $q_i L_{ij}-L_{ij}q_j={\mathcal O}_{ij}$, where ${\mathcal O}_{ij}=\nu(1-\delta_{ij})$.
In the group-theoretical approach to integrable systems \cite{OP}
it is treated as the moment map equation with respect to coadjoint action of ${\rm GL}_N$ Lie group on the space
$T^*{\rm gl}_N$ parameterized by a pair of matrix-valued variables $A,B\in\Mat$. Namely, the moment map is
 $\mu=[A,B]\in{\rm gl}_N^*$, and the equation which fixes its level is
\beq\label{y8}
\displaystyle{
 [A,B]=\mu_0\,,
  }
\eq
where $\mu_0$ is (chosen to be) an element conjugated to  ${\mathcal O}$. The common coadjoint action (i.e. by conjugation) of the group ${\rm GL}_N$ on both sides of (\ref{y8}) can be fixed in two natural ways.
The first one is to make $A$ diagonal and solve (\ref{y8}) with respect to $B$. This leads to equation (\ref{y7}), so that
$A=Q$ and $B=L$. Another possibility is to diagonalize $B$ and solve (\ref{y8}) with respect to $A$. Since $B$ is conjugated to $L$, it turns into $\Lambda$.
Then we come to
equation
\beq\label{y9}
\displaystyle{
 [{\ti L},\Lambda]={\mathcal O}\,,
  }
\eq
so that
\beq\label{y10}
\displaystyle{
 {\ti L}_{ij}=\delta_{ij}{\ti p}_i-\nu\frac{1-\delta_{ij}}{\la_i-\la_j}\,,
  }
\eq
This is also the Lax matrix of the Calogero-Moser model but in different coordinates (and with the different sign of the coupling constant).
It is called dual to (\ref{y4}). The positions of particles are given by the action variables in the initial model,
and the action variables in the dual model are positions of particles $q_1,...,q_N$ (since $A$ is conjugated to $Q$):
\beq\label{y11}
\displaystyle{
 \Psi {\ti L}=Q\Psi\,,\quad Q={\rm diag}(q_1,...,q_N)\,,
  }
\eq
Thus, the duality relates two integrable systems by (anti)canonical map, which interchanges positions of particles and action variables. This relation was first observed and described by S.N.M. Ruijsenaars  \cite{p-q} for integrable many-body systems under consideration. The Ruijsenaars duality is also often called
the p-q duality or action-angle duality. The above example shows that the rational Calogero-Moser model is self-dual in the sense that both dual models are of the same type (although they are different).

The next dual pair of integrable systems is given by the trigonometric Calogero-Moser-Sutherland  and the rational Ruijsenaars-Schneider models. Similarly to (\ref{y8}) they can be obtained by the Hamiltonian reduction. This time the upper phase space is the cotangent bundle $T^*{\rm GL}_N$ to the Lie group ${\rm GL}_N$. The moment map equation takes the form
\beq\label{y12}
\displaystyle{
 A-B^{-1}AB=\mu_0\,.
  }
\eq
Again, by diagonalizing either $A$ or $B$ and solving (\ref{y12}) with respect to $B$ or $A$, one gets
the Lax matrix of the rational Ruijsenaars-Schneider or the trigonometric Calogero-Moser models respectively.

Finally, the trigonometric Ruijsenaars-Schneider model is self-dual similarly to the rational Calogero-Moser model.
It is achieved by the moment map equation arising from the Poisson reduction performed for the Heisenberg double:
\beq\label{y13}
\displaystyle{
 A^{-1}B^{-1}AB=\exp(\mu_0)\,.
  }
\eq

Details  of the dualities are given in \cite{p-q}, and the group-theoretical approach can be found in
\cite{fgnr,Arut}.
To summarize we have the following table of Ruijsenaars dualities:

\beq\label{y14}
\begin{array}{|c|c|c|}
\hline
p\diagdown q & rational & trigonometric
\\
\hline
rat. & \hbox{rational CM} & \hbox{trig. CM}
\\
 & \hbox{(self-dual)}\circlearrowleft & \!\nearrow\hfill
\\
\hline
trig. & \hfill\swarrow\! & \hbox{trig. RS}
\\
 & \hbox{rational RS} & \hbox{(self-dual)}\circlearrowleft
\\
\hline
\end{array}
\eq

Here in the horizontal line we put the type of dependence (of the Lax matrix) on particles coordinates,
and  in the vertical line --   the type of dependence on momenta. Notice that the duality transformation
interchanges these types.

Let us also remark that CM and RS models are naturally extended to elliptic dependence on coordinates. For this purpose
one may choose $U(x)=\wp(x)$ -- the Weierstrass $\wp$-function in (\ref{y1}), and $h(x)=\vth(x)$ -- the theta-function in (\ref{y2}). The dual to elliptic models and the double-elliptic (self-dual) one are studied starting from \cite{fgnr,MM}.
So that the table (\ref{y14}) is a $2\times 2$ block from $3\times 3$ larger table, which includes elliptic types (see \cite{MM}).
Another remark is that all models we discuss here are considered in the case of complex-valued variables. Real reductions provide additional structures, which we do not discuss, see \cite{gr,Feher2}.

\subsubsection*{\underline{Dualities in quantum many-body systems}}

In quantum case we study the stationary Schrodinger equations for the quantized Hamiltonians ${\hat H}\Psi_E=E\Psi_E$.
The quantum version of the Ruijsenaars duality is known as the bispectral problem \cite{grun}.
Consider, for example, the pair of dual models -- the rational Ruijsenaars-Schneider and the trigonometric Calogero-Sutherland models.
 The quantum Hamiltonian of the trigonometric Calogero-Moser model (\ref{y1}) is as follows:
\beq\label{y15}
\displaystyle{
 {\hat H}^{CM}=\frac{\hbar^2}{2}\sum\limits_{k=1}^N\p_{q_i}^2-
 \sum\limits^N_{i<j}\frac{\nu(\nu+\hbar)\om^2}{\sinh^2(\om(q_i-q_j))}\,,
 }
\eq
and the quantum Hamiltonian of the rational Ruijsenaars-Schneider model (\ref{y2}) with coordinates $k_1,...,k_N$ is of the form:
\beq\label{y16}
\displaystyle{
 {\hat H}^{RS}=\sum\limits_{i=1}^N \prod\limits_{j:j\neq i}^N\frac{k_i-k_j+\eta}{k_i-k_j}\,e^{2\om\hbar\p_{k_i}}\,,\quad
 \eta=-2\nu\om\,.
 }
\eq
There are different classes of solutions of quantum problems.
See e.g. \cite{Arut,KhKh} and references therein.
Here we mention that in the class of symmetric functions
the eigenvalue problem for the quantum trigonometric Calogero-Moser model is solved by Jack polynomials, and for the trigonometric Ruijsenaars-Schneider model --  by Macdonald polynomials.

\paragraph{The spectral duality} \cite{grun} means that there exists a common set of eigenfunctions $\Psi_{\vec k}(\vec q)$ for the operators
(\ref{y15})-(\ref{y16}) with the following eigenvalues:
\beq\label{y17}
\begin{array}{c}
\displaystyle{
 {\hat H}^{CM}\Psi_{\vec k}(\vec q)=E\Psi_{\vec k}(\vec q)\,,\quad E=\sum\limits_{i=1}^N\frac{k_i^2}{2}\,,
 }
 \\
 {\hat H}^{RS}\Psi_{\vec k}(\vec q)={\ti E}\Psi_{\vec k}(\vec q)\,,\quad {\ti E}=\sum\limits_{i=1}^N e^{2\om q_i}\,,
 \end{array}
\eq
so that the coordinates are interchanged with spectral parameters (variables) in the wave function for dual models.
For instance, the wave functions of the self-dual trigonometric RS model
are Macdonald functions,  which obey the same
equations with respect to $\vec{q}$ and $\vec{k}$. This reflects its self-duality
at the quantum level and other dual pair can be derived by degeneration
of the Macdonald functions by taking particular limits in its parameters.
In this paper we will discuss some limiting cases. For example, a certain (Inozemtsev) limit provides the open Toda chain from (\ref{y15}). In this way one gets the bispectral problem for Whittaker functions  \cite{Babelon,glo1,SklToda}.

The set of duality relations is described by the classical diagram (\ref{y14}). The eigenvalue problems (\ref{y16}) are naturally extended to all higher Hamiltonians.
%See details in \cite{grun}.

%%%%%%%%%%%%%%%%%%%%%%%%%%%%%%%%%%%%%%%%%%%%%%%%%%%%%%%%%%%%%%%%%%%%%%%%%%%%%%%%%%%%%%%%%%%%%%%%%%
\subsection{Gaudin models, spin chains and spectral dualities}

In contrast to many-body systems, for the Gaudin models (and spin chains) we use the Lax representations with  spectral parameter\footnote{It is a local coordinate on a complex curve $\Sigma$ (Riemann sphere, cylinder or elliptic curve for the rational, trigonometric and elliptic models respectively.  } $z$. This variable is non-dynamical. The Lax equations and other relations hold true identically in $z$ \cite{FT,Skl1}.

\subsubsection*{\underline{Classical Gaudin models and spin chains}}

\paragraph{Classical Gaudin model.} In classical mechanics the rational Gaudin model is given by the following Lax matrix:
  \beq\label{y18}
  \begin{array}{c}
    \displaystyle{
 L(z)=\Lambda+\sum\limits_{k=1}^N\frac{S^k}{z-z_k}\in{\rm Mat}_n\,,\quad S^k=\sum\limits_{a,b=1}^n E_{ab}S^k_{ab},
 }
 \end{array}
 \eq
 where $\Lambda={\rm diag}(\la_1,...,\la_n)$ is a constant diagonal matrix,
  and $S^k\in {\rm Mat}_n$ are residues at marked points (punctures), which are dynamical variables of the model. %coordinates on the phase space
   The Poisson structure is a direct sum of Poisson-Lie brackets:
  \beq\label{y19}
  \begin{array}{c}
    \displaystyle{
 \{S^i_{ab},S^j_{cd}\}=\delta^{ij}(S^i_{ad}\delta_{cb}-S^i_{cb}\delta_{ad})\,,\ \ i,j=1,...,N;\ \ a,b,c,d=1,...,n\,.
 }
 \end{array}
 \eq
The Casimir functions are the set of eigenvalues of $S^k$, so that
the phase space is a direct product of coadjoint orbits related to $S^k$ (i.e. the set of $S^k$ with some fixed eigenvalues). The Poisson structure can be written in $r$-matrix form:
  \beq\label{y20}
  \begin{array}{c}
    \displaystyle{
 \{L_1(z),L_2(w)\}=[r_{12}(z-w),L_1(z)+L_2(w)]\,,
 }
 \end{array}
 \eq
which is linear in $L$. The Gaudin Hamiltonians appear from calculating $\tr(L(z)^2)$, which is now
a generating function (in $z$) of conserved quantities:
  \beq\label{y21}
  \begin{array}{c}
    \displaystyle{
 \frac{1}{2}\tr(L^2(z))=\frac{1}{2}\sum\limits_{k=1}^N \frac{\tr\Big((S^i)^2\Big)}{(z-z_i)^2}
 +\sum\limits_{k=1}^N \frac{H_i}{z-z_i}\,,
 }
 \end{array}
 \eq
where
  \beq\label{y22}
  \begin{array}{c}
    \displaystyle{
 H_i=\tr(S^i\Lambda)+\sum\limits_{k:k\neq i}^N \frac{\tr(S^iS^k)}{z_i-z_k}\,,\quad i=1,...,L\,.
 }
 \end{array}
 \eq
These Hamiltonians commute with respect to the Poisson structure (\ref{y19}) due to existence of the classical $r$-matrix satisfying the classical exchange relations (\ref{y20}). In the rational case it is
$r_{12}(z)=P_{12}/z$, where $P_{12}=\sum_{ab}E_{ab}\otimes E_{ba}\in{\rm Mat}_n^{\otimes 2}$ is the matrix permutation operator.
In order to include into consideration trigonometric (and elliptic) case let us write the Lax matrix
(\ref{y18}) as follows:
  \beq\label{y23}
  \begin{array}{c}
    \displaystyle{
 L(z)=\Lambda+\sum\limits_{k=1}^N \tr_2(r_{12}(z-z_k)S^k_2)\in{\rm Mat}_n\,,
 }
 \end{array}
 \eq
where $\Lambda$ is a symmetry of $r$-matrix: $[\Lambda_1+\Lambda_2,r_{12}(z)]=0$. In fact, one can
substitute to (\ref{y23}) any skew-symmetric $r_{12}(z)=-r_{21}(-z)$ solution of the classical Yang-Baxter equation
  \beq\label{y24}
  \begin{array}{c}
    \displaystyle{
 [r_{12}(z_1\!-\!z_2),r_{23}(z_2\!-\!z_3)]+[r_{12}(z_1\!-\!z_2),r_{13}(z_1\!-\!z_3)]+
 [r_{13}(z_1\!-\!z_3),r_{23}(z_2\!-\!z_3)]=0\,,
 }
 \end{array}
 \eq
and the classical $r$-matrix structure (\ref{y20}) holds true due to (\ref{y24}). The Gaudin Hamiltonians take the form:
  \beq\label{y25}
  \begin{array}{c}
    \displaystyle{
 H_i=\tr(S^i\Lambda)+\sum\limits_{k:k\neq i}^N \tr_{12}(r_{12}(z_i-z_k)S^i_1S^k_2)\,.
 }
 \end{array}
 \eq
For example, by choosing the XXZ classical $r$-matrix one obtains the trigonometric Gaudin model.

\paragraph{Classical spin chains.}  The classical ${\rm GL}_n$ inhomogeneous twisted spin chain on $N$ sites is defined by the monodromy matrix
  \beq\label{y26}
  \begin{array}{c}
    \displaystyle{
 T(z)=VL^N(z-z_N)...L^1(z-z_1)\in {\rm Mat}_n\,,
 }
 \end{array}
 \eq
where $V={\rm diag}(V_1,...,V_n)$ is the twist matrix, and $z_1,...,z_N$ are the inhomogeneities. The Lax matrices
$L^k(z)$ (assigned to $k$-th site) depend on the dynamical variables $S^k\in{\rm Mat}_n$. They can be defined as follows:
  \beq\label{y27}
  \begin{array}{c}
    \displaystyle{
L^k(z)=1_n\frac{\tr(S^k)}{n}+\eta\,\tr_2(r_{12}(z)S^k_2)\,,
 }
 \end{array}
 \eq
where $\eta$ is a constant parameter, $1_n$ is the identity matrix and the classical $r$ matrix is assumed to be ${\rm sl}_n$-valued. For example, in the rational XXX case
  \beq\label{y271}
  \begin{array}{c}
    \displaystyle{
L^k(z-z_k)=1_n\frac{\tr(S^k)}{n}+\eta\,\frac{S^k}{z-z_k}\,.
 }
 \end{array}
 \eq
The Poisson structure of the model is given by the quadratic classical $r$-matrix structure:
  \beq\label{y28}
  \begin{array}{c}
    \displaystyle{
 \{L^i_1(z),L^j_2(w)\}=\delta^{ij}[r_{12}(z-w),L^i_1(z)L^i_2(w)]\,.
 }
 \end{array}
 \eq
 In the general case it provides the classical Sklyanin algebra at each site. The twist matrix is assumed to be a symmetry $[V\otimes V,r_{12}(z)]=0$. Construction of the classical chain is based on the fact that $T(z)$ also satisfies (\ref{y28}):
  \beq\label{y29}
  \begin{array}{c}
    \displaystyle{
 \{T_1(z),T_2(w)\}=[r_{12}(z-w),T_1(z)T_2(w)]\,.
 }
 \end{array}
 \eq
 Similarly to (\ref{y20}) it follows from (\ref{y29}) that $\{\tr\, T^k(z),\tr\, T^m(w)\}=0$. This is why the classical
 transfer-matrix $t(z)=\tr T(z)$ is the generating function of non-local Hamiltonians:
  \beq\label{y3011}
  \begin{array}{c}
    \displaystyle{
  H_i=\res\limits_{z=z_i}t(z)\,,
 }
 \end{array}
 \eq
 which Poisson commute with respect to (\ref{y28}).

 In the limit $\eta\rightarrow 0$ together with substitution $V=1+\eta\Lambda$ the monodromy matrix $T(z)$ reproduces
 the Lax matrix of the Gaudin model (\ref{y18}).

 \paragraph{Local Hamiltonian.} The Hamiltonians (\ref{y3011}) are non-local since they include interaction between all sites. In order to have a true chain (when only neighbour sites interact) one should \cite{FT} put all
 inhomogeneity parameters equal to each other $z_k=0$ (for all $k=1,..,N$), as well as the twist parameters $V=1_n$. Also, the Casimir functions at each site are chosen to be equal to each other, so that there exists a point $z_0$, where all Lax matrices $L^k(z)$ are degenerated: $\det L^k(z_0)=0$ for all $k$. Then at this point all Lax matrices are represented in the form $L^k(z_0)=a^k\otimes b^k$, where $a^k$ is $n$-dimensional column, and $b^k$ is $n$-dimensional row. The monodromy matrix (\ref{y26}) turns into $T(z_0)=a^N\otimes b^N...a^1\otimes b^1$, and the transfer matrix is expressed through scalar products of $(b^i,a^{i-1})$, i.e.
  \beq\label{y301}
  \begin{array}{c}
    \displaystyle{
  t(z_0)=(b^1,a^N)(b^{N},a^{N-1})...(b^2,a^1)\,.
 }
 \end{array}
 \eq
 The local Hamiltonian appears by taking the logarithm
  \beq\label{y302}
  \begin{array}{c}
    \displaystyle{
 \log t(z_0)=\sum\limits_{i=1}^N\log(b^i,a^{i-1})\,,\quad a^0=a^N\,.
 }
 \end{array}
 \eq
 For example, for the rational ${\rm GL}_2$ XXX chain (using some specific symmetries) the Hamiltonian can be finally
 written in the form
  \beq\label{y303}
  \begin{array}{c}
    \displaystyle{
  H^{\rm loc}=\sum\limits_{i=1}^N \log\tr(S^i S^{i-1})\,,\quad S^0=S^N\,.
 }
 \end{array}
 \eq
 To summarize, the local interaction comes from a spin chain of special configuration and corresponds to a special choice of the Hamiltonian flow. In the continuous limit the interaction (\ref{y303}) leads to 1+1 Heisenberg (or Landau-Lifshitz) magnet.

\subsubsection*{\underline{Spectral dualities in classical models}}
Following \cite{adams} consider the special rational Gaudin model (\ref{y18}), where all $S^k$ are matrices of rank one:
  \beq\label{y30}
  \begin{array}{c}
    \displaystyle{
  S^k_{ab}=\xi^k_a\eta^k_b\,.
 }
 \end{array}
 \eq
Notice that the Poisson structure (\ref{y19}) follows from canonical Poisson brackets
$\{\xi^i_a,\eta^j_b\}=\delta^{ij}\delta_{ab}$.
Introduce the dual model. It has the Lax matrix of size $N\times N$ with $n$ simple poles at $\la_1,...,\la_n$
  \beq\label{y31}
  \begin{array}{c}
    \displaystyle{
 {\ti L}(\la)=Z+\sum\limits_{c=1}^n\frac{{\ti S}^c}{\la-\la_c}\in{\rm Mat}_n\,,
 }
 \end{array}
 \eq
where $\lambda$ is a spectral parameter, $Z={\rm diag}(z_1,...,z_N)$ and all the residues ${\ti S}^c$ are rank one matrices
  \beq\label{y32}
  \begin{array}{c}
    \displaystyle{
 {\ti S}^c_{ij}=\xi^i_c\eta^j_c\,.
 }
 \end{array}
 \eq
 Similarly to (\ref{y22}) the dual Gaudin Hamiltonians are of the form:
  \beq\label{y321}
  \begin{array}{c}
    \displaystyle{
 {\widetilde H}_a=\tr({\ti S}^aZ)+\sum\limits_{c:\,c\neq a}^n \frac{ \tr({\ti S}^a {\ti S}^c) }{\la_a-\la_c}\,.
 }
 \end{array}
 \eq

It is easy to show that the following identity holds:
  \beq\label{y33}
  \begin{array}{c}
    \displaystyle{
  \frac{\det\limits_{N\times N}(\la-L(z))}{\det\limits_{N\times N}(\la-\Lambda)}=
   \frac{\det\limits_{n\times n}(z-{\ti L}(\la))}{\det\limits_{n\times n}(z-Z)}\,.
 }
 \end{array}
 \eq
 For any integrable system with Lax matrix the expression $\det(\la-L(z))$ is the generating function of
 Hamiltonians, and the curve $\det(\la-L(z))=0$ in $\mC^2$ (with coordinates $(\la,z)$)  is called the spectral curve.
The
denominators in (\ref{y33}) are non-degenerated and non-dynamical functions. So that (\ref{y33}) means that
the spectral curves of dual models coincide although the are written in different forms.
Since the coefficients of the spectral curve Poisson commute, we conclude that $\{H_i,{\widetilde H}_a\}=0$
(the commutativity $\{H_i,H_j\}=\{{\widetilde H}_a,{\widetilde H}_b\}=0$ is by construction).
A similar phenomenon
is known for the closed Toda chain \cite{FT},
which has $2\times 2$ and $N\times N$ Lax representations.

Following \cite{mmzz}
we call such type relation as spectral duality.
A relation similar to (\ref{y33}) holds true
between the trigonometric ${\rm gl}_N$ Gaudin model with $n+2$ marked points and the XXX ${\rm GL}_n$ spin chain on $N$ sites (\ref{y271}). Two additional marked points in the trigonometric Gaudin model are $z=0$ and $z=\infty$.
The coadjoint orbits attached to these two points are generic, while the rest of orbits are of minimal dimension (those described by rank one matrices).  Finally, the XXZ spin chain turns out to be spectrally self-dual.
Details are given in \cite{mmzz}. To summarize, we obtain the table of spectral dualities:
\beq\label{y34}
\begin{array}{|c|c|c|}
\hline
\{L\!\stackrel{\otimes}{,}\!L\}_r {\bf \diagdown} r_{12} & \phantom{\Bigl|}XXX\phantom{\Bigl|} & XXZ
\\
\hline
linear & \hbox{rational Gaudin} & \hbox{trig. Gaudin}
\\
 & \hbox{(self-dual)}\circlearrowleft & \!\nearrow\hfill
\\
\hline
quadratic & \hfill\swarrow\! & \hbox{XXZ spin chain}
\\
 & \hbox{XXX spin chain} & \hbox{(self-dual)}\circlearrowleft
\\
\hline
\end{array}
\eq
In the horizontal direction we put a type of $r$-matrix (rational or trigonometric), or equivalently,
the anisotropy type (XXX or XXZ). In the vertical line the Gaudin models and spin chains are
distinguished by the type of the classical $r$-matrix structure. It is linear (\ref{y20}) or quadratic (\ref{y28})
respectively.
Notice that the table (\ref{y34}) looks similar to the one for Ruijsenaars dualities in many-body systems (\ref{y14}).

\subsubsection*{\underline{Quantum spin chains and Gaudin models}}
Quantization of Gaudin models and spin chains can be performed in different ways. The most traditional one
 is based on the quantum inverse scattering method and (algebraic) Bethe ansatz \cite{Faddeev}. Let us recall some basic ideas for spin chains since we need them in what follows.
 In quantum case the dynamical variables $S^k_{ab}$ become operators ${\hat S}^k_{ab}$ satisfying some quantum algebra (commutation relations), generated by the quantum version of the exchange relations (\ref{y28}):
  \beq\label{y35}
  \begin{array}{c}
    \displaystyle{
  R_{12}(z-w){\hat L}^i_1(z){\hat L}^i_2(w)={\hat L}^i_2(w){\hat L}^i_1(z)R_{12}(z-w)\,,
  \quad (1-\delta^{ij})[{\hat L}^i_1(z),{\hat L}^j_2(w)]=0\,,
 }
 \end{array}
 \eq
 where the Lax matrix becomes the operator-valued $n\times n$ matrix, and $R_{12}$ is a quantum $R$-matrix satisfying the quantum version of the Yang-Baxter equation (\ref{y24}):
  \beq\label{y36}
  \begin{array}{c}
    \displaystyle{
  R_{12}(z_1-z_2)R_{13}(z_1-z_3)R_{23}(z_2-z_3)=R_{23}(z_2-z_3)R_{13}(z_1-z_3)R_{12}(z_1-z_2)\,.
 }
 \end{array}
 \eq
 The last relation in (\ref{y35}) means that $[{\hat S}^i_{ab},{\hat S}^j_{cd}]=0$ for $i\neq j$. Therefore, the Hilbert space
 $\mathcal H$ is naturally represented as a tensor product of $N$ components (vector spaces)
  \beq\label{y37}
  \begin{array}{c}
    \displaystyle{
  {\mathcal H}={\rm Vect}_1\otimes...\otimes {\rm Vect}_N\,,
 }
 \end{array}
 \eq
 and a set of operators ${\hat S}^i_{ab}$ at each site (with fixed $i$) acts nontrivially on the $i$-th component only.
 The quantum version of the Lax matrix is given by the quantum $R$-matrix
  \beq\label{y38}
  \begin{array}{c}
    \displaystyle{
  {\hat L}_0^i(z-z_i)=R_{0i}(z-z_i)\,,
 }
 \end{array}
 \eq
 where $0$ is the (auxiliary) matrix space ${\rm Mat}_n$ and $i$ is the (quantum) representation space. The relation (\ref{y35}) is fulfilled due to the Yang-Baxter equation (\ref{y36}).
 For example, in the fundamental representation of ${\rm GL}_n$ (then in (\ref{y37}) ${\rm Vect}_k=\mC^n$ for all $k$)
 the Lax operator (\ref{y271}) for the XXX chain takes the following form:
  \beq\label{y381}
  \begin{array}{c}
    \displaystyle{
  R_{12}(z)=1_n\otimes 1_n +\frac{\hbar}{z}\,P_{12}\,.
 }
 \end{array}
 \eq
 where operator $P_{12}$ is the permutation operator between two spaces.
 Similarly to the classical case (\ref{y26}) the quantum monodromy matrix is defined as
 \beq \label{y39}
\displaystyle{
\hat{T}_0(z) =V_0{\hat L}_0^1(z-z_N)...{\hat L}^N_0(z-z_1) =V_0R_{01}(z-z_N)...R_{0N}(z-z_1)\,,
}
\eq
where the twist matrix $V$ is a symmetry of $R$-matrix: $[V_1V_2,R_{12}(z)]=0$.
It follows  from (\ref{y35}) that the monodromy matrix satisfies the same quantum algebra (RTT-relations)
\beq \label{y40}
\displaystyle{
R_{12}(z-w){\hat T}_1(z){\hat T}_2(w)={\hat T}_2(w){\hat T}_1(z)R_{12}(z-w)\,,
}
\eq
which is the quantum analogue of (\ref{y29}). Since $R$-matrix is non-dynamical (its elements commute with all $S^i_{ab}$) and non-degenerate, then  (\ref{y40}) provides commutativity
\beq \label{y41}
\displaystyle{
[{\hat t}(z),{\hat t}(w)]=0
}
\eq
 of the quantum transfer-matrix
 \beq \label{y42}
\displaystyle{
{\hat t}(z)=\tr\hat{T}(z) =\tr_0\Big(V_0R_{01}(z-z_N)...R_{0N}(z-z_1)\Big)
}
\eq
for different values of the spectral parameter. Therefore, the quantum Hamiltonians
 \beq \label{y43}
\displaystyle{
 {\hat H}_i=\res\limits_{z=z_i}{\hat t}(z)
}
\eq
commute as well, and we may formulate the common eigenvalue problem for these operators, i.e. to find $\Phi\in{\mathcal H}$, which solves the system of $N$ equations
 \beq \label{y44}
\displaystyle{
 {\hat H}_i\Phi=H_i\Phi\,,\ \ i=1...N
}
\eq
and its spectrum $H_k$. Assuming $\res\limits_{z=0}R(z)=\hbar P_{12}$ (this is in agreement with (\ref{y28}) and (\ref{y381})) we may write down the Hamiltonians (\ref{y43}) explicitly:
 \beq \label{y441}
\displaystyle{
 {\hat H}_i=R_{i,i-1}(z_i-z_{i-1})...R_{i1}(z_i-z_1)V^{(i)}R_{in}(z_i-z_N)...R_{i,i+1}(z_i-z_{i+1})\,.
}
\eq
The nested algebraic Bethe ansatz assumes a special substitution
$\Phi^{\rm off}$ for $\Phi$ (the off-shell Bethe vector) into the eigenvalue problem ${\hat t}(z)\Phi=t(z)\Phi$, which contains all equations
(\ref{y44}). The substitution is constructed by using some combination(s) of operators ${\hat T}_{ab}(\mu_\ga^c)$
(at some points $\mu_\ga^c$) applied to the vacuum vector in $\mathcal H$. This vector in $\mathcal H$ depends explicitly
on the set of new variables $\mu_\ga^c$ (the Bethe roots) and is called the off-shell Bethe vector. Using the quantum algebra of
exchange relations (\ref{y40}) one can show that ${\hat t}(z)\Phi^{\rm off}$ contains the wanted term (proportional to $\Phi^{\rm off}$) together with a set of unwanted terms, which should vanish. Vanishing of unwanted terms is formulated through
a system of equations (for Bethe roots), which are called the Bethe equations.
A solution of Bethe equations being substituted into $\Phi^{\rm off}$ provides
$\Phi$ (the on shell Bethe vector) a true solution of the eigenvalue problem.

For example, for the XXX spin chain  (\ref{y381}) the set of Bethe roots $\mu^b_\be$ are numbered as
$b\!=\!1\,,...\,,n\!-\!1$, $\be\!=\!1\,,...\,,N_b$, where $N\geq N_1\geq N_2\geq\ldots\geq N_{n-1}\geq 0$. We also put
$N_0=N_{n} = 0$. The integers $N_k$ means the number of excitations. For instance, when $n=2$, the number $N_1$
is number of spins up (if the vacuum vector consists of all spins down).  The Bethe equations
(the total number of equations is
$\sum\limits_{b=1}^{n-1}N_b$) are of the form:
\beq\label{y45}
  BE_{1}: \ \
{V_1}\prod\limits_{k=1}^N \frac{\!\mu^1_\be-q_k+\hbar}{\,\,\mu^1_\be-q_k}
={V_{2}}\prod\limits_{\ga\neq \be}^{N_{1}}\frac{\mu^1_\be-\mu_\ga^{1}+
\hbar }{\mu^1_\be-\mu_\ga^{1}-\hbar}
\prod\limits_{\ga=1}^{N_{2}}
\frac{\mu^1_\be-\mu_\ga^{2}-\hbar }{\mu^1_\be-\mu_\ga^{2}}\,,
  \eq
 \beq\label{y46}
%\begin{array}{c}
%\hbox{for}\ b=2\,,...,\,n\!-\!2:
% \\
BE_{\, b}:\ \
{V_b}
 \prod\limits_{\ga=1}^{N_{b\!-\!1}}\frac{\mu^b_\be-\mu_\ga^{b\!-\!1}+
 \hbar}{\mu^b_\be-\mu_\ga^{b\!-\!1} }
={V_{b\!+\!1}}\prod\limits_{\ga\neq \be}^{N_{b}}
\frac{\mu^b_\be-\mu_\ga^{b}+\hbar }{\mu^b_\be-\mu_\ga^{b}-\hbar}
\prod\limits_{\ga=1}^{N_{b\!+\!1}}
\frac{\mu^b_\be-\mu_\ga^{b\!+\!1}-\hbar}{\mu^b_\be-\mu_\ga^{b\!+\!1}}
  \eq
for $b=2\,,...,\,n\!-\!2$ and
 \beq\label{y47}
  BE_{\, n\! -\! 1}:\ \
{V_{n-1}}
\prod\limits_{\ga=1}^{N_{n\!-\!2}}
\frac{\mu^{n\!-\!1}_\be-\mu_\ga^{n\!-\!2}+\hbar}{\mu^{n\!-\!1}_\be-
\mu_\ga^{n\!-\!2} }
={V_{n}}\prod\limits_{\ga\neq \be}^{N_{n\!-\!1}}\frac{\mu^{n\!-\!1}_\be-\mu_\ga^{n\!-\!1}+\hbar
}{\mu^{n\!-\!1}_\be-\mu_\ga^{n\!-\!1}-\hbar}\,.
  \eq
The eigenvalues $H_i$ of the Hamiltonians ${\hat H}_i$ take the form
\beq \label{y48}
\displaystyle{
\frac{1}{\hbar}H_i = V_1
\prod\limits_{k=1}^N\frac{z_i-z_k+\hbar}{z_i-z_k}\prod\limits_{\ga
=1}^{N_1}\frac{z_i-\mu^1_{\ga}-\hbar}{z_i-\mu^1_{\ga}}\,.
}
\eq
Having a solution for the system of Bethe equations one should substitute it into the off-shell Bethe vector $\Phi^{\rm off}$. Then it becomes the on-shell Bethe vector -- a true solution of the quantum problem (\ref{y44}) with the eigenvalues
(\ref{y48}).

The limit to Gaudin model is obtained by redefining $\hbar\rightarrow\epsilon\hbar$, $V=1_n+\epsilon\Lambda$
and $\epsilon\rightarrow 0$. In particular, the spin chain Hamiltonians (\ref{y441}) provide the Gaudin Hamiltonians
\beq \label{y481}
\displaystyle{
 {\hat H}_i=\Lambda^{(i)}+\hbar\sum\limits_{k:l\neq i}^N r_{ik}(z_i-z_k)\,,\quad i=1,...,N\,.
}
\eq
\paragraph{Local Hamiltonian.} Similarly to the classical case (\ref{y301})-(\ref{y303}) local interaction (between neighbour sites only) in quantum model appears in the special case when
all $z_k=0$ and $V=1_n$. The quantum version of the Hamiltonian (\ref{y303}) has the form (up to identity operator)
\beq \label{y482}
\displaystyle{
 {\hat H}^{\rm loc}=\hbar\sum\limits_{i=1}^N P_{i,i-1}\,.
}
\eq
It comes from the transfer matrix (\ref{y42}) as
\beq \label{y483}
\displaystyle{
 {\hat H}^{\rm loc}= \frac{d {\hat t}(z)}{dz} \Big({\hat t}(z)\Big)^{-1}\Big|_{z=z_0}\,,
}
\eq
where (as in the classical case) $z_0$ is a special point, which is chosen depending on the $R$-matrix normalization.

\subsubsection*{\underline{ Spectral dualities in quantum models}}
 A natural way to describe spectral dualities at quantum level is to compare quantum problems (\ref{y44}) of a pair of dual models. See e.g. the rank-size duality in \cite{BaS}. We discuss it below in a more general case -- for Knizhnik-Zamolodchikov equations. Here we mention another approach, which can be viewed
 as straightforward quantization of the classical formulation.

 It is known \cite{SoV} that the pair of variables $(\la,z)$ in the spectral curve $\Gamma:\ \det(\la-L(z))=0$ can be considered as a pair of canonically conjugated separation variables. For example in ${\rm sl}_2$ rational Gaudin model the set of separated canonical variables can be defined as $\la_\al=L_{11}(z_\al),z_\al$, where $z_\al$ are zeros of $L_{12}(z)$.  The Poisson bracket between a pair of separated variables $(\lambda_\al,z_\be)$ is of the form
 \begin{equation}\label{y49}
 \{\lambda_\al,z_\be\}=h_\al(\lambda_\al,z_\al)\delta_{\al\be},\ \ \{\lambda_\al,\lambda_\be\}=\{z_\al,z_\be\}=0\,.
   \end{equation}
where a choice of functions $h_\al$ takes into account different possibilities such as
 $\{\lambda_\al,z_\be\}=1$ or $\{\log(\lambda_\al),z_\be\}=1$ or $\{\log(\lambda_\al),\log(z_\be)\}=1$. It is fixed by the Seiberg-Witten (SW) differential. For example, in the XXX ${\rm GL}_2$ chain on $N$ sites with the spectral curve
  $\Gamma^{\hbox{\tiny{Heisen}}}(w,x):\det_{2\times 2}(w-T(x))=0$,
  this differential is
$\hbox{d}S^{\hbox{\tiny{Heisen}}}(w,x)=x\frac{\hbox{d}w}{w}$. The dual model is a special Gaudin model with
the spectral curve $\Gamma^{\hbox{\tiny{Gaudin}}}(y,z):\det_{N\times N}(y-L(z))=0$ the SW differential
$\hbox{d}S^{\hbox{\tiny{Gaudin}}}(y,z)=y\hbox{d}z$.
 The spectral duality at the classical
level can be formulated as follows. The change of variables $z=w$, $y=x/w$ relates the spectral  curves and the SW differentials of two models as
  \begin{equation}\label{y50}
   \begin{array}{c}
  \Gamma^{\hbox{\tiny{Gaudin}}}(y,z)=\Gamma^{\hbox{\tiny{Heisen}}}(w,x)\,,
\ \ \ \hbox{d}S^{\hbox{\tiny{Gaudin}}}(y,z)=\hbox{d}S^{\hbox{\tiny{Heisen}}}(w,x)\,.
   \end{array}
  \end{equation}
Then, the quantum version of  the duality comes from exact semiclassical quantization of the spectral curves (based on the
corresponding SW differentials). Namely, the quantum spectral curves (which can be treated as the Baxter equations) are of the form
  \begin{equation}\label{y51}
  \begin{array}{c}
{\hat\Gamma}^{\hbox{\tiny{Heisen}}}(z,\hbar z\partial_z)\psi_{\hbox{\tiny{Heisen}}}(z)=0\,,\ \ \
\hat\Gamma^{\hbox{\tiny{Gaudin}}}(\hbar
\partial_z,z)\psi_{\hbox{\tiny{Gaudin}}}(z)=0
  \end{array}
  \end{equation}
and the duality states
  \begin{equation}\label{y52}
  \begin{array}{c}
{\hat\Gamma}^{\hbox{\tiny{Heisen}}}(z,\hbar z\partial_z)\sim\hat\Gamma^{\hbox{\tiny{Gaudin}}}(\hbar
\partial_z,z).
   \end{array}
  \end{equation}
and, finally,
  \begin{equation}\label{y53}
  \begin{array}{c}
\psi_{\hbox{\tiny{Heisen}}}(z)=\psi_{\hbox{\tiny{Gaudin}}}(z)\,.
   \end{array}
  \end{equation}
 See details in \cite{mmzz}.
In this way  the table  (\ref{y34}) is reproduced at quantum level:
\beq\label{y343}
\begin{array}{|c|c|c|}
\hline
\hbox{quant. alg.} {\bf \diagdown} R_{12} & \phantom{\Bigl|}XXX\phantom{\Bigl|} & XXZ
\\
\hline
linear: & \hbox{rational Gaudin} & \hbox{trig. Gaudin}
\\
Lie\ algebras & \hbox{(self-dual)}\circlearrowleft & \!\nearrow\hfill
\\
\hline
quadratic: & \hfill\swarrow\! & \hbox{XXZ spin chain}
\\
RLL\ relations & \hbox{XXX spin chain} & \hbox{(self-dual)}\circlearrowleft
\\
\hline
\end{array}
\eq
Here in the horizontal line we again put a type of $R$-matrix (rational or trigonometric). The $R$-matrix
is classical for Gaudin model and is quantum for spin chains. In vertical line the type of commutation
relations (quantum algebra) is placed.

%Sklyanin-Toda bispectral...

\subsection{Dualities in Knizhnik-Zamolodchikov equations}

\subsubsection*{\underline{KZ and qKZ}} Consider a system of $N$ equations \cite{KZ}
\beq \label{y55}
\displaystyle{
\kappa\p_{z_i}{\bf\Psi}=\Big(\Lambda^{(i)}+\hbar\sum\limits_{k:l\neq i}^N r_{ik}(z_i-z_k)\Big){\bf\Psi}\,,\quad i=1,...,N\,,\quad {\bf\Psi}\in{\mathcal H}\,,
}
\eq
where the space ${\mathcal H}$ is the same as the Hilbert space in integrable models (\ref{y37}), and  $r_{ij}$ are the classical ${\rm gl}_n$-valued $r$-matrices.
It is the system of Knizhnik-Zamolodchikov equations. These equations are compatible for skew-symmetric $r$-matrices
(i.e. $r_{ij}(z)=-r_{ji}(-z)$) due to the classical Yang-Baxter equation (\ref{y24}). Following N. Reshetikhin \cite{Re} we treat these equations as quantum version of the Schlesinger system. The latter is the non-autonomous version
of the Gaudin models. That is (\ref{y55}) takes the form of non-stationary Schrodinger equations
\beq \label{y56}
\displaystyle{
\kappa\p_{z_i}{\bf\Psi}={\hat H}_i{\bf\Psi}\,,\quad i=1,...,N\,.
}
\eq
with the Gaudin
Hamiltonians (\ref{y481}) and the time variables being identified with the marked points $z_k$. Notice that equations contain two ''Planck constants''. The first one $\hbar$ is the one entering the Gaudin Hamiltonians, and the second
one is $\kappa$.

The quantum Knizhnik-Zamolodchikov (qKZ) equations are defined as follows  \cite{FR}:
\beq \label{y57}
\displaystyle{
 e^{\kappa\hbar\p_{z_i}}{\bf\Psi}=K^{(\kappa)}_i{\bf\Psi}\,,\quad i=1,...,N\,.
}
\eq
with the operators $K^{(\kappa)}_i$
 \beq \label{y58}
\displaystyle{
K^{(\kappa)}_i={\bf R}_{i,i-1}(z_i-z_{i-1}+\kappa\hbar)...{\bf R}_{i1}(z_i-z_1+\kappa\hbar)V^{(i)}
{\bf R}_{in}(z_i-z_N)...{\bf R}_{i,i+1}(z_i-z_{i+1})\,.
}
\eq
The system is compatible
 \beq \label{y581}
\displaystyle{
\Bigl (e^{\hbar \kappa \p_{x_i}}{ K}_{j}^{(\kappa )}\Bigr )
{ K}_{i}^{(\kappa )}=
\Bigl (e^{\hbar \kappa \p_{x_j}}{ K}_{i}^{(\kappa )}\Bigr )
{ K}_{j}^{(\kappa )}
}
\eq
due to the quantum Yang-Baxter equation (\ref{y36}) for $R$-matrices ${\bf R}_{ij}$, which are also assumed in (\ref{y58}) to be unitary, i.e. they are normalized with the condition
${\bf R}_{ij}(z){\bf R}_{ji}(-z)={\rm id}$. For example, in the rational XXX case we have
 \beq \label{y59}
\displaystyle{
 {\bf R}_{12}(z)=\frac{z 1_n\otimes 1_n+\hbar P_{12}}{z+\hbar}\,.
}
\eq
Comparing with (\ref{y381}) we get
 \beq \label{y60}
\displaystyle{
 R_{12}(z)=\frac{z+\hbar}{z}\,{\bf R}_{12}(z)\,.
}
\eq
When $\kappa=0$ the operators (\ref{y58}) take the form similar to the spin chain Hamiltonians (\ref{y441}). Using
(\ref{y60}) we write this relation as follows:
 \beq \label{y61}
\displaystyle{
 {\hat H}_i=K^{(0)}_i \prod\limits_{k\neq i}^N\frac{z_i-z_k+\hbar}{z_i-z_k}\,.
}
\eq

\subsubsection*{\underline{Dualities for (q)KZ equations}}

 We begin with another quantum version of spectral duality. Consider the KZ equations (\ref{y55})-(\ref{y56}) for ${\rm gl}_n$ XXX $r$-matrices. For simplicity
we also choose the fundamental representation of ${\rm gl}_n$, i.e. ${\mathcal H}=(\mC^n)^{\otimes N}$ and
 \beq \label{y62}
\displaystyle{
 r_{ij}(z_i-z_j)=\frac{P_{ij}}{z_i-z_j}=\sum\limits_{a,b=1}^n \frac{ E_{ab}^{(i)}E_{ba}^{(j)} }{z_i-z_j}\,,
}
\eq
where $E_{ab}$ is the standard matrix basis in ${\rm Mat}_n$.
Then the Gaudin Hamiltonians take the form\footnote{In fact, (\ref{y63}) keeps its form in any representation if
we denote $E_{ab}$ to be the ${\rm gl}_n$ Lie algebra basis.}
 \beq \label{y63}
\displaystyle{
 {\hat H}_i=\sum\limits_{a=1}^n E_{aa}^{(i)}\la_a+\hbar\sum\limits_{j:j\neq i}^N\sum\limits_{a,b=1}^n \frac{ E_{ab}^{(i)}E_{ba}^{(j)} }{z_i-z_j}\,.
}
\eq
Consider the spectrally dual Gaudin model (\ref{y31})-(\ref{y32}). The quantum version of the dual Hamiltonians
(\ref{y321}) is as follows:
  \beq\label{y64}
  \begin{array}{c}
    \displaystyle{
 {\hat{\widetilde H}}_a=\sum\limits_{i=1}^N{\hat{\ti S}^a_{ii}}z_i +
 \hbar\sum\limits_{c:\,c\neq a}^n \frac{ \tr({\hat{\ti S}^a} {\hat{\ti S}^c}) }{\la_a-\la_c}\,,
 }
 \end{array}
 \eq
 and in the fundamental representation of ${\rm gl}_N$ it takes the form:
  \beq\label{y65}
  \begin{array}{c}
    \displaystyle{
 {\hat{\widetilde H}}_a=\sum\limits_{i=1}^N{e}^{(a)}_{ii}z_i +
 \hbar\sum\limits_{c:\,c\neq a}^n \sum\limits_{i,j=1}^N\frac{  e^{(a)}_{ij} e^{(c)}_{ji}  }{\la_a-\la_c}\,,
 }
 \end{array}
 \eq
 where $e_{ij}$ is the matrix basis in ${\rm Mat}_N$. This Hamiltonian naturally acts on the Hilbert space of the dual model
 $\ti{\mathcal H}=(\mC^N)^{\otimes n}$. Similarly to (\ref{y56}) may also define the dual KZ equation as
\beq \label{y66}
\displaystyle{
\kappa\p_{\la_a}{\widetilde{\bf\Psi}}={\hat{\widetilde H}}_a{\widetilde{\bf\Psi}}\,,\quad a=1,...,n\,,
\quad {\widetilde{\bf\Psi}}\in \ti{\mathcal H}\,.
}
\eq
Recall that at the classical level the Hamiltonians $H_i$ and the dual Hamiltonians ${\widetilde H}_a$ Poisson commute.
It is then natural to expect commutativity of operators (\ref{y63}) and (\ref{y65}). For this purpose we need to define the action of these operators on the same space, say ${\mathcal H}$. Although  ${\mathcal H}\cong\ti {\mathcal H}$, these Hilbert spaces have different tensor structure. In the case of generic representations
the dual spaces are identified through the Howe duality.

Consider first the classical dual Hamiltonian (\ref{y321}). Plugging the parametrization ${\ti S}^a_{ij}=\xi^i_a\eta^j_b$ (\ref{y32}) into (\ref{y321}) one can easily rewrite $\ti { H}_a$ in terms of initial variables $S^i_{ab}$:
 \beq\label{y67}
 \begin{array}{c}
  \displaystyle{
 {\widetilde H}_a=\sum\limits_{i=1}^k\sum\limits_{a=1}^l z_i\xi_a^i\eta_a^i+\sum\limits_{i,j=1}^k\sum\limits_{b:b\neq
 a}^l\frac{\xi^i_a\eta^j_a\xi^j_b\eta^i_b}{\la_a-\la_b}=
 }
 \\ \ \\
  \displaystyle{
 =\sum\limits_{i=1}^k\sum\limits_{a=1}^l z_i S^i_{aa}+\sum\limits_{i,j=1}^k\sum\limits_{b:b\neq
 a}^l\frac{S^i_{ab}S^j_{ba}}{\la_a-\la_b}
 =\sum\limits_{i=1}^k\sum\limits_{a=1}^l z_i S^i_{aa}+\sum\limits_{b:b\neq
 a}^l\frac{T_{ab}T_{ba}}{\la_a-\la_b}\,,\quad T_{ab}=\sum\limits_{i=1}^kS^i_{ab}\,.
 }
 \end{array}
 \eq
Its quantum version is obtained in the same way:
 \beq\label{y68}
 \begin{array}{c}
  \displaystyle{
 {\hat{\widetilde H'}}_a=\sum\limits_{i=1}^N z_i E_{aa}^{(i)}+\hbar\sum\limits_{b:b\neq a}^n
 \frac{ {\bf E}_{ab}{\bf E}_{ba}-{\bf E}_{aa} }{\la_a-\la_b}\,,\quad
 {\bf E}_{ab}=\sum\limits_{i=1}^N E_{ab}^{(i)}\,.
  }
 \end{array}
 \eq
 The commutativity $[{\hat{\widetilde H'}}_a, {\hat H}_i]=0$ is verified straightforwardly. Together with
  $\p_{\la_a}{\hat H}_i=\p_{z_i}{\hat{\widetilde H'}}_a$ we come to compatible $N+n$ equations
  for ${\bf\Psi}\in{\mathcal H}$:
 \beq\label{y69}
 \begin{array}{l}
   \displaystyle{
   \kappa\p_{z_i}{\bf\Psi} = {\hat H}_i{\bf\Psi}\,,
   }
 \\
  \displaystyle{
 \kappa\p_{\la_a}{\bf\Psi}={\hat{\widetilde H'}}_a{\bf\Psi}\,.
  }
 \end{array}
 \eq
 Then we may formulate the problem as to find a set of equations (differential or difference) compatible with a given system of (q)KZ equations \cite{FMTV}.
Being written in the form (\ref{y69}) the dual KZ equations (the second line in (\ref{y69}))
 are also called the dynamical equations.  Notice that this formulation of duality depends on a choice of the Hilbert space. The system (\ref{y69})
 is obtained for ${\bf\Psi}\in{\mathcal H}$. Alternatively, we could start with (\ref{y66}), and transform
 ${\hat H}_i$ from (\ref{y56}) to obtain ${\hat H'}_i$ acting on the space $\ti{\mathcal H}$, and then
 come to the compatible equations for
 ${\widetilde{\bf\Psi}}$ with the Hamiltonians  ${\hat{\widetilde H}}_a$ and ${\hat H'}_i$.

 \paragraph{$({\rm gl}_N, {\rm gl}_M)$ duality} (in our case $M=n$)
 establishes exact matching between dual (q)KZ connections depending on the choice of representation space
 of corresponding Lie algebra. These results are  extended to similar (but more complicated) relations
 between the trigonometric KZ and rational qKZ equations and those between the trigonometric qKZ equations.
 This approach was developed mainly by Mukhin, Tarasov and Varchenko. See \cite{mukhin} for details.
 %The $({\rm gl}_N, {\rm gl}_M)$ duality transforms ${\hat{\widetilde H}}_a$ into ${\hat{\widetilde H'}}_a$.

 To summarize, we have the table similar to (\ref{y34}):
\beq\label{y345}
\begin{array}{|c|c|c|}
\hline
{\rm (q)KZ} {\bf \diagdown} R_{12} & \phantom{\Bigl|}XXX\phantom{\Bigl|} & XXZ
\\
\hline
\hbox{differential} & \hbox{rational KZ} & \hbox{trig. KZ}
\\
 & \hbox{(self-dual)}\circlearrowleft & \!\nearrow\hfill
\\
\hline
\hbox{difference} & \hfill\swarrow\! & \hbox{trig qKZ}
\\
 & \hbox{rational qKZ} & \hbox{(self-dual)}\circlearrowleft
\\
\hline
\end{array}
\eq
 It follows from the above that each dual pair has two possible forms in ${\mathcal H}$ or in $\ti{\mathcal H}$.

 % At classical level -- isomon. deformations - only the first box is clear .... Harnad
 %To find dualities between discrete Schlesinger systems is an interesting problem

\paragraph{Cherednik's DAHA approach.}
Let us also mention the Cherednik's approach to (q)KZ equations based on Hecke algebras and KZ equations
Cherednik \cite{Ch1,Ch2}.  The most general is the double affine
Hecke algebra. It provides algebraic description of dual trigonometric qKZ equations. Degenerations of DAHA provide the rest of dual pairs.
%
%Toledano  Stokman
%
Details can be found in \cite{Toledano}. This approach is especially useful for studies of polynomial representations since
the Hecke algebras have natural polynomial representations using
operators of permutation of variables.

%%%%%%%%%%%%%%%%%%%%%%%%%%%%%%%%%%%%%%%%%%%%%%%%%%%%%%%%%%%%%%%%%%%%%%%%%%%%%%%%%%%%%%%%%%%%%%%%%%
\subsection{Quantum-classical and quantum-quantum dualities}

In the previous subsections we described two families of integrable systems and dualities acting inside each family. The first family is many-body systems (of Calogero-Ruijsenaars type). The dualities are the Ruijsenaars' one (or ''p-q'' or ''action-angle'') at classical level and its quantum analogue -- the bispectral duality. The second family consists of Gaudin models and spin chains. They are related by spectral duality, which is then lifted to $({\rm gl}_N,{\rm gl}_M)$
duality (or compatible dynamical KZ equation) at the level of (q)KZ equations. In this subsection we discuss  another set of dualities, which relates both families.

\subsubsection*{\underline{Quantum spin chains and classical many-body systems}}
The quantum Gaudin models or spin chains on one side and the classical Calogero-Moser or Ruijsenaars-Schneider systems
on the other side can be related by a certain identification of their data. Namely, we describe relations in four pairs
called quantum-classical (QC) dualities:
 \beq\label{y691}
 \begin{array}{c}
 \hbox{\underline{Quantum-classical dualities:}}
 \\ \ \\
 \begin{array}{ccc}
%\ \hbox{\underline{Quantum-classical dualities:}} \
% \\
 %  \displaystyle{
   \hbox{class. rat. Calogero-Moser} & \longleftrightarrow & \hbox{quant. rat. Gaudin}
 %  }
 \\
%  \displaystyle{
   \hbox{class. trig. Calogero-Moser} & \longleftrightarrow & \hbox{quant. trig. Gaudin}
%  }
  \\
 %    \displaystyle{
   \hbox{class. rat. Ruijsenaars-Schneider} & \longleftrightarrow & \hbox{quant. XXX chain}
 %  }
   \\
 %    \displaystyle{
   \hbox{class. trig. Ruijsenaars-Schneider} & \longleftrightarrow & \hbox{quant. XXZ chain}
 %  }
 \end{array}
  \end{array}
 \eq
Below we give an example for the rational classical Ruijsenaars-Schneider model and the quantum XXX spin chain (for the rest of pairs it is very similar)
\beq \label{y70}
%\displaystyle{
%\left\{
\begin{array}{l}
\displaystyle{
\eta=\hbar\,,
}
\\
\displaystyle{
q_i = z_i\,,
}
\\
\displaystyle{
\dot{q}_i = H_i/\hbar\,,
}
\end{array}
%\right.
%}
\eq
so that the coupling constant in the Ruijsenaars-Schneider model $\eta$ (\ref{y5}) is identified with the Planck constant,
the positions of RS particles are the inhomogeneities in the quantum chain, and the classical velocities are
the eigenvalues of quantum Hamiltonians (\ref{y43}).

The relations of type (\ref{y70}) were observed in different contexts for different examples \cite{GK,AKLTZ,mtv2,GaK,GZZ}. Below
we explain that this relation is a simple consequence of the Matsuo-Cherednik map together with the semiclassical
limit of (q)KZ equations. Here, following \cite{GZZ} we formulate the detailed version of (\ref{y70}) as follows.

Given the Lax matrix of RS system (\ref{y5}) make the substitution (\ref{y70})
Then the eigenvalues of the Lax matrix are determined by the twist parameters:
\beq \label{y71}
\displaystyle{
\mathrm{Spec} L = \left(\underbrace{V_1,...,V_1}_{N-N_1},\underbrace{V_2,...,V_2}_{N_1-N_2}, ..., \underbrace{V_n,...,V_n}_{N_{n-1}} \right),
}
\eq
where the multiplicities numbers $M_1=N-N_1$, $M_2=N_1-N_2$,..., $M_n=N_{n-1}$ are defined by the numbers of Bethe roots at the levels
of (the hierarchy of) the nested Bethe ansatz. In fact, these multiplicities are eigenvalues of the operators
\beq \label{y72}
\displaystyle{
{\hat M}_a=\sum\limits_{k=1}^N E_{aa}^{(k)}\,,
}
\eq
which commute with the spin chain Hamiltonians.

Recall that the eigenvalues of the classical Lax matrix are the action variables. That is the quantum spin chain
corresponds to intersection of two Lagrangian submanifolds in the phase space of the classical many-body system. The first Lagrangian submanifold is given by the set of fixed positions $q_i=z_i$, and the second -- by fixation of the action variables as in (\ref{y71}). Moreover, the multiplicities in the set of action variables are given in terms
of the quantum occupation numbers. Details of the relations (\ref{y70}) and (\ref{y71}) for all pairs of models (\ref{y691}) can be found in \cite{GZZ,BLZZ}.

\paragraph{Extension to correspondence.} Let us also remark on the extension of dualities (\ref{y70}) to the correspondence \cite{TsuboiZZ}, when on the r.h.s. of
(\ref{y70}) we deal with not a single spin chain (or Gaudin model) but with a set of $n+1$ SUSY models constructed by means of Lie superalgebras
$$
{\rm gl}(n|0)\,, {\rm gl}(n-1|1)\,, ... \,,{\rm gl}(0|n)\,.
$$
The QC-duality holds true for any
of these models. It is important by the following reason. The identifications (\ref{y70}) and (\ref{y71}) allows (in principle) to solve the quantum eigenvalue problem for a spin chain without using the Bethe ansatz. Indeed,  (\ref{y70})-(\ref{y71}) means that in order to
find the spectrum $H_i$ of a quantum spin chain one could fix the level of action variables in the classical Ruijsenaars-Schneider model as in (\ref{y71}). That is $\tr L^k=\sum_i M_i V_i^k$. Then, by expressing velocities through the coordinates ($q_i=z_i$) and these action variables we find the spectrum of spin chain as $H_i=\hbar {\dot q}_i$. Expressing velocities through coordinates and action variables lead to a system of algebraic  equations, which can be solved. However, a given solution of this system may correspond to the spectrum for any of $n+1$ models with ${\rm gl}(n-i|i)$, and to know which
model it is becomes a complicated combinatorial problem.

\subsubsection*{\underline{Matsuo-Cherednik projection}}
The quantum-quantum duality extends the quantum-classical one to the case of quantum many-body systems.
It is, in fact, given by the Matsuo-Cherednik map \cite{M1,Ch2} from solutions of (q)KZ equations to
solutions of (stationary) Schrodinger equations for quantum many-body systems.

\paragraph{Quantum Calogero-Moser model from KZ equations.} We follow the  Felder-Veselov approach \cite{FeVe,ZZ1,ZZ2}.
Consider KZ equations (\ref{y55}) with the rational XXX $r$-matrix $r_{ij}=P_{ij}/(z_i-z_j)$. It is also convenient to use the Dirac ''bra''-''cket'' notations here, so that we denote solution of (\ref{y55}) as ${\bf\Psi}=\Bigl | \Psi \Bigr >\in{\mathcal H}$:
\beq\label{y73}
\kappa \p_{z_i}\Bigl | \Psi \Bigr >=\left ({ V}^{(i)}+
\hbar \sum_{j\neq i}^n \frac{{ P}_{ij}}{z_i-z_j}\right )
\Bigl | \Psi \Bigr >.
\eq
Consider the decomposition of the Hilbert space into a sum of eigenspaces of the operator (\ref{y72}):
\beq\label{y74}
\displaystyle{
{\mathcal H}=(\mC^n)^{\otimes N}=
\bigoplus_{M_1, \ldots , M_n} \!\! \!\! {\mathcal V}(\{M_a \})\,.
}
\eq
Basis vectors in the space ${\cal V}(\{M_a \})$ are
$
\Bigl |J\Bigr > =e_{j_1}\otimes e_{j_2}\otimes \ldots \otimes e_{j_n},
$
where the number of indices $j_k$ with condition $j_k =a$ is equal to $M_a$ for all
$a=1, \ldots , n$.\footnote{In the underlying Gaudin model a choice of ${\mathcal V}(\{M_a \})$ corresponds
to fixation of numbers of Bethe roots at all levels of the nested Bethe ansatz.}
Decompose solution of (\ref{y73})
\beq\label{y75}
\displaystyle{
\Bigl |\Psi \Bigr >=\sum_J \Psi_J \Bigl |J\Bigr >.
}
\eq
Then it is verified that
\beq\label{y76}
\left ( \kappa^2 \sum_{i=1}^n \p_{z_i}^2-\sum_{i\neq j}^n
\frac{\hbar (\hbar -\kappa )}{(z_i-z_j)^2}\right )\Psi = E\Psi\,,
\eq
where the wave-function $\Psi$ and eigenvalue $E$ are as follows:
\beq\label{y77}
\Psi =\sum_J \Psi_J\,,\quad E=\sum_{a=1}^n M_a V_a^2\,.
\eq
The easiest way to see it is to introduce
\beq\label{y78}
\Bigl < \Omega \Bigr |=\sum_J \Bigl < J \Bigr |\in ({\cal V}(\{M_a\}))^*\,,
\eq
which has the property
%
%\beq\label{y79}
$
\Bigl < \Omega \Bigr |P_{ij}=\Bigl < \Omega \Bigr |\,.
$
%\eq
%
%
Then the wave function (\ref{y77}) is then obtained as a projection of $\Bigl | \Psi \Bigr >$ on (\ref{y78}):
\beq\label{y80}
\Psi =\Bigl <\Omega \Bigr | \Psi \Bigr >\,,
\eq
and the eigenvalue $E$ arises as
\beq\label{y81}
\sum_{i=1}^N\Bigl <\Omega \Bigr |(V^{(i)})^2\Bigl |\Psi \Bigr >=\left (
\sum_{a=1}^n M_aV_a^2\right )\Psi=E\Psi\,.
\eq
For higher Calogero Hamiltonians the eigenvalues are $E_k=\sum_a M_a V_a^k$.

\paragraph{Quantum Ruijsenaars-Schneider model from qKZ equations.}  The relation between the quantum Ruijsenaars-Schneider model and the rational qKZ is almost the same.
Consider the rational XXX qKZ equation (\ref{y57})-(\ref{y58}). Then
\beq\label{y82}
e^{\kappa \hbar \p_{z_i}}\Bigl <\Omega \Bigr |\Psi \Bigr >=
e^{\kappa \hbar \p_{z_i}}\Psi =
\Bigl <\Omega \Bigr |   { K}_{i}^{(\kappa )}\Bigl | \Psi \Bigr >=
\Bigl <\Omega \Bigr |   { K}_{i}^{(0 )}\Bigl | \Psi \Bigr >.
\eq
Then, multiplying both parts by $\displaystyle{\prod_{j\neq i}^n \frac{z_i-z_j +\hbar}{z_i-z_j}}$
and summing up over $i$, we may use the relation (\ref{y61}) to obtain
\beq\label{y83}
  \begin{array}{c}
    \displaystyle{
\sum_{i=1}^N \left ( \prod_{j\neq i}^N \frac{z_i-z_j +\hbar}{z_i-z_j}\right )
e^{\kappa \hbar \p_{z_i}}\Psi \,\, =\,\,
\sum_{i=1}^N  \prod_{j\neq i}^N \frac{z_i-z_j +\hbar}{z_i-z_j}\,
\Bigl <\Omega \Bigr |   { K}_{i}^{(0 )}\Bigl | \Psi \Bigr >=
}
\\ \ \\
    \displaystyle{
=\, \sum_{i=1}^N \Bigl <\Omega \Bigr |   {\hat H}_{i}\Bigl | \Psi \Bigr > \,\, =
\,\, \sum_{i=1}^N \Bigl <\Omega \Bigr |   {V}^{(i)}\Bigl | \Psi \Bigr >\,\, =\,\,
\sum_{a=1}^n V_a \Bigl <\Omega \Bigr |   {\hat M}_a\Bigl | \Psi \Bigr >\,\, =\,\,
\left (\sum_{a=1}^n V_a M_a\right )\Psi\,,
}
 \end{array}
 \eq
where the following operator identity is also used:
 \beq\label{y84}
\sum_{i=1}^{N}{\hat H}_i= \sum_{i=1}^N {V}^{(i)}=
\sum_{a=1}^{n}V_a {\hat M}_a\,.
 \eq
For trigonometric models the calculations are more complicated but the general idea is the same. See details in \cite{ZZ1,ZZ2}. To summarize, we have the following  set of quantum-quantum dualities (Matsuo-Cherednik projections):
 \beq\label{y85}
 \begin{array}{c}
 \hbox{\underline{Quantum-quantum dualities:}}
 \\ \ \\
 \begin{array}{ccc}
%\ \hbox{\underline{Quantum-classical dualities:}} \
% \\
 %  \displaystyle{
   \hbox{quant. rat. Calogero-Moser} & \longleftrightarrow & \hbox{rational KZ}
 %  }
 \\
%  \displaystyle{
   \hbox{quant. trig. Calogero-Moser} & \longleftrightarrow & \hbox{trigonometric KZ}
%  }
  \\
 %    \displaystyle{
   \hbox{quant. rat. Ruijsenaars-Schneider} & \longleftrightarrow & \hbox{rational qKZ}
 %  }
   \\
 %    \displaystyle{
   \hbox{quant. trig. Ruijsenaars-Schneider} & \longleftrightarrow & \hbox{trigonometric qKZ}
 %  }
 \end{array}
  \end{array}
 \eq
 It is similar to (\ref{y691}).

Let us remark on SUSY version. Similarly to the quantum-classical duality,  the Matsuo-Cherednik projection holds for a set of qKZ equations
constructed by means of Lie superalgebras ${\rm gl}(n|0)$, ${\rm gl}(n-1|1)$, ... ${\rm gl}(0|n)$, see \cite{TsuboiZZ,VZZ}.

%%%%%%%%%%%%%%%%%%%%%%%%%%%
%%%%%%%%%%%%%%%%%%%%%%%%%%%
%%%%%%%%%%%%%%%%%%%%%%%%%%%
To summarize the discussion about the QC duality between quantum integrable spin chains and integrable systems of many-body type we present a table of correspondences between the data on two sides:
\begin{center}
\begin{tabular}{|c | c|}
 \hline
 {\bf Quantum spin chains} &  {\bf Integrable many-body system}  \\ [0.5ex]
 \hline\hline
  Inhomogeneity parameters $\{z_i\}_{i=1}^N$ &  Coordinates of particles $\{q_i\}_{i=1}^N$ $\phantom{\prod}$  \\
 \hline
   Twist parameters $\{V_i\}_{i=1}^n$ & Spectrum of Lax matrix $L$  \\
  & (higher Hamiltonians) \\
 \hline
 Energy level of non-local Hamiltonians $\{H_i\}_{i=1}^N$ & Velocities of particles $\{\dot{q}_i \}_{i=1}^N$ \\
 \hline
 Numbers of Bethe roots $\{N_i\}_{i=1}^{n-1}$ & The degeneracy of spectrum   \\
 & of the Lax matrix $L$
 \\
 \hline
 Plank constant $\hbar$ & Coupling constant of the model $\eta$ \\
 \hline
\end{tabular}
\end{center}
This table is the repetition of formulas (\ref{y70}), (\ref{y71}) presented earlier. The positions of particles in classical models are mapped to inhomogeneities, while the velocities are identified with eigenvalues of non-local quantum Hamiltonians and the twist parameters account for the corresponding values of all classical higher Hamiltonians. The correspondences for the quantum-quantum version of duality are almost the same. The only difference is that on the many-body side we consider a quantum mechanical problem of diagonalizing certain differential (difference) operators (\ref{y76}) of Calogero-Ruijsenaars family and on the spin chain side we no longer study eigenproblem for non-local Hamiltonians (\ref{y73}), (\ref{y82}) but study the (q)KZ equations connected to them.
%%%%%%%%%%%%%%%%%%%%%%%%%%%%%%%%%
%%%%%%%%%%%%%%%%%%%%%%%%%%%%%%%%%
%%%%%%%%%%%%%%%%%%%%%%%%%%%%%%%%%

\subsection{Interrelations between dualities and limiting cases}

\subsubsection*{\underline{Quasi-classical limit of Matsuo-Cherednik projection}}
As we mentioned above, the relations (\ref{y70}) for quantum-classical duality were found in different contexts and examples. Here we show a simple derivation of (\ref{y70}) coming from the Matsuo-Cherednik projection in the quasi-classical limit $\kappa\rightarrow 0$, which brings the set of quantum-quantum (QQ) dualities
(\ref{y85}) to the set of quantum-classical QC relations (\ref{y691}):
\beq \label{y855}
\displaystyle{
\begin{tikzcd}
\text{quant. many-body} \arrow{d}{\kappa \rightarrow 0} \arrow[r, leftrightarrow, "\text{QQ}"]
& \text{(q)KZ} \arrow{d}{\kappa \rightarrow 0} \\
\text{class. many-body} \arrow[r, leftrightarrow, "\text{QC}"]
& \text{Gaudin/spin chain}
\end{tikzcd}
}
\eq
The quasi-classical limit of (q)KZ equations was described in \cite{ReVa}. The starting point is the integral representation for solutions of the (q)KZ equation
\cite{SchV}. It is given as an integral over off-shell Bethe vector
$\Phi^{\rm off}\in\mathcal H$ -- the ansatz for solution of the eigenvalue problem (\ref{y44}) but the Bethe roots $\mu$ are not restricted to solutions of Bethe equations yet:
\beq\label{y86}
\displaystyle{
\Psi=\int_C e^{S/\kappa}\Phi^{\rm off} d\mu\,,
}
\eq
where $S=S(z,\mu)$ is a certain multivalued function, which becomes a single-valued on the integration cycle $C$. It is also known as the Yang-Yang function. In the limit $\kappa\rightarrow 0$ the integral is localized at critical points of $S$, define by the set of equations
\beq\label{y87}
\displaystyle{
\p_{\mu_a}S=0\,.
}
\eq
These are the Bethe equations, which restrict the off-shell vector $\Phi^{\rm off}$ to the on-shell Bethe vector $\Phi$.
Consider KZ equations (\ref{y55}). As $\kappa\rightarrow 0$ we have the following asymptotic form for its solution
\beq\label{y88}
\displaystyle{
\Psi=\Big( \Psi_0+\kappa\Psi_1+... \Big)e^{S/\kappa}\,,
}
\eq
where $S$ satisfies (\ref{y87}) and
\beq\label{y89}
\displaystyle{
\Psi_0=\Phi\,.
}
\eq
The substitution (\ref{y88}) into (\ref{y55}) provides the eigenvalue problem
\beq\label{y90}
\displaystyle{
{\hat H}_i\Psi_0=H_i\Psi_0\,,\quad H_i=\p_i S\,,\quad i=1...N
}
\eq
for the Gaudin model. Similarly to the Matsuo-Cherednik projection, one can
can consider the sum of squared equations (\ref{y90}) and project it to
an appropriate $\langle\Omega|$. Then the eigenvalues $H_i=\p_i S$ are identified as velocities in classical many-body system.

To summarize, the quantum-classical duality provides a certain
relation between inhomogeneous
spin chains (Gaudin models) and the classical integrable many-body
systems from Ruijsenaars-Schneider (Calogero-Moser) family. There is the complete matching of
parameters and the corresponding objects at two sides of
the relation. The coupling constant at many-body side
gets mapped into the Planck constant at the spin side.
%, while
%the relativization parameter $\beta$ in the RS system gets
%mapped into the anizotropy parameter in XXZ spin chain.
The inhomogeneities at the spin chain side become the
coordinates of particles, while the eigenvalues
of twist matrix in the spin chains gets mapped into the
eigenvalues of Lax operator in the RS-CM family. Non-local
Hamiltonians in the spin chains become the velocities in the
RS-CM family.

As we mentioned above, at the classical many-body side we deal
with  intersection of two
Lagrangian submanifolds in the phase space. The first is $q_i=z_i$
and the second is defined by fixing the action variables as in (\ref{y71}).
The relation (\ref{y71}) holds on-shell the Bethe equations.
In this respect the Bethe ansatz equations in the spin chains
define the equations for intersection of two
Lagrangian submanifolds and the Yang-Yang function, whose
extremization yields the Bethe ansatz equations corresponds
to the generating function for the canonical transformations.

We remark that similar questions were considered in \cite{NRS}.
Also, the geometry underlying quantum-classical and quantum-quantum dualities were
studied in the context of K-theory and Nakajima varieties \cite{koroteev17,Koroteev:2018azn,koroteev21}. Finally, we mention that the quantum-classical duality is extended to triality, which third ingredient
is integrable hierarchy (of KP type) and its tau-function \cite{AKLTZ,TsuboiZZ,KV}.

\subsubsection*{\underline{Consistency of dualities}}
Recall again that we deal with two families of integrable systems. The first is the many-body systems endowed with the Ruijsenaars (or spectral) dualities at both -- classical and quantum levels. The second family consists of Gaudin models and spin chains, which are related by spectral dualities. The latter can be lifted to the level of dual (q)KZ equations. We also described the (QC) quantum-classical and (QQ) quantum-quantum dualities relating both families. Here we discuss consistency of these relations.

First, the spectrally dual classical many-body systems are related (QC dual) to spectrally dual spin chains (or Gaudin models). This is represented by the following commutative diagram:
\beq \label{y91}
\displaystyle{
\begin{tikzcd}
\text{spin chain/Gaudin} \arrow{d}{\text{QC}} \arrow[r, rightarrow, "\text{spectral}"]
& \text{dual spin chain/Gaudin} \arrow{d}{\text{QC}} \\
\text{class. many-body} \arrow[r, leftarrow, "\text{spectral}"]
& \text{dual class. many-body}
\end{tikzcd}
}
\eq
Indeed, the spectral duality interchanges positions of particles with action variables. They are respectively identified with the inhomogenuities and the twist parameters in the spin chains (or Gaudin models). And the
spectral duality interchanges the inhomogenuities and the twist parameters themselves. Put it differently, the QC duality deals
with the intersection of two Lagrangian submanifolds. These two
submanifolds are the same on the right and the left hand sides of the diagram. At the same time these two
submanifolds are interchanged by either the upper or the lower maps.

Similar diagram appears at quantum level:
\beq \label{y92}
\displaystyle{
\begin{tikzcd}
\text{(q)KZ equations} \arrow{d}{\text{QQ}} \arrow[r, rightarrow, "\text{duality in KZ}"]
& \text{dual (q)KZ equations} \arrow{d}{\text{QQ}} \\
\text{quant. many-body} \arrow[r, leftarrow, "\text{spectral}"]
& \text{dual quant. many-body}
\end{tikzcd}
}
\eq
Instead of two Lagrangian submanifolds we deal with the wave function at quantum level of many-body systems. The wave function obeys two sets of
 equations
\beq \label{y93}
\begin{array}{c}
\displaystyle{
{\hat H}_i\Psi_\la(z)=h_i(\la)\Psi_\la(z)\,,
}
\\ \ \\
\displaystyle{
{\hat {\widetilde H}}_i\Psi_\la(z)={\ti h}_i(z)\Psi_\la(z)\,,
}
\end{array}
\eq
This is a quantum counterpart
of the intersection of two Lagrangian submanifolds at the
classical level.

It often happens in the studies of bispectral problem
that the original Hamiltonian has discrete spectrum.
Then the dual Hamiltonian is an operator providing
recurrence relations for the wave functions. In this respect
the bispectral problem itself contains a
quantum counterpart
of two Lagrangian submanifolds.

\subsubsection*{\underline{Some limiting cases}}

It is interesting to study different limits of the described above models.
For example, one can take the limit $\hbar \rightarrow 0$  in the spin chains (more, precisely $\hbar:=\epsilon\hbar$, $\epsilon\rightarrow 0$) and obtain the
Gaudin models, while for the classical many-body systems the non-relativistic limit $\eta \rightarrow 0$ (more, precisely $\eta:=\epsilon\nu$, $\epsilon=1/c\rightarrow 0$) provides
the Calogero-Moser systems from the Ruijsenaars-Schneider. These limits may be described by the following diagram:
\beq \label{D1}
\displaystyle{
\begin{tikzcd}
XXX \text{ spin chain} \arrow{d}{\hbar \rightarrow 0} \arrow[r, leftrightarrow, "\text{QC}"]
& \text{Ruijsenaars-Schneider} \arrow{d}{\eta \rightarrow 0} \\
\text{Gaudin model} \arrow[r, leftrightarrow, "\text{QC}"]
& \text{Calogero-Moser}
\end{tikzcd}
}
\eq
%\begin{array}{ccc}
%{\rm XXX\ spin\ chain}  &  \stackrel{\rm QC}{\longleftrightarrow}  &  {\rm Ruijsenaars-Schneider\ model}
%\\
%\downarrow \hbar \rightarrow 0 & & \downarrow \nu \rightarrow 0
%\\
%{\rm Gaudin\ model}  &  \stackrel{\rm QC}{\longleftrightarrow}  &  {\rm Calogero-Moser\ model}
 %\end{array}
%\eq
%
These limits are well-known and well studied. However, there exists a more
non-trivial limit $\hbar,\eta \rightarrow \infty$ which we are going to study in
this paper from the perspective of quantum-classical duality. The limit $\eta
\rightarrow \infty$ in the classical RS systems is easily obtained and one gets the
goldfish models \cite{Calogero},\cite{Feher}. Unfortunately, this limit is quite
singular meaning that the Lax matrix for goldfish models obtained by this limit from the
Lax matrices of RS models become degenerate, thus, not providing enough integrals of
motion for integrability. Instead of the Lax representation one can deal with a set of Hamiltonians. The limit $\hbar \rightarrow
\infty$ for the quantum models is more complicated. To do such a limit it is convenient to
consider ASEP integrable stochastic model. From the point of view of
quantum integrable models it is exactly the XXZ spin chain (up to the conjugation of
$R$-matrix). The limit of the ASEP model is available $\hbar \rightarrow
\infty$ (in some notations it corresponds to $q=e^{\pm \hbar} \rightarrow 0$ or $\infty$). It yields the well studied TASEP system, which describes the jumps of particles only in
one direction.  The latter system is used to describe different physical models ranging from
biology \cite{CSN} to stochastic integrable processes \cite{L}. The diagrams which we
are going to study in this paper are the following:
\beq \label{D2}
\displaystyle{
\begin{tikzcd}
XXZ \text{ spin chain} \arrow{d}{\hbar \rightarrow \infty} \arrow[r, leftrightarrow, "\text{QC}"]
& \text{Trigonometric RS} \arrow{d}{\eta \rightarrow \infty} \\
\text{TASEP} \arrow[r, leftrightarrow, "\text{QC}"]
& \text{Trigonometric goldfish}
\end{tikzcd}
}
%\begin{array}{ccc}
%{\rm XXZ\ spin\ chain}  &  \stackrel{\rm QC}{\longleftrightarrow}  &  {\rm Trigonometric\ RS}
%\\
%\downarrow \hbar \rightarrow \infty & & \downarrow \nu \rightarrow \infty
%%\phantom{\Bigg[}
%\\
%{\rm TASEP}  &  \stackrel{\rm QC}{\longleftrightarrow}  &  {\rm Trigonometric\ goldfish}
%\end{array}
\eq
and
\beq \label{D3}
\displaystyle{
\begin{tikzcd}
XXX \text{ spin chain} \arrow{d}{\hbar \rightarrow \infty} \arrow[r, leftrightarrow, "\text{QC}"]
& \text{rational RS} \arrow{d}{\eta \rightarrow \infty} \\
\text{Rational 5-vertex model} \arrow[r, leftrightarrow, "\text{QC}"]
& \text{Rational goldfish}
\end{tikzcd}
}
%\begin{array}{ccc}
%{\rm XXX\ spin\ chain}  &  \stackrel{\rm QC}{\longleftrightarrow}  &  {\rm rational\ RS}
%\\
%\downarrow \hbar \rightarrow \infty & & \downarrow \nu \rightarrow \infty
%\\
%{\rm rational\ 5-vertex\ model}  &  \stackrel{\rm QC}{\longleftrightarrow}  &  {\rm rational\ goldfish}
%\end{array}
\eq
The limit $\hbar \rightarrow \infty$ from XXX spin chain to the five-vertex model is not so well understood due to our knowledge. It is not easy to
obtain the $R$ -matrix for the rational 5-vertex model from the Yang's
$R$-matrix as a proper limit, however, the eigenvalues and Bethe
equations for the 5-vertex model are easily obtained by the limit $\hbar
\rightarrow \infty$ from the corresponding eigenvalues and Bethe equations for
the XXX spin chain. We call the multispecies TASEP (\ref{q24}) and the rational quantum model (\ref{q10}) as five-vertex models because in the lowest rank case the $R$-matrices are five vertex.

Later we shall discuss in details the QQ duality between
the Goldfish models and multi-species periodic TASEP.

Another example of duality, which we shall discuss later
concerns the fact that rational and trigonometric
Calogero's Goldfish models are p-q (Ruijsenaars) dual to
non-relativistic and relativistic open Toda chains \cite{Feher}, \cite{Ruij}.
At the quantum level, it was proved that Whittaker functions, which solve the quantum open Toda chain problem are simultaneously the eigenfunctions for the rational Goldfish model in dual variables \cite{Babelon,glo1,glo2,SklToda}.
\beq \label{D4}
\displaystyle{
\begin{tikzcd}
  \text{XXX chain} \arrow[d,leftrightarrow,"\text{spectral duality}"] \arrow[r, leftrightarrow, "\text{QC}"]
& \text{class. rational RS} \arrow[d,leftrightarrow," \text{spectral duality}"] \\
\text{trig. Gaudin } \arrow[r, leftrightarrow, "\text{QC}"]
& \text{class. trig. Calogero}
\end{tikzcd}
}
\eq

\noindent Thus, one may hope for the following diagram to exist
\beq \label{D5}
\displaystyle{
\begin{tikzcd}
 \text{5-Vertex models} \arrow[d,leftrightarrow,"\text{spectral duality}"] \arrow[r, leftrightarrow, "\text{QC}"]
& \text{goldfish models} \arrow[d,leftrightarrow,"\text{spectral duality}"] \\
\:\:\;\;\;\;\;\;\;\text{?}\;\;\;\;\;\;\;\:\: \arrow[r, leftrightarrow, "\text{QC}"]
& \text{Toda systems}
\end{tikzcd}
}
%\begin{array}{ccc}
%{\rm 5-vertex\ models}  &  \stackrel{\rm QC}{\longleftrightarrow}  & % {\rm goldfish}
%\\
%\downarrow {\rm spectral\ duality} & & \downarrow  {\rm spectral\ %duality}
%\\
%{\rm ?}  &  \stackrel{\rm QC}{\longleftrightarrow}  &  {\rm Toda}
%\end{array}
\eq
One may try to search for the "question mark" system which will be spectrally dual to 5-vertex models and quantum-classically dual to open Toda chains. One
attempt to find such a system corresponding to Toda system was done in
\cite{GK} using quantum cohomology.

In the appendix we present a classification of five-vertex $R$-matrices which
was inspired by the attempt to find the $R$-matrix which corresponds to the
quantum integrable model which will be quantum-classically dual to the
rational Calogero's goldfish. An interesting task would be to find the quantum-classically dual classical models for all $R$-matrices from appendix,
supposedly, all such models would be dual to Goldfish models.

\subsubsection*{\underline{Cherednik's approach, local tKZ equation and stochastic models}}
The (q)KZ equations can be equivalently formulated in terms of generators of Hecke algebras (including its extension of affine and double affine types) \cite{Ch1,Ch2}. As we mentioned previously, the bispectral duality is described in this way. It interchanges the Cherednik-Dunkl and Demazure-Lustig operators. At the same time the Matsuo-Cherednik projection is also formulated in terms of Hecke algebras. Following \cite{Kasatani} one can define the tKZ family of solutions of polynomial (q)KZ equations. Then the Matsuo-Cherednik projection is just a symmetrization of these (non-symmetric) solutions. A crucial role in this construction plays the local KZ equations, which states
$\hat{s}_i{\bf\Psi}=P_{i,i+1}R_{i,i+1}{\bf \Psi}$, so that the action of $R$-matrix in some vector (spin) representation is given by permutation of variables $z_i$ and $z_{i+1}$ by ${\hat s}_i$ (related to polynomial representation). The local tKZ equation is equivalently written in the form of Zamolodchikov algebra $R_{12}A_1A_2=A_2A_1$. Together with some additional conditions this algebra can be used as main link to stochastic type models \cite{rag2}. Then it becomes possible to study the stochastic dualities in the context of (q)KZ equations \cite{chen,cantini,garbali}. In the continuation of this paper we will consider these contractions in detail and explain interrelations between stochastic dualities and dualities in many-body systems using the above arguments.

\section{ Integrable probabilities }

In this Section we  comment on higher spin six-vertex-ASEP-TASEP  and Schur-Macdonald families of
non-equilibrium stochastic models. Certainly our comments
are very brief and we recommend
the reader to consult the reviews \cite{rag,cor12,bprev}.

\subsection{Six-vertex-ASEP-TASEP family}
The large family of non-equilibrium stochastic processes involves
ASEP, TASEP and their numerous versions (PushTASEP etc). The higher spin
stochastic 6-vertex model is at the top of this family.
The six-vertex models are
the space-time discretization of KPZ equation and
ASEP model can be considered as the continuous time limit of stochastic 6-vertex model,
which describes
the system of  particles on the line with  particular rules
of the  evolution.
%In this 'case
%the height functions and corresponding "`polymer partition sum Z"'
%can be introduced as well, for example, for ASEP
%\beq\label{rr6}
%Z_{ASEP} =\exp (h_{ASEP})
%\eq
The integrable many-body spin systems emerge if we present the Markov
operator as the operator
for Hamiltonian evolution \cite{spohn2,asep}. For instance, the Markov
evolution for joint  probabilities $P$ for ASEP reads as follows \cite{Gwa}:
\beq\label{rr7}
\frac{dP}{dt}=M P\,,
\eq
where the Markov matrix $M$  for homogeneous  periodic lattice of size $L$
in the spin representation reads as
\beq\label{rr8}
M=\sum_{i=1}^{L}(sS_i^{-}S_{i+1}^{+}+ rS_i^{+}S_{i+1}^{-} + \frac{1}{4}S_i^{z}S_{i+1}^{z}) -\frac{L}{4}\,.
\eq
Here $s$ and $r$ are real parameters which parameterize the  jump rates to the left and to the right respectively,
and $S^{-}_i,S^{+}_i,S^{z}_i$ are three generators of $SU(2)$ algebra acting on i-th site.
Upon the gauge transformation $M$ gets mapped into the Hamiltonian of  periodic $SU(2)$ XXZ Heisenberg (quantum) spin chain
\beq\label{rr9}
H_{XXZ}= \sum_{i=1}^{L}(S_i^{x}S_{i+1}^{x}+ S_i^{y}S_{i+1}^{y} + \frac{\alpha + \alpha^{-1}}{2}S_i^{z}S_{i+1}^{z})
+\alpha^{L}S_L^{-}S_1^{+} + \alpha^{-L}S_L^{+}S_1^{-} + \frac{\alpha + \alpha^{-1}}{2}S_L^{z}S_{1}^{z}\,,
\eq
where
\beq\label{rr10}
\alpha=\sqrt{\frac{s}{r}}\,, \qquad S^{+}_{L+1}=\alpha^{L}S^{+}_{1}\,, \qquad S^{-}_{L+1}=\alpha^{-L}S^{-}_{1}\,.
\eq
To get the TASEP model we consider the limit $s=0$, that is directed Markov evolution. For symmetric case (SEP) one selects $s=r=1/2$ and
the XXX Hamiltonian (\ref{y482}) emerges.
The stationary solution to the (\ref{rr8}) with $E=0$ corresponds to the steady state whose eigenfunction can be presented in the matrix product form.

%\subsection{Inhomogeneous stochastic models}

In this study we shall be interested in inhomogeneous
stochastic ASEP model and its TASEP limit  when the jump rates for each particle
are different. In the spin chain language we are considering
the inhomogeneous  XXZ spin chain with  local spectral parameters.
Instead of the local Hamiltonian which governs a Markovian
evolution of the probability of the homogeneous models in
continuous time, the inhomogeneous models are described in terms
of the transfer matrix
\beq
T(x|\vec{z})=tr_0\Big[R_{0,L}(\frac{x}{z_L})R_{0,L-1}(\frac{x}{z_{L-1}})\dots
R_{0,1}(\frac{x}{z_1})\Big],
\eq
where $R$ is  $R$-matrix of the corresponding model. It is considered in the
particular gauge respecting a unitarity which is necessary
for the probabilistic interpretation. The spectral parameter
$x$ generically is a complex variable and the inhomogeneities
$z_i$ are also arbitrary ${\mathbb C}$-numbers. Transfer matrix commutes
for different values of the spectral parameter $x$ due to the
Yang-Baxter equation.

Instead of a continuous time evolution for the homogeneous
TASEP only the discrete-time evolution is available now
with the update dictated by the Markov matrix:
\beq
P(t+1) = M(x|\vec{z})P(t)\,,
\eq
where the Markov matrix $M(x|\vec{z})$ is the normalized transfer matrix
\beq
M(x|\vec{z})= T(x|\vec{z})T(z_1|\vec{z})^{-1}\,.
\eq
In the limit $z_i\rightarrow 1$ the continuous time TASEP model
is restored and the Markov matrix in this case reads as $M'(1)$.

The analogue of the steady state can be defined in the
inhomogeneous case as well. It corresponds to the state
\beq
M|S>=|S>\,.
\eq
If one introduces
the vector
\beq \label{vec}
\displaystyle{
v(z)=\left( \begin{array}{cccc}
z \\
1
\end{array}\right)\,,
}
\eq
which satisfies the associative Zamolodchikov-Faddeev quadratic algebra
\beq
R_{12}(z_2/z_1)v_1(z_1)v_2(z_2)= v_2(z_2)v_1(z_1)\,.
\eq
Then the steady state can be written as
\beq
|S>=v_1(z_1)v_2(z_2)\dots v_L(z_L)\,.
\eq
The steady state for a generic open lattice with boundary conditions
can be presented in a similar matrix product form
\beq
P_{steady}\propto <W|A(z_1) \dots A(z_N)|V>\,,
\eq
where $A(z)$ is the vector
\beq \label{vec2}
\displaystyle{
A(z)=\left( \begin{array}{cccc}
E(z) \\
D(z)
\end{array}\right)
}
\eq
obeying the quadratic Faddeev-Zamolodchikov
algebra
\begin{equation}
    [E(z_1),E(z_2)]=0, \qquad  [D(z_1),D(z_2)]=0, \qquad \frac{z_1}{z_2}D(z_1)E(z_2)= D(z_2)E(z_1)
    \end{equation}
and $z_i$ parameterize a local jump rate.

The inhomogeneous ASEP and TASEP models are solved via the corresponding
Bethe ansatz equations. Let us emphasize that due to the
inhomogeneities we can build the non-local conservation laws
in the stochastic processes similarly to the non-local conservation
laws in the inhomogeneous spin chains. We shall derive them for TASEP  model as the
residues of the transfer matrix at the inhomogeneities.
It is these non-local Hamiltonians, which will play an prominent role
in our study.

\subsection{Schur-Macdonald family}
\paragraph{Schur process}

The Schur process has been introduced in \cite{okoun1}. It
 generalizes the Schur  measure \cite{okoun2} $M_{\lambda}(x,y)$ on the partition $\lambda$:
$$M_{\lambda}(x,y) =s_{\lambda}(x)s_{\lambda}(y)\,,$$ where $(x,y)$ are two
sets of variables.
The Schur process generalizes the Schur measure for the time-dependent weight.
It is defined on sequences of partitions $\lambda(t)$:
\beq
Prob(\lambda(t))\propto \prod S^{(t)}(\lambda(t),\lambda(t+1))\,,
\eq
where $S^{(t)}(\lambda,\mu)$ is a generalization of  the Schur skew  function.

To fix the specialization of the Schur skew function one introduces
the function $\phi(z)$ non-vanishing at a unit circle
with the Wiener-Hopf factorization $\phi(x) =\phi^{+}(z)\phi^{-}(z)$,
where
\beq
\phi^{\pm}= 1 +\sum_k \phi_k^{\pm}z^k\,.
\eq
The skew Schur function   is defined as
\beq
\det(\phi^{+}_{\lambda_i -\mu_j +j-i})= s_{\lambda/\mu}(\phi^{+})\,.
\eq
Then the transition weight is  as follows:
\beq
S_{\phi}=\sum_{\nu}  s_{\lambda/\nu}(\phi^{-})  s_{\mu/\nu}(\phi^{+})\,,
\eq
and the probabilities for the Schur process are given by
\beq
Prob(\{ \lambda(t)\}) =Z^{-1} \prod_{m\in Z+1/2}S_{\phi(m)}(\lambda(m-1/2),\lambda(m+1/2))\,,
\eq
where $Z$-is a normalization factor. The Schur process is Markovian and
at the particular specialization of the Schur polynomials it yields the 3D Young diagram with the $q^{\pi}$ measure, which amounts to the MacMachon
generating function \cite{okoun1} .

\paragraph{Macdonald process}

The Macdonald process was suggested in \cite{borodin11}. It is at the top of this family of integrable probabilities. It generalizes the Macdonald measure written in terms of Macdonald polynomials
depending on two sets of variables as
\begin{equation}
M_{MM}(a,b) = \frac{P_{\lambda}(a) Q_{\lambda}(b)}{\Pi(a,b)}\,,
\end{equation}
where $Q_{\lambda}$ form the dual basis for Macdonald polynomials.
We will be mainly interested in ascending Macdonald process
$M_{as}(a_1,\dots ,a_N; \rho:q,t)$ providing the probability measure
on interlacing partitions
\beq
0 \prec \lambda^1 \prec  \lambda^2 \dots \prec  \lambda^ N\,,
\eq
where the number of non-zero elements in $\lambda^m$ is at most $m$. The sequence
of partitions $\lambda^i$ is interlaced which for two partitions
$\lambda$ and $\mu$ means that $\mu_i\prec \lambda_i \prec \mu_{i-1}$.
Such interlacing triangular arrays of non-integer integers represent
the Gelfand-Tsetlin (GT) patterns. The interlacing condition
is the generalization of the interlacing condition for the
corner GUE process.

The Macdonald process on the GT patterns is based on the branching
rule for the Macdonald polynomials
\begin{equation}
  P_{\lambda^N}(a_1,a_2,\dots a_N)= \sum_{\lambda^{N-1} \prec \lambda^N}
  P_{\lambda^{N-1}}(a_1,a_2,\dots ,a_{N-1})P_{\lambda^{N}/\lambda^{N-1}}(a_N)\,,
\end{equation}
where the sum is over all partitions $\lambda^{N-1}$ which
interlace with $\lambda^N$. The branching rule yields
the discrete Markovian evolution from the level $N$ to the level $N-1$
with the kernel
\begin{equation}
    \Lambda^N_{N-1}(\lambda^N, \lambda^{N-1})=
    \frac{P_{\lambda^{N-1}}(a_1,a_2,\dots ,a_{N-1})P_{\lambda^{N}/\lambda^{N-1}}(a_N)}
    {P_{\lambda^N}(a_1,a_2,\dots a_N)}\,.
\end{equation}

The Macdonald process defined via the product of the Markov kernels is
indexed by positive variables $a_1,\dots a_N$ and non-negative
specialization $\rho$:
\beq
M_{as}(a_1,\dots ,a_N; \rho)(\lambda^1 \dots \lambda^N)
=\frac{P_{\lambda^1}(a_1)P_{\lambda^2/\lambda^1}(a_2) \dots
P_{\lambda^N/\lambda^{N-1}}(a_N)Q_{\lambda^N}(\rho)}
{\Pi(a_1,\dots a_N;\rho)}\,.
\eq
The normalization factor is defined as
\beq
\sum_{\lambda,l(\lambda)\leq n}P_{\lambda} (a_1,\dots ,a_N)
Q_{\lambda}(\rho)= \Pi(a_1,\dots _N;\rho)\,,
\eq
where $P$ and $Q$ are symmetric Macdonald functions.

The ascending process can be considered as a peculiar
specialization of the generic Macdonald process when
two different discrete Markovian dynamics are possible
\cite{borodin11}. The second Markov kernel acts at the
fixed level of GT scheme and can be considered as the $(q,t)$ deformation
of the Brownian motion. For general Macdonald process
the combination of two kernels amounts to the evolution
rules on the whole GT pattern.

If  $\lambda^k=\lambda, k=1 \dots N$ the ascending Macdonald
process reduces to the Macdonald measure. If $t=0$ the
Macdonald process gets reduced to the q-Whittaker process,
which further reduces to the Schur process at $q=0$ or to the
Whittaker process at $q=1$.
For instance, the q-Whittaker
Markovian dynamics on GT scheme with partitions $\lambda_k^{n} $
involves the local coordinate depending jump rate, which reads as
\beq
a_n\frac{(1- q^{\lambda^{(n-1)}_{k-1}- \lambda^{(n)}_{k}})
(1- q^{\lambda^{(n)}_{k}- \lambda^{(n)}_{k+1} +1})}
{(1- q^{\lambda^{(n)}_{k}- \lambda^{(n-1)}_{k} +1})}\,.
\eq
Let us emphasize that from the Hamiltonian viewpoint the Macdonald
polynomial $P_{\lambda}(x)$ is the wave function of the trigonometric RS system. Hence,
the evolution of the partition $\lambda(t)$ and the corresponding measure,
which occurs in the  Macdonald process should be treated as the evolution in the space of integrals of motion in the integrable system. The trigonometric
RS system with $N$-particles naturally emerges in the field theory and has been identified in the $SU(N)$ CS theory on $T^2\times R$  at level $k$ perturbed by  Wilson loop and Polyakov line  \cite{gn}. The corresponding phase space is the
moduli space of $SU(N)$ flat connections on the torus with the marked
point. The monodromy around the marked point defines the RS coupling
constant $\nu$ corresponding to the one-raw representation of $SU(N)$. The path
integral suggested in \cite{gn} provides the representation
of the Macdonald polynomial as the twisted character for the
quantum group
\beq
P_{\lambda}(\vec{x})= Tr_{\lambda}(V_{\nu} e^{i\vec{t}\vec{x}})\,,
\eq
where $V_{\nu}$ is the intertwiner in $M_{\lambda}\leftarrow{ M_\lambda\times O_{\nu}}$ between the Verma module and the finite-dimensional representation.

Remark also that $N$-particle trigonometric RS model also describes the $N$-soliton sector in sin-Gordon model \cite{bernard}.
This representation is naturally related with the matrix product
representation of the Macdonald polynomials and provides the wave function
of $N$ solitons upon quantization.

%\section{Wave function $\rightarrow$ measure $\rightarrow$ process. Toy %example}

\section{Dualities in stochastic processes versus dualities in integrable many-body systems}
\subsection{QQ duality}
In this Section we shall argue that  dualities in integrable systems we have
reviewed in the previous Sections have a clear-cut counterpart
in the world of integrable probabilities. We  claim that a couple
of results concerning the relations between the seemingly unrelated
objects in the integrable probability framework can be collected under
the single roof of the duality
between the families of many-body integrable systems. The following results
are of importance for us:
\begin{itemize}
    \item The non-symmetric Macdonald polynomial gets identified with
    the partition function of the ensemble of the colored paths on the
    lattice cylinder with inhomogeneities,
    nontrivial boundary conditions and 6-vertex $R$-matrix \cite{BW19}.
    \item There is the identity of the  expectation value of particular observables evaluated
    with the Macdonald measure at one side and evaluated with the measure
    on the non-abelian stochastic non-homogeneous  6-vertex model \cite{borodin} on the other side.
    There is analogous identity of the  expectation value of  observables evaluated
    with the measure on the processes in the $q=0$ \cite{BBW} and $t=0$
    \cite{orr}
    limits of Macdonald process and particular processes at 6-vertex side.
    \item The partition functions of the inhomogeneous stochastic 6-vertex
    model obey the qKZ(KZ) difference equations with respect to the inhomogeneities \cite{chen}.
    \item Both Macdonald polynomials \cite{cantini} and the partition function of inhomogeneous
    stochastic model \cite{rag} admit the matrix product representation.

    \end{itemize}
To the best of our knowledge these results have not been related with the
dualities between the families of the integrable many-body systems so far.

\subsubsection{Wave functions}
First, let us argue   that QQ duality
in integrable systems relating symmetric Macdonald
polynomials, which are wave functions of trigonometric
Ruijsenaars model,
with the solutions to qKZ equations
supplemented with the Matsuo-Cherednik projection
has clear-cut counterpart in the integrable probabilities.
The elements of the mapping can be summarized as follows:
\begin{itemize}
    \item

    The identification of inhomogeneities.

The mapping of inhomogeneities and arguments of the Macdonald polynomials
polynomial can be seen in different representations
at the probabilistic side. In particular, it is clear in  representation of non-symmetric Macdonald polynomials in terms of colored paths on the cylinder and
solutions to qKZ equations for higher spin six-vertex models \cite{BW19}.
Similarly one can use the
"matrix product representation" developed in \cite{ cantini,garbali}.
It allows to present the Macdonald polynomial
in the form
\beq
P_{\lambda}(x_1,\dots,x_n|q,t)\propto <C(x_1)\dots C(x_n)>_{\lambda}\,,
\eq
where $C(x_i)$ is the element of the auxiliary transfer matrix
taken at some point. Comparing with the matrix product representation of the probability in inhomogeneous ASEP-like model
we immediately identify the inhomogeneities and the arguments of the Macdonald polynomial.

Note that the similar matrix product representation
for the Jack polynomials has been developed in \cite{cardy}.
In this case the operators $C(z_i)$ get identified with the
operators $\Psi_{2,1}(z_i)$ from the 2d conformal models
with peculiar
central charges and the relation with the
radial SLE stochastic process has been clarified. We briefly
review this point in Appendix B.

    \item Interpretation of twists.

    To recognize the proper identification of twists let us
take use of the representation of Macdonald polynomials
as sums over the coloured discrete paths on the cylinder of
size $L\times n$, where $U(\hat{sl(n+1)})$ is involved
\cite{BW19} as the underlying algebraic structure.
The parameter $q$ governs the wrappings around the cylinder
and for $q=0$  wrappings are forbidden. In the picture
of the Macdonald polynomial as the sum over colored paths the spectral
argument $\lambda$ of $P_{\lambda}(x)$ provides
the boundary conditions for the colored paths as exactly
expected from the QQ duality.

    \item Eigenvalues of the Markov operator.

    One more element of mapping between
     dualities concerns the selection of the
eigenvalues of the non-local Hamiltonians,
or equivalently, the eigenvalue of the transfer-matrix
of the inhomogeneous chain.
The steady state at the stochastic Markov spin chain side
plays a special role.
In this case  the eigenvalues of the non-local Hamiltonians
are identical  and the
eigenfunctions allow the factorizable representation
in terms of operators of the Faddeev-Zamolodchikov algebra.
At the Macdonald side the steady state corresponds to
the situation when the velocities of all degrees of
freedom are identical, hence the corresponding state
is highly coherent.

However QQ and QC duality in integrable models
are not restricted to the specific values
of non-local Hamiltonians, hence we can consider
the QQ duality  for integrable probabilities  beyond
the steady states as well.

    \item qKZ, Matsuo-Cherednik and semiclassics.

    The last question concerns the analogue of the Matsuo-Cherednik
    projection in integrable probability framework. The point is that the colored path
representation of the Macdonald polynomials \cite{BW19} yields
the non-symmetric Macdonald's, hence to get
the symmetric Macdonald's one has to apply the
symmetrization procedure, which is precise counterpart of the
Matsuo-Cherednik projection.

Remark that the wave
functions at the Macdonald side get mapped into the
solutions to qKZ equations at the stochastic six-vertex side
\cite{chen} involving R-matrix but there are subtleties
concerning the choice of $R$-matrix at the inhomogeneous
spin chain (ASEP,TASEP) side. It has to be chosen in
the proper gauge to ensure the conservation of unitarity, which
is required for the probabilistic interpretation.

The mapping of QQ duality and stochastic duality allows
to fix properly the mapping of parameters of quantization. The Planck
constant at Macdonald side $\hbar_{RS}$ gets mapped into
the parameter of qKZ equation, while the Planck constant
at the 6-vertex-ASEP side $\hbar$ gets identified with the
coupling constant in the Macdonald operator. The semiclassical
expansion at $\hbar_{RS}\rightarrow 0$ amounts to the
expression of the RS wave function, or equivalently, solution
to qKZ to the integral form involving solution to the BA equations.

\end{itemize}

\subsubsection{Measures}

The duality between the Macdonald polynomial
and the partition function of the stochastic colored inhomogeneous model \cite{BW19}
was extended to the evaluation of the particular
observables with the corresponding measures. The
most general relation reads for any $0\leq l\leq N$ \cite{borodin}
as follows:
\beq\label{rr11}
(-1)^lE_{6v}(\prod_{i=1}^{l} \frac{Q^{h(M,N)} - Q^{i-1}}{1-Q^i})=
E_{MM}(e_l(q^{\lambda_1}t^{n-1}, q^{\lambda_2}t^{n-2},\dots, q^{\lambda_n}))\,,
\eq
where $e_l$ is the symmetric function, the observables are evaluated
with the Macdonald measure $M(\rho_1,\rho_2)$ and measure for the 6-vertex
inhomogeneous model at the quadrant $(M,N)$  respectively.
A height function assigns to each vertex $(M,N)$ the number
$h(M,N)$ of paths in ensemble that pass through or to the right
of this vertex. The  expectation value at the 6-vertex side is expressed
as the multiple contour integral:
\begin{equation}
\begin{array}{l}
\displaystyle{
 E_{6v}\Big(\prod_{i=1}^{l} (Q^{h(M,N)} -Q^{i-1})\Big)=
 }
 \\ \ \\
 \displaystyle{
 Q^{\frac{l(l-1)}{2}} \oint \frac{dw_i}{2\pi i}
 \dots \oint \frac{dw_l}{2\pi i} \prod_{1\leq a<b\leq l}\frac{w_a-w_b}{w_a-Qw_b} \times
 \prod _{i=1}^{l}\Big( w_i^{-1} \prod_{x=1}^{M-1} \frac{1-s_x\chi_x^{-1}w_i}{1-s_x^{-1}\chi_x^{-1}w_i}
\prod_{y=1}^N\frac{1-Qu_y}{1-Qu_y}\Big)\,.
}
\end{array}
\end{equation}
 The expectation value at 6-vertex side  depends on the inhomogeneities in the
 lattice path model $u_i,s_x,\chi_x$ and parameter $Q$, which enter
 the local statistical weights. The mapping of the 6-vertex parameters
 and parameters involved in the expectation value at the Macdonald side goes as follows.
 First, $Q=t, n=N$,  one specialization
 of the Macdonald measure $M(\rho_1,\rho_2)$  $\rho_1$  gets mapped to inhomogeneities $u_i$
 $\rho_1=(x_1,\dots x_n) =(u_i^{-1},\dots u_n^{-1}) $ while the
 second specialization $\rho_2$ corresponds to $s_i,\chi_i$. Hence,
 similar to the case of wave function the arguments (specializations) of the pair of
 Macdonald polynomials involved into the measure provide the local weights for statistical model.

Similar relation for the $t=0$ limit has been found in \cite{orr} that is
the following relation for the joint probabilities holds \cite{bcgs}:
\beq\label{rr111}
E_{q-TASEP}(\prod_{j=1}^{k} q^{x_{n_i} +n_i})=
E_{q-Whit}(\prod_{j=1}^{k} q^{\lambda_{n_i}} )=
\frac{D_1^k \Pi(a)}{\Pi(a)}\,,
\eq
where at the l.h.s. the joint probability is evaluated for q-TASEP, while at the r.h.s. the joint
probability is evaluated for the q-Whittaker process,$D_1$ is the
$t=0$ limit of Macdonald difference operator acting on $a_i$ variables
\beq
D_1^N P_{\lambda_N^N}(a)= q^{x_N}P_{\lambda_N^N}(a)\,,
\eq
where again  $\lambda_N^N$ are the boundary variables in Gelfand-Tsetlin scheme.
Hence variables $a_i$ are the inhomogeneities in q-TASEP and
simultaneously are the arguments of the probability in q-Whittaker process
as it should be according to QQ duality.

What is the meaning of these identities in the context of QQ duality
between the quantum integrable many-body systems? As we know in the wave functions
for RS model the arguments correspond to the inhomogeneities involved in solution
to qKZ equations, while the spectral arguments $(\lambda_1,\dots ,\lambda_n)$ correspond
to the twists in statistical model, or equivalently, the boundary conditions. Hence, the quantum mechanical expression for some function
averaged through the spectrum  gets translated to the
integral over the boundary twists in the statistical model.

\subsubsection{Processes}

The identification of QQ duality for the Macdonald wave function and Macdonald
measure in the framework of integrable probability can be extended to the
stochastic processes.
The available examples of  stochastic duality for processes relate the Hall-Littlewood
process and the higher spin (SU(N)) six-vertex model \cite{BBW,BW19} as well as the
q-Whittaker process and the q-TASEP model \cite{orr,borodin}. Consider first the
$q=0$ limit of the Macdonald process corresponding to the Hall-Littlewood process.
It is defined as
\beq
W_{\vec{a},\vec{b}}\propto P_{\lambda^1}(a_1) P_{\lambda^2/\lambda^1}(a_2) \dots
P_{\lambda^M/\lambda^{M-1}}(a_M)Q_{\lambda^M}(b_1,\dots b_N)\,,
\eq
where $P,Q$ are Hall-Littlewood polynomials. The probability of the process
$\lambda^i=\lambda$ for all i reads as $P_{\lambda}(a_1,\dots a_M)Q_{\lambda}(b_1,\dots b_N)$.

It was argued in \cite{BBW,BW19} that the proper observables in the Hall-Littlewood
process and higher spin six-vertex model with inhomogeneities $z_i$  at the quadrant $N\times M$ get mapped
to each other upon the identifications
\beq
z_i \Longleftrightarrow a_i \qquad v_i \Longleftrightarrow b_i
\eq
The parameter $t$ in the Hall-Littlewood polynomial gets identified with the
parameter Q in the six-vertex model.

Let us make a comment concerning the process in terms of the
matrix product representation. The Macdonald polynomial
has the structure
\begin{equation}
    P_{\lambda}(a_1,\dots a_N)\propto <\lambda|A(a_1)\dots A(a_N)|0>\,,
\end{equation}
which allows the representation in terms of the product of matrix elements
via insertion of the full basis of states
\begin{equation}
    P_{\lambda}(a_1,\dots a_N)\propto \sum_{\lambda_1 \dots \lambda_N}
    <\lambda|A(a_1)|\lambda_1> <\lambda_1|A(a_2)|\lambda_2> \dots <\lambda_N| A(a_N)|0>\,.
\end{equation}
That is, we sum over all paths in the space of states. However, when
we consider the process we select the specific part in the Hilbert space
when the initial and final states are fixed.

We make just a few general remarks concerning the dual
interpretation of the processes at the spin chain side
postponing the detailed analysis for further coming publication.
The chain of polynomials at the Macdonald side corresponds
to the chain of  solutions to qKZ equation for the inhomogeneous
spin chain of the TASEP type.
The dual array of spin chains is defined on the Gelfand-Tsetlin scheme.
Each spin chain in the process is inhomogeneous however  with only
one non-vanishing inhomogeneity $a_i \neq 0$ at each i-th factor.

The interlacing condition for partitions $\lambda^i$ in GT at
Macdonald side gets mapped via QQ duality into interlacing condition for the
boundary conditions for XXZ spin chains. Therefore at the discrete time
step of the dual XXZ process the single non-trivial inhomogeneity gets
shifted $a(t_i)\rightarrow a(t_I+1)$ and the boundary condition
gets modified respecting interlacing property.

In the next Section  we consider as our main example the
stochastic duality for the large coupling limit $\nu \rightarrow \infty$
of Macdonald polynomials without performing the Inozemtsev rescaling
of coordinates. In this peculiar limit at the Macdonald side we get new process corresponding to the Goldfish integrable many-body system.
At the dual six-vertex side  we get the inhomogeneous multi-species
periodic TASEP model with nontrivial twists.

\subsection {QC duality in stochastic processes}

The QC duality for the classical $\hbar_{RS} \rightarrow 0$ limit of the Macdonald
polynomials to the classical actions of the
trigonometric RS model is quite standard.
The wave function written in the peculiar Givental-like form
amounts to the condition for intersection of Lagrangian submanifolds
while at the spin-ASEP side a similar semiclassical limit
of the solution to qKZ equation is considered which involves
the product of Bethe on-shell vectors depending on the solutions
to the BA equations.

For a process each factor $<\lambda|A(z_i)|\mu>$ yields not
an intersection of two Lagrangian manifolds indexed by $(z_i,\lambda)$
but a intersection of three Lagrangian submanifolds indexed by
$(z_i,\lambda,\mu)$. The Lagrangian submanifolds indexed
by $\lambda_i$ obey the interlacing condition. At the spin chain -ASEP
side the saddle point solution for the integral representation of
each involved solution to qKZ yields the hierarchy of BA equations.
The interlacing structure for the boundary conditions yields
a kind of nesting structure for BA hierarchy.

To illustrate the semiclassical limit at the ASEP side let us
consider the simplest limit of ASEP model when the Markov Hamiltonian
gets identified with the transfer matrix for the twisted Gaudin model,
whose wave functions obey the KZ equations with respect
to inhomogeneities. It goes as follows.
Recall the non-local $SU(2)$  Gaudin Hamiltonians
\begin{equation}
    \hat{H_i}= 2\Lambda\hat{s^z_i} - \sum_{i\neq j}\frac{\vec{{\hat s}}_j\vec{\hat{s}}_i}{z_i-z_j}
\end{equation}
and introduce the function
\begin{equation}
\Psi(\vec{\lambda},\vec{z})= \prod_{i=1}^{M}L^{+}(\lambda_i)|0>\,,\qquad
L^{+}(\lambda)=\sum_{j=1}^N \frac{{\hat s}_j^+}{\lambda -z_j}\,,
\end{equation}
where $|0>$ is the minimal weight state and $z_i$ are inhomogeneities
and $\Lambda$ is diagonal twist.
The function $\Psi(\vec{\lambda},\vec{z})$ is the so-called off-shell
Bethe state. The Gaudin Hamiltonians act on the off-shell Bethe state
as follows
\begin{equation}
    {\hat H}_j\Psi =h_j\Psi + \sum_i \frac{ g_i s_j\Psi_i}{\lambda_i -z_j}
\end{equation}
and
%where $\Psi_i$ is vector derived from $\Psi$ removing %$L^+(\lambda_i)$.
\begin{equation}
    h_i= -2\Lambda s_i - \sum_{i\neq j}\frac{s_is_j}{z_j-z_k} + \sum_j\frac{s_j}{z_i-\lambda_j}\,, \end{equation}
    \begin{equation}
    g_i=2\Lambda -  \sum_j\frac{s_j}{z_i-\lambda_j} -
    \sum_{j\neq j}\frac{1}{\lambda_i-\lambda_j}\,.
\end{equation}
At the next step the Yang-Yang function $W$ is introduced via relations
\begin{equation}
    \frac{\partial W}{\partial z_i}=h_i\,, \qquad
    \frac{\partial W}{\partial \lambda_i }=g_i\,,
\end{equation}
which can be written  explicitly
\begin{equation}
W(\vec{\lambda},\vec{z})= -2\Lambda \sum_iz_is_i + 2\Lambda \sum_i \lambda_i
+ \sum_i\sum_j s_i ln(z_i-\lambda_j) -
\end{equation}
$$
-\frac{1}{2} \sum_i\sum_j s_j s_i ln(z_i-z_j)
-\frac{1}{2}\sum_i\sum_j  ln(\lambda_i-\lambda_j)\,.
$$
With these notations
the solution to the KZ equations can be written in terms of
a off-shell Bethe vector and Yang-Yang function as follows
\begin{equation}
\Psi_{KZ}(\vec{z})= \oint d \vec{\lambda} \exp(\frac{iW}{\kappa}) \Psi(\vec{\lambda},\vec{z})\,,
\end{equation}
where $\kappa=\hbar_{Calogero}$ is the Planck constant in Calogero model.
The semiclassical limit at
Calogero side provides the corresponding limit at ASEP side
when the integral is done via the saddle point and the off-shall
Bethe vector degenerates to the on-shell Bethe vector when
the Bethe roots obey the BA equations. Hence, the semiclassical
limit of probabilistic measures and processes have the clear-cut
interpretation at the 6-vertex- ASEP side.

\section{Classical Calogero's Goldfish models}
\subsection{Rational Calogero's Goldfish model}
The rational Calogero's Goldfish model \cite{Calogero},\cite{Ruij} is the classical integrable many-body system with the following Hamiltonian
\beq \label{q1}
H^{\rm{R}} = \sum\limits_{k=1}^{N}e^{p_k}\prod\limits_{j \neq k}^{N}\frac{1}{q_k-q_j}\,,
\eq
so that
\beq \label{q101}
{\dot q}_k=\{H^{\rm{R}},q_k \} =e^{p_k}\prod\limits_{j \neq k}^{N}\frac{1}{q_k-q_j}\,,\qquad
{\dot p}_k=\{H^{\rm{R}},p_k \} =-\p_{q_k}H^{\rm{R}}\,.
\eq
Equations of motion can be written in the Newtonian form:
\beq \label{q2}
\displaystyle{
\ddot{q}_i = 2\sum\limits_{k \neq i}^{N}\frac{\dot{q}_i\dot{q}_k}{q_i-q_k}\,.
}
\eq
The system (\ref{q1}) is $\nu \rightarrow \infty$ limiting case  of the rational Ruijsenaars-Schneider model \cite{RS}. Its Hamiltonian (\ref{y2}) should be first rescaled by $\nu^{N-1}$:
\beq \label{q3}
\displaystyle{
H^{\rm{RS}} = \sum\limits_{k=1}^{N}e^{p_k}\prod\limits_{j \neq k}^{N}\frac{q_k-q_j+\nu}{\nu(q_k-q_j)}\,.
}
\eq
Then the limit $\nu \rightarrow \infty$ of (\ref{q3}) yields (\ref{q1}). The Newtonian form for equations of motion
\beq \label{q4}
\displaystyle{
\ddot{q}_i = -2\sum\limits_{k \neq i}^{N}\frac{\nu^2 \dot{q}_i\dot{q}_k}{(q_i-q_k)((q_i-q_k)^2-\nu^2)}
}
\eq
is independent of the rescaling factor. The limit $\nu \rightarrow \infty$ in (\ref{q4}) gives (\ref{q2}).
%Notice that equations of motion (\ref{q2}) can be easily obtained from (\ref{q4}) by sending $\nu \rightarrow \infty$ %in the right hand side of (\ref{q4}).
Let us also
comment on other origins of system (\ref{q1}).

Besides the direct limit from the Ruijsenaars-Schneider model the system (\ref{q1})
can be derived by analyzing the dynamics of poles for the rational solutions to the Novikov-Veselov equation
\cite{KrZ}. What is more important for us, is that this system is p-q dual to the open
non-relativistic Toda chain \cite{Ruij}, \cite{Feher}. In particular,
it was proved in those papers that the rational Calogero's Goldfish model (\ref{q1}) is a completely
integrable system with the pairwise Poisson commuting independent integrals of motion
\beq \label{I1}
\displaystyle{
H_k^{\rm{R}}(\{\dot{q}\},\{q\}) =(-1)^{\frac{k(k-1)}{2}} \sum\limits_{I_k}\dot{q}_{i_1}\,...\,\dot{q}_{i_k}\prod\limits_{l<j}^{k}(q_{i_l}-q_{i_j})^2\,,
}
\eq
where $I_k$ is a $k$-tuple $(i_1,..,i_k)$ such that $1 \leq i_1<i_2<..<i_k \leq N$. The Hamiltonians
(\ref{I1}) depend on positions of particles $\{q\}=\{q_1,...,q_N\}$ and their velocities $\{{\dot q}\}=\{{\dot q}_1,...,{\dot q}_N\}$, which can be expressed through momenta via (\ref{q101}) and the term $(-1)^{\frac{k(k-1)}{2}}$ is used for further convenience.

\subsection{Trigonometric Calogero's Goldfish model}
To study the trigonometric Goldfish model we first recall basic facts about Macdonald-Ruijsenaars Hamiltonians. Macdonald-Ruijsenaars Hamiltonians (on the classical level) form an integrable many-body system with Poisson commuting Hamiltonians
\beq
\label{MR}
H_r^{MR} = \sum\limits_{I_r} \prod\limits_{i \in I_r, j \notin I_r}\frac{t x_i-x_j}{x_i-x_j} \prod\limits_{m \in I_r}\Theta_m,\;\; 1 \leq r \leq N,
\eq
where $I_r$ is a $r$-tuple $(i_1,..,i_r)$ such that $1 \leq i_1 < \cdots < i_k \leq N$. The symplectic form is given by
\beq
\label{sympl}
\omega = \sum\limits_{i=1}^N \frac{d\Theta_i}{\Theta_i} \wedge \frac{d x_i}{x_i}.
\eq
In the logarithmic coordinates $x_i=e^{q_i}$, $\Theta_i = e^{p_i}$ Hamiltonians (\ref{MR}) and symplectic form (\ref{sympl}) are
\beq
\label{MR1}
\begin{array}{c}
\displaystyle{
H_r^{MR} = \sum\limits_{I_r} \prod\limits_{i \in I_r, j \notin I_r}\frac{t e^{q_i}-e^{q_j}}{e^{q_i}-e^{q_j}} \prod\limits_{m \in I_r}e^{p_m},\;\; 1 \leq r \leq N,
}
\\ \ \\
\displaystyle{
\omega = \sum\limits_{i=1}^N dp_i \wedge dq_i.
}
\end{array}
\eq
In the quantum case joint eigenfunctions of (\ref{MR}) are the Macdonald polynomials. Let us now consider limit $t \rightarrow 0$ of (\ref{MR}), which will be called the trigonometric Goldfish model
\beq
\label{TG}
H_r^{TG}= (-1)^{r(N-r)}\sum\limits_{I_r} \prod\limits_{i \in I_r, j \notin I_r}\frac{e^{q_j}}{e^{q_i}-e^{q_j}} \prod\limits_{m \in I_r}e^{p_m},\;\; 1 \leq r \leq N.
\eq
We are specifically interested in the flow generated by the first Goldfish Hamiltonian
\beq
\label{TG1}
H_1^{TG} = (-1)^{N-1} \sum\limits_{i=1}^N \prod\limits_{j \neq i}^N\frac{e^{q_j}}{e^{q_i}-e^{q_j}} e^{p_i}\,.
\eq
Then
\beq
\label{qdot}
\dot{q}_k =\frac{\partial H_1^{TG}}{\partial p_k} = (-1)^{N-1}\sum\limits_{i=1}^N \prod\limits_{j \neq i}^N \frac{e^{q_j}}{e^{q_k}-e^{q_j}} e^{p_k}\,,
\eq
and the Newtonian equations of motion take the form
\beq
\label{eqmot}
\ddot{q}_k = 2\sum\limits_{j \neq k} \dot{q}_k \dot{q}_j \coth{\left(\frac{q_k-q_j}{2} \right)}
\eq
and the Hamiltonians (\ref{TG}) can be rewritten in terms of velocities (\ref{qdot}) in the following form:
\beq
\label{ham}
H^{TG}_r = (-1)^{\frac{r(r-1)}{2}} 2^{r(r-1)} \sum\limits_{I_r} \left(\prod\limits_{i \in I_r} \dot{q}_i \right) \prod\limits_{m<l}^r \sinh^2{(\frac{q_{i_m} - q_{i_l}}{2})}.
\eq
We call the system (\ref{TG}) the trigonometric Goldfish model mostly due to the fact that the Newtonian equations of motion (\ref{eqmot}) in the rational limit produce equations of motion for the rational Calogero's Goldfish model (\ref{q2}). Also, it is worth mentioning that in quantum case it was proven that system (\ref{TG}) is bispectrally dual to the relativistic quantum Toda chain \cite{glo1}.
\section{Five-Vertex models and their higher rank extensions}

Some basic facts concerning the quantum inverse scattering method and the algebraic Bethe ansatz were described in (\ref{y35})-(\ref{y483}). Below we use these approaches.
The starting point is a quantum $R$-matrix -- solution of the
quantum Yang-Baxter equation:
\beq \label{q9}
\displaystyle{
R_{12}(u_1-u_2)R_{13}(u_1-u_3)R_{23}(u_2-u_3)=R_{23}(u_2-u_3)R_{13}(u_1-u_3)R_{12}(u_1-u_2)\,.
}
\eq
We are looking for solution of joint eigenvalue problem
\beq \label{q161}
\begin{array}{c}
\displaystyle{
\hat{H}_i\psi=h_i(\{z\},\{\mu\})\psi\,,\quad \psi\in{(\mC^n)^{\otimes N}}\,,\quad i=1,...,N\,,
}
\end{array}
\eq
for $N$ commuting operators
\beq \label{q16}
\begin{array}{c}
\displaystyle{
\hat{H}_i = \res\limits_{u=z_i}\hat{t}(u) = R_{i,i-1}(z_i-z_{i-1})...R_{i,1}(z_i-z_1)V_iR_{i,N}(z_i-z_N)...R_{i,i+1}(z_i-z_{i+1}),
}
\end{array}
\eq
where $V=\mathrm{diag}\left(V_1,..,V_n \right)$ is a diagonal twist
matrix. The eigenvalues $h_i(\{z\},\{\mu\})$ depend on the set $\{z\}=\{z_1,...,z_N\}$ and the additional set of variables $\{\mu\}$ satisfying
the algebraic (nested) Bethe ansatz equations.

\subsection{Rational 5-vertex quantum integrable model}
%where $R$ is a map $R:\mathbb{C} \rightarrow \mathrm{End}(\mathbb{C}^n)\otimes
%\mathrm{End}(\mathbb{C}^n)$ and lower subscripts account for the tensor spaces in
%which the $R$-matrix acts nontrivially.
\paragraph{R-matrix.} In this paper we consider the following
solution for quantum Yang-Baxter equation \cite{W}
\beq \label{q10}
\displaystyle{
R(u) = \frac{P}{u}+\sum\limits_{\al > \be}^{n}E_{\al \al}\otimes E_{\be \be}\,,\qquad
P=\sum\limits_{\al,\be}^{n}E_{\al \be}\otimes E_{\be \al}\,,
}
\eq
where $P$ is the permutation
operator and $E_{\al \be}$ are the standard matrix units (standard basis in ${\rm Mat}_n$) with only one non-zero
element in the $\al$-th row and $\be$-th column. For example, for $n=2$ (\ref{q10}) takes the form
\beq \label{q11}
\displaystyle{
R(u)=\left( \begin{array}{cccc}
\frac{1}{u} & 0 & 0 & 0\\
0 & 0 & \frac{1}{u} & 0\\
0 & \frac{1}{u} & 1 & 0\\
0 & 0 & 0 & \frac{1}{u}
\end{array}\right).
}
\eq
It satisfies the unitarity
condition
\beq  \label{a1}
\displaystyle{
%\begin{array}{cc}
R_{12}(u)R_{21}(-u) = -\frac{1}{u^2}\, 1_n\otimes1_n\,.
%\end{array}
}
\eq

\paragraph{Bethe ansatz.} The algebraic Bethe ansatz deals with a certain substitution for the Bethe vector $\psi$
(\ref{q161}), which depends on the parameters -- the Bethe roots $\mu^{i}_a$, $i=1,...,n-1$, $a=1,...,M_{i}$, where
 $M_1\,,...\,,M_{n-1}$ -- are the numbers of excited states (occupation numbers). We assume $N\geq M_1\geq ... \geq M_{n-1}\geq 0$. The eigenvalue of the transfer matrix $\hat{t}(u)$ is obtained in the following form:
\beq \label{q18}
\displaystyle{
\Lambda(u) =V_1\prod\limits_{j=1}^{N}\frac{1}{u-z_j}\prod\limits_{l=1}^{M_1}\frac{1}{\mu^1_{l}-u}+ \sum\limits_{i=2}^{n}V_i\prod\limits_{l=1}^{M_{i-1}}\frac{1}{u-\mu^{i-1}_l}\prod\limits_{k=1}^{M_i}\frac{1}{\mu^i_k-u}\,.
}
\eq
To find the eigenvalues of $\hat{H}_i$ (\ref{q161}) one should compute the residues of (\ref{q18}):
\beq \label{q22}
\displaystyle{
h_i(\{z\},\{\mu\}) =\res\limits_{u=z_i}\Lambda(u)= V_1\prod\limits_{j \neq i}^{N}\frac{1}{z_i-z_j}\prod\limits_{l=1}^{M_1}\frac{1}{\mu^1_l-z_i}\,.
}
\eq
The Bethe roots satisfy the following Bethe equations:
\beq \label{q19}
\displaystyle{
V_1\prod\limits_{k=1}^{N}\frac{1}{\mu^1_m-z_k} = (-1)^{M_1-1}V_2\prod\limits_{k=1}^{M_2}\frac{1}{\mu^2_k-\mu^1_m}\,,
}
\eq
\beq \label{q20}
\displaystyle{
V_i\prod\limits_{l=1}^{M_{i-1}}\frac{1}{\mu^i_m-\mu^{i-1}_l} = (-1)^{M_i-1}V_{i+1}\prod\limits_{k=1}^{M_{i+1}}\frac{1}{\mu^{i+1}_k-\mu^i_m}
}
\eq
for $2 \leq i \leq n-2$, and
\beq \label{q21}
\displaystyle{
V_{n-1}\prod\limits_{l=1}^{M_{n-2}}\frac{1}{\mu^{n-1}_m-\mu^{n-2}_l} = (-1)^{M_{n-1}-1}V_n\,.
}
\eq
Total number of equations is equal to the number of Bethe roots, i.e. to $\sum_{a=1}^{n-1} M_a$.

\paragraph{Local Hamiltonian.} The Hamiltonians (\ref{q16}) are non-local since they contain the interaction terms between all sites.
In order to construct the local Hamiltonian one should put all inhomogeneity parameters to zero $z_i=0$,
set the twist matrix to identity $V=1_n$ and
consider the operator
\beq \label{q17}
\displaystyle{
\hat{H}^{\rm loc} = \left. \frac{d\left(u^{N}\hat{t}(u)\right)}{d u}\left(u^N\hat{t}(u)\right)^{-1} \right|_{u=0}\,.
}
\eq
In $n=2$ case it takes the form
\beq \label{E1}
\hat{H}^{\rm loc}=\sum\limits_{i=1}^{N}\sigma^{+}_i\sigma^{-}_{i+1}\,,
\eq
where $\sigma^{\pm}$ are standard notations for the Pauli matrices and $N+1 = 1 \; (\mathrm{mod} \; N)$.

\subsection{Trigonometric 5-vertex and TASEP model}
In this Section we use multiplicative form for dependence on spectral parameters.
The Yang-Baxter equation (\ref{q9}) is written as
\beq \label{q23}
\displaystyle{
R_{12}\left(x_1/x_2\right)R_{13}\left(x_1/x_3\right)R_{23}\left(x_2/x_3\right)=
R_{23}\left(x_2/x_3\right)R_{13}\left(x_1/x_3\right)R_{12}\left(x_1/x_2\right)\,,
}
\eq
which is equivalent to (\ref{q9}) under change $u_i=\exp(x_i)$.
Similarly to the previous section we construct the transfer matrix
\beq \label{q26}
\displaystyle{
\hat{t}(x) = \tr_{0}\left(R_{0N}(x/z_N)..R_{01}(x/z_1)V_0 \right)\,,
}
\eq
where again $V=\mathrm{diag}\left(V_1,..,V_n\right)$ and $z_i \in \mathbb{C}$ are inhomogeneity parameters. The transfer matrix (\ref{q26}) provides commuting family of operators
\beq \label{q27}
\begin{array}{c}
\displaystyle{
\hat{H}_i = \frac{1}{z_i}\res\limits_{x=z_i}\hat{t}(x)=
R_{i,i-1}(z_i/z_{i-1})...R_{i,1}(z_i/z_1)V_iR_{i,N}(z_i/z_N)...R_{i,i+1}(z_i/z_{i+1})\,,
}
\\ \ \\
\displaystyle{ \left[ \hat{H}_i,\hat{H}_j\right]=0\,.
}
\end{array}
\eq

\paragraph{R-matrix.} We deal with the following solution of (\ref{q23}), which is connected to celebrated multispecies TASEP model \cite{AB}, \cite{AKSS}
\beq \label{q24}
\begin{array}{c}
\displaystyle{
R(x) =\frac{1}{1-x} \left( \sum\limits_{a=1}^{n}E_{aa}\otimes E_{aa} + (1-x)\sum\limits_{a > b}^{n}E_{aa}\otimes E_{bb} + x\sum\limits_{a<b}^{n}E_{ab}\otimes E_{ba} \right.
}
\\ \ \\
\displaystyle{
\left.+ \sum\limits_{a>b}^{n}E_{ab}\otimes E_{ba} \right).
}
\end{array}
\eq
For $n=2$ $R$-matrix simplifies to TASEP $R$-matrix
\beq \label{q25}
\displaystyle{
R(x) =\frac{1}{1-x} \left(\begin{array}{cccc}
1 & 0 & 0 & 0\\
0 & 0 & x & 0\\
0 & 1 & 1-x &0\\
0 & 0 & 0 & 1
\end{array} \right).
}
\eq
The $R$-matrix (\ref{q24}) satisfies the unitarity condition
\beq \label{a2}
\displaystyle{
R_{12}(x)R_{21}(x^{-1}) = -\frac{x}{(x-1)^2} 1_n\otimes 1_n\,.
}
\eq

Recall that according to \cite{AB}, \cite{AKSS} the N- species ASEP and TASEP
models are the following generalization of the conventional ASEP and TASEP models.
At each site of the lattice we introduce $(N-1)$-vector characterizing the state
at the site. Hence, the local jumps correspond to the interchange of these
vectors. Since we are interested in inhomogeneous model the local lumps are
position dependent. The periodic boundary condition with the  $SU(N)$ valued twist is
imposed. The $N=2$ case corresponds to the conventional ASEP and TASEP.

\paragraph{Bethe ansatz.}
The eigenvalues for  the non-local Hamiltonians (\ref{q27}) are found through the nested Bethe ansatz \cite{AKSS}:
\beq \label{q29}
\displaystyle{
h_i(\{z\},\{\mu\}) = V_{1}\prod\limits_{k \neq i}^{N}\frac{z_k}{z_k-z_i}
\prod\limits_{l=1}^{M_1}\frac{z_i}{z_i-u^1_l}\,,
}
\eq
where the Bethe roots $\mu^i_{a}$ ($i=1,...,n-1$, $a=1,...,M_i$)  satisfy the following Bethe equations:
\beq \label{q30}
\displaystyle{
V_1\prod\limits_{k=1}^{N}\frac{z_k}{z_k-u^1_\be}=(-1)^{M_1-1}V_2\prod\limits_{\ga \neq \be}^{M_1}\frac{u^1_\ga}{u^1_\be}\,\prod\limits_{\ga=1}^{M_2}\frac{u^1_\be}{u^1_\be-u^2_\ga}\,,
}
\eq
\beq \label{q31}
\displaystyle{
V_i \prod\limits_{k=1}^{M_{i-1}}\frac{u^{i-1}_k}{u^{i-1}_k-u^i_\be} = (-1)^{M_i-1}V_{i+1}\prod\limits_{\ga \neq \be}^{M_i}\frac{u^i_\ga}{u^i_\be}\,\prod\limits_{n=1}^{M_{i+1}}\frac{u^i_\be}{u^i_\be-u^{i+1}_n}\,,
}
\eq
for $2 \leq i \leq n-2$,
\beq \label{q32}
\displaystyle{
V_{n-1}\prod\limits_{\ga=1}^{M_{n-2}}\frac{u^{n-2}_\ga}{u^{n-2}_\ga-u^{n-1}_\be}=(-1)^{M_{n-1}-1}V_{n}\prod\limits_{\ga \neq \be}^{M_{n-1}}\frac{u^{n-1}_\ga}{u^{n-1}_\be}\,.
}
\eq
For comparison we write the Bethe equations for trigonometric 6-vertex $R$-matrix (see \cite{BLZZ})
\beq \label{eq1}
\begin{array}{c}
\displaystyle{
V_b \prod\limits_{\ga=1}^{M_{b-1}} \frac{\left(\mu_{\be}^{b} \right)^2 q^2 - (\mu^{b-1}_{\ga})^2}{q \left((\mu^b_{\be})^2- (\mu_{\ga}^{b-1})^2\right)} = V_{b+1}\prod\limits_{\ga \neq \be}^{M_b}\frac{\left(\mu_{\be}^{b} \right)^2 q^2 - (\mu^{b}_{\ga})^2}{(\mu^b_{\be})^2 - (\mu^b _{\ga})^2 q^2} \prod\limits_{\ga=1}^{M_{b+1}}\frac{(\mu^b_{\be})^2-q^2(\mu^{b+1}_{\ga})^2}{q \left((\mu^b_{\be})^2-(\mu^{b+1}_{\ga})^2 \right)}
}
\end{array}
\eq
for $b=1,...,n-1$, where we used the same notations for $M_0$ and $M_n$ as in (\ref{q19})-(\ref{q21}).

\paragraph{Local Hamiltonian.} To obtain the Hamiltonian describing local interaction from the transfer matrix (\ref{q26}) we
put all inhomogeneity parameters to unit $q_i=1$ and set the twist matrix to identity $V=1_n$. Then the local Hamiltonian is given by the operator
\beq \label{q28}
\displaystyle{
\hat{H}^{\rm loc} = \left. \frac{d\Big((1-x)^{N}\hat{t}(x)\Big)}{d x}\Big((1-x)^N\hat{t}(x)\Big)^{-1} \right|_{x=1}\,,
}
\eq
which in the $n=2$ case reproduces the Markov matrix for TASEP model:
\beq \label{E2}
\hat{H}^{\rm loc}=\sum\limits_{i=1}^{N}\Big((E_{11})_{i+1} (E_{22})_{i}- (E_{21})_{i+1}(E_{12})_{i}\Big)\,,
\eq
where we assume that $N+1 = 1 \; (\mathrm{mod} \; N)$.

\section{Quantum-classical duality for higher rank five vertex models}

In this Section we describe the quantum classical duality between the five-vertex spin chains and the
Goldfish models in the rational and trigonometric cases respectively.

\subsection{Duality between rational 5-vertex  and Goldfish models}
Let us first formulate the main statement.

%\noindent{\bf Theorem.} {\em

\begin{theor}\label{th1} Let us identify the inhomogeneity
parameters $z_i$ in the quantum rational ${\rm GL}(n)$ spin chain (\ref{q10})-(\ref{q21}) on $N\geq n$ sites with the positions of particles $q_i$ in the rational Calogero's Goldfish model (\ref{I1})
 \beq\label{q33}
    \displaystyle{
 z_j=q_j\,,\quad j=1\,, \ldots \,,N
 }
 \eq
 and make a substitution of the
 eigenvalues (\ref{q22})\footnote{These are the eigenvalues of quantum Hamiltonians (\ref{q16}) of the rational five-vertex model (\ref{q10}).}
 \beq\label{q34}
    \displaystyle{
\dot q_j= h_j(\{q\},\{\mu\})
 }
 \eq
 into the classical higher integrals of motion $H^{\rm{R}}_k(\{\dot{q}\},\{q\})$ (\ref{I1}).
  %Here $H_j\left(\{q\},\{\mu\}\right)$ are eigenvalues of the rational five-vertex model Hamiltonians (\ref{q22}).
If, for any $N \geq M_1 \geq M_2 \geq ... \geq M_{n-1} \geq 0$, the Bethe roots $\{\mu^i_a\}$ satisfy the Bethe
equations (\ref{q19})-(\ref{q21}), i.e.,
$h_j\left(\{q\},\{\mu\} \right)$ belong to the spectrum
of the rational five-vertex model,
then the conserved quantities of the classical rational Calogero's Goldfish system are equal to
\beq \label{q35}
\displaystyle{
H_k^{\rm{R}}\Big(\{h\left(\{q\},\{\mu\}\right)\},\{q\}\Big) = \left\{ \begin{array}{l}
V_1^{N-M_1}V_2^{M_1-M_2}...V_{n-1}^{M_{n-2}-M_{n-1}}V_{n}^{M_{n-1}},\;\; k=N\,,\\ \\
V_{i+1}^{M_i-M_{i+1}}V_{i+2}^{M_{i+1}-M_{i+2}}...V_{n}^{M_{n-1}}, \;\; k=M_i\,,\\ \\
0, \;\; \mathrm{otherwise}\,,
\end{array} \right.
}
\eq
$k=1,...,N$, so that only $n$ of $N$ classical Hamiltonians are non-zero.
\end{theor}

Before proceeding to the proof consider two simple but important examples of (\ref{q35}).

%\noindent {\bf Example 1.}
\begin{example} Let us write down (\ref{q35}) for the rank $n=2$ (${\rm gl}_2$ model). In this case there is only one level of Bethe roots ($M=M_1$) and two twist parameters.  Then, as a result of substitution (\ref{q34}), only two classical Hamiltonians are non-zero:
\beq \label{e1}
\begin{array}{c}
\displaystyle{
H_k^{\rm{R}}\Big(\{h\left(\{q\},\{\mu\}\right)\},\{q\}\Big) = \left\{ \begin{array}{l}
V_1^{N-M}V_2^{M},\;\; k=N,\\\\
V_{2}^{M}, \;\; k=M,\\\\
0, \;\; \mathrm{otherwise}.
\end{array} \right.
}
\end{array}
\eq
\end{example}

\begin{example} Consider the most non-degenerate case $n=N$ and $M_i = N-i$. Then all $N$ classical Hamiltonians are non-zero:
\beq \label{e2}
\displaystyle{
H_k^{\rm{R}}\Big(\{h\left(\{q\},\{\mu\}\right)\},\{q\}\Big) = \prod\limits_{j=0}^{k-1} V_{n-j}\,.
}
\eq
\end{example}

\subsubsection*{\underline{Auxiliary Lemmas}}

In \cite{GZZ} the proof of quantum-classical duality between the XXX spin chain and the rational Ruijsenaars-Schneider models was based on factorization of the classical Lax matrix \cite{TsuboiZZ},\cite{VZ} and determinant identities. Here we do not have Lax matrix with such properties. Instead, we formulate a similar statement in terms of higher
Hamiltonians.

The first step is to prove (\ref{q35}) for the case when all $M_i=0$.
%(for the quantum integrable model this is the case of the highest weight vector).

\begin{lemma}\label{L1} For the case when all $M_i=0$ the substitution $\dot q_j= h_j(\{q\},\{\emptyset\})$ into (\ref{I1})
provides
\beq \label{q36}
\displaystyle{
H_k^{\rm{R}}\Big(\{h\left(\{q\},\{\emptyset\}\right)\},\{q\}\Big) =
 \left\{ \begin{array}{l}
V_1^{N}\,,\;\;k=N\,,
\\\\
0, \;\; \mathrm{otherwise}\,.
\end{array} \right.
}
\eq
\end{lemma}

\noindent\underline{\em{Proof:}}\quad
 Consider first $k=N$. Write down explicit expression for the l.h.s. of (\ref{q36}):
\beq \label{q37}
\begin{array}{c}
\displaystyle{
H_N^{\rm{R}}\Big(\{h\left(\{q\},\{\emptyset\} \right)\},\{q\}\Big) = (-1)^{\frac{N(N-1)}{2}}\prod\limits_{i=1}^{N}h_{i}(\{q\},\{\emptyset \})\prod\limits_{l < j}^{N}(q_l-q_j)^2 =
}
\\ \ \\
\displaystyle{
=(-1)^{\frac{N(N-1)}{2}}V_1^{N}\prod\limits_{k \neq i}^{N}\frac{1}{q_i-q_k}\prod\limits_{l < j}^{N}(q_l-q_j)^2=V_1^{N}\,.
}
\end{array}
\eq
For all others $H_k^{\rm{R}}$ we shall analyze the pole structure in variable $q_i$. Consider the behaviour in $q_1$.
%\footnote{It is enough to study behaviour for $q_1$ since (\ref{I1}) is a ??? symmetric function of  %$\{q\}_{i=1}^{N}$.}.
Expressions (\ref{q36}) have only simple poles at $q_1=q_i$. Let us evaluate residues of (\ref{q36}) at $q_1=q_2$.  The simple pole at $q_1=q_2$ may come from only those terms in the sum (\ref{I1}), where $i_1=1$ and $i_2 \neq 2$, or $i_1=2$. Thus, we obtain by direct calculation:
\beq \label{q38}
\begin{array}{c}
\displaystyle{
\res\limits_{q_1=q_2}H_k^{\rm{R}}\Big(\{h\left(\{q\},\{\emptyset\}\right)\},\{q\} \Big) = V_1 \left[ H^{\rm{R}}_{k-1} \Big(\{h\left(\{q_m\}_{m=3}^N,\{\emptyset\} \right)   \},\{q_j\}_{j=3}^N \Big)\prod\limits_{l \neq 1,2}^N\frac{1}{q_2-q_l}\right.-
}
\\ \ \\
\displaystyle{
-\left. \left.H^{\rm{R}}_{k-1} \Big(\{h\left(\{q_m\}_{m=3}^N,\{\emptyset\} \right)   \},\{q_j\}_{j=3}^N \Big)\prod\limits_{l \neq 1,2}^N\frac{1}{q_1-q_l}\right|_{q_1=q_2} \right] =0.
}
\end{array}
\eq
In a similar way one may show the absence of poles at all possible $q_i=q_j$.
The behaviour at infinity in any $q_j$ is also easily calculated. For $q_1$ we have
\beq \label{q39}
\displaystyle{
\lim\limits_{q_1 \rightarrow \infty}H_k^{\rm{R}}\Big(\{h\left(\{q\},\{\emptyset \}\right)\},\{q\} \Big) = 0\,.
}
\eq
 Therefore, the l.h.s. of (\ref{q36}) has no poles and is equal to zero at infinity. We conclude
\beq \label{q40}
\displaystyle{
H_k^{\rm{R}}\Big(\{h\left(\{q\},\{\emptyset \}\right)\},\{q\} \Big) = 0\quad \mathrm{for}\quad k<N,
}
\eq
which completes the proof. $\blacksquare$

Let us study the pole structure of the l.h.s (\ref{q35}). The following important property could be proved similarly to the previous lemma.

\begin{lemma}\label{L2} The left hand side of (\ref{q35})
\beq \label{q41}
\displaystyle{
H_k^{\rm{R}}\Big(\{h\left(\{q\},\{\mu\}\right)\},\{q\} \Big) = (-1)^{\frac{k(k-1)}{2}}\sum\limits_{I_k}h_{i_1}\left(\{q\},\{\mu\} \right)\,...\,h_{i_k}\left(\{q\},\{\mu\}\right)\prod\limits_{l < j}^{k}\left( q_{i_l}-q_{i_j}\right)^2
}
\eq
does not have poles at $q_i=q_j$, all poles appear at $q_i=\mu_j$ only.
\end{lemma}

\noindent\underline{\em{Proof:}}\quad For the case $k=N$ all poles at $q_i=q_j$ of $h_i\left(\{q\}, \{\mu\} \right)$ are annihilated by the Vandermonde determinant $\prod\limits_{l<j}^N(q_l-q_j)^2$. When $k<N$ one should perform the same calculation as in (\ref{q38}).  $\blacksquare$

In what follows we will be interested in calculating the residues of $H_k^{\rm{R}}\Big(\{h\left(\{q\},\{\mu\}\right)\},\{q\} \Big)$ at $q_i=\mu_j$. Lemma \ref{L2} allows us to formulate the main technical tools for the proof of the Theorem \ref{th1}. In what follows we call them determinant identities due to  similarity  with identities from \cite{GZZ,TsuboiZZ}. The main idea is that one can more or less change the role of coordinates $q$ and the Bethe roots $\mu$ in the l.h.s. of (\ref{q35}), to be more precise the following Proposition holds.

\begin{predl}\label{Pr1} Introduce the following sets of functions:
\beq \label{q42}
\begin{array}{c}
\displaystyle{
G_i\left(\{q\},\{\mu\} \right) = V\prod\limits_{j \neq i}^N \frac{1}{q_i-q_j}\prod\limits_{l=1}^M\frac{1}{\mu_l-q_i}, \quad 1\leq i\leq N,
}
\\ \ \\
\displaystyle{
F_i\left(\{\mu\},\{q\} \right)=V\prod\limits_{j \neq i}^M\frac{1}{\mu_j-\mu_i}\prod\limits_{l=1}^N\frac{1}{\mu_i-q_l}, \quad 1 \leq i \leq M,
}
\end{array}
\eq
depending on the set of $N$ variables $q_i$, the set of $M$ variables $\mu_j$ and parameter $V$.
Then the following identities hold true:
\beq \label{q43}
\begin{array}{c}
\displaystyle{
H_k^{\rm{R}}\Big(\{G\left(\{q\},\{\mu\}\right)\},\{q\} \Big) = H_k^{\rm{R}}\Big(\{F\left(\{\mu\},\{q\}\right)\},\{\mu\}\Big) \quad \mathrm{for} \quad k\leq M\,,
}
\\ \ \\
\displaystyle{
H_k^{\rm{R}}\Big(\{G\left(\{q\},\{\mu\}\right)\},\{q\} \Big) = 0 \quad \mathrm{for} \quad M < k < N\,,
}
\\ \ \\
\displaystyle{
H_k^{\rm{R}}\Big(\{G\left(\{q\},\{\mu\}\right)\},\{q\} \Big) = V^{N-M}H_M^{\rm{R}}\Big(\{F\left(\{\mu\},\{q\}\right)\},\{\mu\}\Big) \quad \mathrm{for} \quad k=N\,.
}
\end{array}
\eq
\end{predl}

\noindent\underline{\em{Proof:}}\quad
The Proposition is proved by induction on the number of Bethe roots $M$. For $M=0$ the Proposition holds true due to Lemma \ref{L1} (\ref{q36}). Suppose (\ref{q43}) is true for the number of Bethe roots $M-1$. We need to prove for $M$. Compute the residues of the both sides of (\ref{q43}) at $\mu_1=q_1$. Verification of all other poles at $q_i=\mu_j$
is performed in a similar fashion.
%\footnote{The expressions in (\ref{q43}) are symmetric in $\{q\}$ and $\{\mu\}$},

\underline{\noindent 1. For the case $k\leq M$} the residue of the l.h.s. is equal to
\beq \label{q44}
\begin{array}{c}
\displaystyle{
\res\limits_{\mu_1=q_1}H_k^{\rm{R}}\Big(\{G\left(\{q\},\{\mu\}\right)\},\{q\} \Big) =
}
\\ \ \\
\displaystyle{
V\prod\limits_{j \neq 1}^N\frac{1}{q_1-q_j}\prod\limits_{l \neq 1}^M \frac{1}{\mu_l-q_1}H^{\rm{R}}_{k-1}\Big(\{G\left(\{q\}/\{q_1\},\{\mu\}/\{\mu_1\}\right)\},\{q\}/\{q_1\} \Big)\,,
}
\end{array}
\eq
which comes by direct calculations from definitions (\ref{q42}) and (\ref{q41}). On the other hand we compute the r.h.s of the first equality from (\ref{q43}):
\beq \label{q45}
\begin{array}{c}
\displaystyle{
\res\limits_{\mu_1=q_1}H_k^{\rm{R}}\Big(\{F\left(\{\mu\},\{q\}\right)\},\{\mu\}\Big)=
}
\\ \ \\
\displaystyle{
=V\prod\limits_{ j\neq 1}^{M}\frac{1}{\mu_j-q_1}\prod\limits_{l \neq 1}^{N}\frac{1}{q_1-q_l} \, H_{k-1}^{\rm{R}}\Big(\{F\left(\{\mu\}/\{\mu_1\},\{q\}/\{q_1\}\right)\},\{\mu\}/\{\mu_1\} \Big)\,.
}
\end{array}
\eq
By the induction assumption
\beq \label{q46}
\begin{array}{c}
\displaystyle{
H^{\rm{R}}_{k-1}\Big(\{G\left(\{q\}/\{q_1\},\{\mu\}/\{\mu_1\}\right)\},\{q\}/\{q_1\} \Big)=
}
\\ \ \\
\displaystyle{
H_{k-1}^{\rm{R}}\Big(\{F\left(\{\mu\}/\{\mu_1\},\{q\}/\{q_1\}\right)\},\{\mu\}/\{\mu_1\} \Big)\,,
}
\end{array}
\eq
so that
\beq \label{q47}
\displaystyle{
\res\limits_{\mu_1=q_1}H_k^{\rm{R}}\Big(\{G\left(\{q\},\{\mu\}\right)\},\{q\} \Big)=\res\limits_{\mu_1=q_1}H_k^{\rm{R}}\Big(\{F\left(\{\mu\},\{q\}\right)\},\{\mu\}\Big)\,.
}
\eq
Finally, the behaviour at infinity is as follows:
\beq \label{q48}
\begin{array}{c}
\displaystyle{
\lim\limits_{\mu_1 \rightarrow \infty}H_k^{\rm{R}}\Big(\{G\left(\{q\},\{\mu\}\right)\},\{q\}\Big) = 0\,,
}
\\ \ \\
\displaystyle{
\lim\limits_{\mu_1 \rightarrow \infty }H_k^{\rm{R}}\Big(\{F\left(\{\mu\},\{q\}\right)\},\{\mu\} \Big)=0, \quad M \geq 1\,.
}
\end{array}
\eq
In this way we proved the first line of the Proposition (\ref{q43}).

\underline{\noindent 2. Consider the case $M < k < N$.} This part is again proved by the induction on the number of Bethe roots $M$, the base of induction is formula (\ref{q36}). In this case we have
\beq \label{q49}
\begin{array}{c}
\displaystyle{
\res\limits_{\mu_1=q_1}H_k^{\rm{R}}\Big(\{G\left(\{q\},\{\mu\}\right)\},\{q\} \Big)=
}
\\ \ \\
\displaystyle{
=V\prod\limits_{j \neq 1}^N\frac{1}{q_1-q_j}\prod\limits_{l \neq 1}^M \frac{1}{\mu_l-q_1}\,H^{\rm{R}}_{k-1}\Big(\Big\{G(\{q\}/\{q_1\},\{\mu\}/\{\mu_1\})\Big\},\{q\}/\{q_1\} \Big)=0\,,
}
\end{array}
\eq
where the last equality follows from the induction assumption.
The behaviour at the infinity is as follows:
\beq \label{q52}
\displaystyle{
\lim\limits_{\mu_1 \rightarrow \infty}H_k^{\rm{R}}\Big(\{G\left(\{q\},\{\mu\}\right)\},\{q\}\Big) = 0\,,
}
\eq
so we proved the second line of (\ref{q43}).

\underline{\noindent 3. Consider the case $k=N$}, for which we have
\beq \label{q53}
\displaystyle{
H_N^{\rm{R}}\Big(\{G\left(\{q\},\{\mu\}\right)\},\{q\} \Big) =V^N \prod\limits_{k=1}^{N}\prod\limits_{l=1}^{M}\frac{1}{\mu_k-q_l}\,.
}
\eq
For the residue at $q_1=\mu_1$ we get:
\beq \label{q531}
\begin{array}{c}
\displaystyle{
\res\limits_{\mu_1=q_1}H_N^{\rm{R}}\Big(\{G\left(\{q\},\{\mu\}\right)\},\{q\} \Big)=V^N\prod\limits_{j \neq 1}^N\frac{1}{q_1-q_j}\prod\limits_{m \neq 1}^{M}\frac{1}{\mu_m-q_1}\prod\limits_{k \neq 1}^{M}\prod\limits_{l \neq 1}^{N}\frac{1}{\mu_k-q_l}=
}
\\ \ \\
\displaystyle{
=V\prod\limits_{j \neq 1}^N\frac{1}{q_1-q_j}\prod\limits_{m \neq 1}^{M}\frac{1}{\mu_m-q_1}H_{N-1}^{\rm{R}}\Big(\Big\{G(\{q\}/\{q_1\},\{\mu\}/\{\mu_1\})\Big\},\{q\}/\{q_1\} \Big)\,.
}
\end{array}
\eq
On the other hand, similar computation yields
\beq \label{q532}
\begin{array}{c}
\displaystyle{
\res\limits_{\mu_1=q_1}V^{N-M}H_M^{\rm{R}}\Big(\{F\left(\{\mu\},\{q\}\right)\},\{\mu\}\Big)=
}
\\ \ \\
\displaystyle{
= V^{N-M+1}\prod\limits_{j \neq 1}^N\frac{1}{q_1-q_j}\prod\limits_{m \neq 1}^{M}\frac{1}{\mu_m-q_1}H_{M-1}^{\rm{R}}\Big(\Big\{F\left(\{\mu\}/\{\mu_1\},\{q\}/\{q_1\}\right)\Big\},\{\mu\}/\{\mu_1\} \Big)\,.
}
\end{array}
\eq
Therefore,  using also the induction hypothesis we conclude
\beq \label{q533}
\displaystyle{
\res\limits_{\mu_1=q_1}H_N^{\rm{R}}\Big(\{G\left(\{q\},\{\mu\}\right)\},\{q\} \Big) = \res\limits_{\mu_1=q_1}V^{N-M}H_M^{\rm{R}}\Big(\{F\left(\{\mu\},\{q\}\right)\},\{\mu\}\Big)\,.
}
\eq
Finally, we compare the behaviour at the infinity
\beq \label{q534}
\begin{array}{c}
\displaystyle{
\lim\limits_{\mu_1 \rightarrow \infty}H_N^{\rm{R}}\Big(\{G\left(\{q\},\{\mu\}\right)\},\{q\} \Big) = 0\,,
}
\\ \ \\
\displaystyle{
\lim\limits_{\mu_1 \rightarrow \infty }V^{N-M}H_M^{\rm{R}}\Big(\{F\left(\{\mu\},\{q\}\right)\},\{\mu\}\Big)=0\,,
}
\end{array}
\eq
which completes the proof of the third line of (\ref{q43}). $\blacksquare$

\subsubsection*{\underline{Proof of Theorem \ref{th1}}}

Now we can prove the main Theorem (\ref{q35}) of this Section. Let us comment on the general idea of the proof. One should apply the Proposition \ref{Pr1} to rewrite the l.h.s of (\ref{q35}) and apply the Bethe equations to the obtained expression. After doing so the dependence on coordinates $\{q\}$ disappears and one is left with only $\{\mu^1\}$ and $\{\mu^2\}$. Then one should repeat this scheme to be left with only $\{\mu^{n-1}\}$, for which the statement follows from Lemma \ref{L1} (\ref{q36}).

%Let us prove (\ref{q35}).
The l.h.s. of (\ref{q35}) is equal to
\beq \label{q54}
\displaystyle{
H_k^{\rm{R}}\Big(\{h\left(\{q\},\{\mu^1\}\right)\},\{q\}\Big)\,.
}
\eq
By applying the determinant identities (\ref{q43}) to (\ref{q54}) one obtains
\beq \label{q55}
\begin{array}{c}
\displaystyle{
H_k^{\rm{R}}\Big(\{h\left(\{q\},\{\mu^1\}\right)\},\{q\}\Big) = H_k^{\rm{R}}\Big(\{F^{(1)}\left(\{\mu^1\},\{q\}\right)\},\{\mu^1\}\Big) \quad \mathrm{for} \quad k\leq M\,,
}
\end{array}
\eq
\beq \label{q56}
\displaystyle{
H_k^{\rm{R}}\Big(\{h\left(\{q\},\{\mu^1\}\right)\},\{q\}\Big)=
V^{k-M}H_M^{\rm{R}}\Big(\{F^{(1)}\left(\{\mu^1\},\{q\}\right)\},\{\mu^1\}\Big) \quad \mathrm{for} \quad M < k \leq N\,,
}
\eq
where
\beq \label{q57}
\displaystyle{
F^{(1)}_i\left(\{\mu^1\},\{q\}\right) = V_1\prod\limits_{j \neq i}^{M_1}\frac{1}{\mu^1_j-\mu^1_i}\prod\limits_{l=1}^N\frac{1}{\mu^1_i-q_l}
}
\eq
and $\{h\left(\{q\},\{\mu\}\right)\}$ are eigenvalues (\ref{q22}). Next, apply the Bethe equations (\ref{q19}) to  (\ref{q57}):
\beq  \label{q58}
\begin{array}{c}
\displaystyle{
F^{(1)}_i\left(\{\mu^1\},\{q\}\right) = V_1\prod\limits_{j \neq i}^{M_1}\frac{1}{\mu^1_j-\mu^1_i}\prod\limits_{l=1}^N\frac{1}{\mu^1_i-q_l}
=V_2\prod\limits_{ j \neq i}^{M_1}\frac{1}{\mu^1_i-\mu^1_j} \prod\limits_{l=1}^{M_2}\frac{1}{\mu^2_l-\mu^1_i}=
}
\\ \ \\
\displaystyle{
=\frac{V_2}{V_1}\,h_i\left(\{\mu^1\},\{\mu^2\} \right)\,,
}
\end{array}
\eq
so that dependence on coordinates $\{q\}$ disappears.
Notice that the role of Bethe roots, coordinates and twists have changed by the following rule. Comparing to expression (\ref{q22}): $q \rightarrow \mu^1$, $\mu^1 \rightarrow \mu^2$, $V_1 \rightarrow V_2$. Using the determinant identities (\ref{q43}) and the first system of Bethe equations (\ref{q30}), we obtain
\beq \label{q59}
\begin{array}{c}
\displaystyle{
H_k^{\rm{R}}\Big(\{h\left(\{q\},\{\mu^1\}\right)\},\{q\}\Big) = H_k^{\rm{R}}\Big(\Big\{\frac{V_2}{V_1}\,h\left(\{\mu^1\},\{\mu^2\} \right)\Big\},\{\mu^1\}\Big) \quad \mathrm{for} \quad 1\leq k \leq M\,,
}
\\ \ \\
\displaystyle{
H_k^{\rm{R}}\Big(\{h\left(\{q\},\{\mu^1\}\right)\},\{q\}\Big)=0 \quad \mathrm{for} \quad M <k <N\,,
}
\\ \ \\
\displaystyle{
H_N^{\rm{R}}\Big(\{h\left(\{q\},\{\mu^1\}\right)\},\{q\}\Big) = V_1^{N-M}H_k^{\rm{R}}\Big(\Big\{\frac{V_2}{V_1}\,h\left(\{\mu^1\},\{\mu^2\} \right)\Big\},\{\mu^1\}\Big)\,.
}
\end{array}
\eq
Then we proceed with applying determinant identities together with Bethe equations. In this way one can easily obtain that some of the Hamiltonians are set zero\footnote{Recall that $N \geq M_1 \geq M_2 \geq ... \geq M_{n-1} \geq 0$.}:
\beq \label{q60}
\displaystyle{
H_k^{\rm{R}}\Big(\{h\left(\{q\},\{\mu^1\}\right)\},\{q\}\Big)=0 \quad \mathrm{if} \quad k \neq M_i \quad \forall \quad i=0,...,n-1,
}
\eq
where we set for convenience $M_0=N$. After applying determinant identities and Bethe equations for other Hamiltonians one obtains
\beq \label{q61}
\begin{array}{c}
\displaystyle{
H_{M_i} = V_{i+1}^{M_i-M_{i+1}}V_{i+2}^{M_{i+1}-M_{i+2}}...V_{n-1}^{M_{n-2}-M_{n-1}} H_{M_{n-1}}^{\rm{R}}\Big(\Big\{\frac{V_n}{V_1}\, h\left(\{\mu^{n-1}\},\{\emptyset \} \right)\Big\},\{\mu^{n-1}\}\Big)=
}
\\ \ \\
\displaystyle{
=V_{i+1}^{M_i-M_{i+1}}V_{i+2}^{M_{i+1}-M_{i+2}}\,...\,V_{n-1}^{M_{n-2}-M_{n-1}}V_{n}^{M_{n-1}}\,,
}
\end{array}
\eq
where the last equality follows from Lemma \ref{L1}. This completes the proof of the Theorem \ref{th1}. $\blacksquare$
\subsection{Duality for trigonometric 5-vertex model (TASEP)}
The results of the previous subsection are mostly repeated for trigonometric $R$-matrix (\ref{q24})-(\ref{q25}) with some small technical changes.

\begin{theor}\label{th2}
Identify the inhomogeneity parameters
$z_i$ in the quantum trigonometric ${\rm GL}(n)$ spin chain (\ref{q27})-(\ref{q32}) on $N\geq n$ sites with the positions of particles $q_i$ in the trigonometric Calogero's Goldfish model
particles
 \beq\label{q62}
    \displaystyle{
 z_j=e^{q_j}\,,\quad j=1\,, \ldots \,,N
 }
 \eq
 and make a substitution of the
 eigenvalues (\ref{q29})\footnote{These are the eigenvalues of quantum Hamiltonians (\ref{q27}) of the trigonometric five-vertex model (\ref{q24}).}
 \beq\label{q63}
    \displaystyle{
\dot q_j= h_j(\{z\},\{\mu\})
 }
 \eq
 into the classical higher integrals of motion $H^{\rm{T}}_k(\{\dot{q}\},\{q\})$ (\ref{ham}).
 If, for any $N \geq M_1 \geq M_2 \geq ... \geq M_{n-1} \geq 0$, the Bethe roots $\{\mu^i_a\}$ satisfy the Bethe
equations (\ref{q30})-(\ref{q32}), i.e.,
$h_j\left(\{q\},\{\mu\} \right)$ belong to the spectrum
of the trigonometric five-vertex model,
then the conserved quantities of the classical trigonometric Calogero's Goldfish system are equal to
\beq \label{q64}
\displaystyle{
H^{\rm{T}}_{k}\Big(\{h\left(\{z\},\{\mu\}\right) \},\{z\} \Big) = V_{1}^{k-[M_{1}]_k} V_{2}^{[M_{1}]_{k}-[M_{2}]_{k}}.. V_{n-1}^{[M_{n-2}]_{k}-[M_{n-1}]_k} V_{n}^{[M_{n-1}]_k} ,
}
\eq
where
\beq  \label{q65}
\displaystyle{
[M]_k = \left\{ \begin{array}{l}
M,\;\;\; M \leq k,\\
k,\;\;\; M>k.
\end{array}\right.
}
\eq
\end{theor}

%\noindent Again before proving the theorem (\ref{q64}) we consider one example

\begin{example}
  Consider the most non-degenerate case $n=N$ and $M_i = N-i$. The level set of classical Hamiltonians takes the form:
\beq \label{e3}
\displaystyle{
H_k^{\rm{T}}\Big(\{h(\{z\},\{\mu\})\},\{z\}\Big) = \prod\limits_{j=0}^{k-1} V_{n-j}\,.
}
\eq
\end{example}

\subsubsection*{\underline{Auxiliary Lemmas}}

To prove the theorem (\ref{q64}) we follow the same way as in the previous section. First, we prove the theorem when all the Bethe roots are absent.

\begin{lemma}\label{L3}
For the case when all $M_i=0$ the substitution (\ref{q63}) provides
\beq \label{q66}
\displaystyle{
H^{\rm{T}}_{k}\Big(\{h\left(\{z\},\{\emptyset \}\right) \},\{z\} \Big) = V_1^{k}\,.
}
\eq
\end{lemma}

\noindent\underline{\em{Proof:}}\quad In the absence of Bethe roots the eigenvalues of the TASEP model is of the form
\beq \label{q67}
\begin{array}{c}
\displaystyle{
h_i(\{z\},\{\emptyset \}) = V_1 \prod\limits_{k \neq i}^N \frac{z_k}{z_k-z_i}\,.
}
\end{array}
\eq
Plugging it into the Hamiltonians of the trigonometric Calogero's trigonometric Goldfish model (\ref{ham}) one obtains
\beq \label{q68}
\begin{array}{c}
\displaystyle{
H^{\rm{T}}_k\Big(\{h\left(\{z\},\{\emptyset \} \right)\}, \{z\} \Big) =
}
\\ \ \\
\displaystyle{
=(-1)^{\frac{k\left(k-1\right)}{2}}\sum\limits_{I_k}h_{i_1}\left(\{z\},\{ \emptyset \}\right)\,...\,h_{i_k}\left(\{z\},\{\emptyset \} \right)\prod\limits_{l< j}^k \frac{\left(z_{i_l}-z_{i_j}\right)^2}{z_{i_k}z_{i_l}}\,.
}
\end{array}
\eq
Similarly to the previous subsection we study the structure of poles. It follows from (\ref{q67})-(\ref{q68})
that the l.h.s. of (\ref{q66}) may have only simple poles at $z_i=z_j$.
Let us compute the residue at $z_1=z_2$, all other poles are verified in a similar way.  Using the same arguments as in (\ref{q38}) one obtains
\beq \label{q69}
\displaystyle{
\res\limits_{z_1=z_2}H^{\rm{T}}_k\Big(\{h\left(\{z\},\{\emptyset \} \right)\},\{z\} \Big) = 0\,.
}
\eq
and the behaviour at the infinity is as follows:
\beq \label{q70}
\displaystyle{
\lim\limits_{z_1 \rightarrow \infty}h_i\left(\{z\},\{\emptyset\}\right) = h_i\left(\{z\}/\{z_1\},\{\emptyset\}\right) \quad \mathrm{for} \quad i \neq 1,
}
\eq
so that one gets
\beq \label{q71}
\begin{array}{c}
\displaystyle{
\lim\limits_{z_1 \rightarrow \infty} H^{\rm{T}}_k\Big(\{h\left(\{z\},\{\emptyset \} \right)\},\{z\} \Big) =
}
\\ \ \\
\displaystyle{
=V_1 H^{\rm{T}}_{k-1}\Big(\Big\{h\left(\{z\}/\{z_1\},\{\emptyset \} \right)\Big\},\{z\}/\{z_1\} \Big) \quad \mathrm{for} \quad 2 \leq k \leq N\,.
}
\end{array}
\eq
from which we obtain
\beq \label{q72}
\displaystyle{
H^{\rm{T}}_k\Big(\Big\{h\left(\{z\},\{\emptyset \} \right) \Big\},\{z\} \Big)=V_1\,H^{\rm{T}}_{k-1}\Big(\Big\{h\left(\{z\}/\{z_1\},\{\emptyset \} \right)\Big\},\{z\}/\{z_1\} \Big)
}
\eq
for $\quad 2 \leq k \leq N$. By applying relation (\ref{q72}) we get
\beq \label{q73}
H^{\rm{T}}_k\Big(\Big\{h\left(\{z\},\{\emptyset \} \right) \Big\},\{z\} \Big)=V_1^{k-1}H^{\rm{T}}_{1}\Big(\Big\{h\left(\{z\}/\{z_1,..,z_{k-1}\},\{\emptyset \} \right) \Big\},\{z\}/\{z_1,..,z_{k-1}\} \Big)\,.
\eq
Again, we proceed with computation of the behaviour at infinity:
\beq \label{q74}
\displaystyle{
\lim\limits_{z_1 \rightarrow \infty}H^{\rm{T}}_{1}\Big(\{h\left(\{z\},\{\emptyset \} \right) \},\{z\} \Big) = H^{\rm{T}}_{1}\Big(\Big\{h\left(\{z\}/\{z_1 \},\{\emptyset \} \right) \Big\},\{z\}/\{z_1\} \Big)\,.
}
\eq
The problem is now reduced to computation of the first Hamiltonian depending on two coordinates only:
\beq \label{q75}
\displaystyle{
H^{\rm{T}}_{1}\Big(\Big\{h\left(\{z_1,z_2\},\{\emptyset \} \right) \Big\},\{z_1,z_2\} \Big)=V_1 \left(\frac{z_2}{z_2-z_1}+\frac{z_1}{z_1-z_2} \right)=V_1\,.
}
\eq
Combining together (\ref{q73})-(\ref{q75}) we obtain the statement of Lemma:
\beq \label{q76}
\displaystyle{
H^{\rm{T}}_{k}\Big(\{h\left(\{z\},\{\emptyset \}\right) \},\{z\} \Big) = V_1^{k}.\quad  \blacksquare
}
\eq
Following the ideas from the previous section we claim that the l.h.s. of (\ref{q64}) does not have poles in $z_i=z_j$ and $z_i=0$ even when the Bethe roots are present.

\begin{lemma}\label{L4} The l.h.s. of (\ref{q64})
\beq \label{q77}
\displaystyle{
H_k^{\rm{T}}\Big(\{H\left(\{z\},\{\mu\}\right)\},\{z\}\Big) = (-1)^{\frac{k(k-1)}{2}}\sum\limits_{I_k}h_{i_1}\left(\{z\},\{\mu\} \right)...h_{i_k}\left(\{z\},\{\mu\}\right)\prod\limits_{l < j}^{k}\frac{\left(z_{i_l}-z_{i_j}\right)^2}{z_{i_l} z_{i_j}}
}
\eq
does not have poles at $z_i=z_j$ and $z_i=0$, all poles appear at $z_i=\mu_j$ only.
\end{lemma}

\noindent\underline{\em{Proof:}}\quad  Poles in $z_i=0$ do not appear due to the form of eigenvalues of the TASEP model (\ref{q29}). To prove the absence of poles at $z_i=z_j$ one should perform similar computations as in the Lemma \ref{L3} (\ref{q69}).  $\blacksquare$

Let us formulate the trigonometric analogue of Proposition \ref{Pr1} (\ref{q43}).

\begin{predl}\label{Pr2}
Introduce the set of functions
\beq \label{q78}
\begin{array}{c}
\displaystyle{
G_i\left(\{z\},\{\mu\}\right) = V\prod\limits_{j \neq i}^N\frac{z_j}{z_j-z_i}\prod\limits_{l=1}^M\frac{z_i}{z_i-\mu_l}\,,
}
\\ \ \\
\displaystyle{
F_{i}\left(\{\mu\},\{z\} \right) = V\prod\limits_{l \neq i}^M \frac{\mu_i}{\mu_i-\mu_l} \prod\limits_{j=1}^N\frac{z_j}{z_j-\mu_i}\,.
}
\end{array}
\eq
Then the following identities hold true:
\beq \label{q79}
\begin{array}{c}
\displaystyle{
H_k^{\rm{T}}\Big(\{G\left(\{z\},\{\mu\}\right)\},\{z\} \Big) = H_k^{\rm{T}}\Big(\{F\left(\{\mu\},\{z\}\right)\},\{\mu\} \Big) \quad \mathrm{for} \quad 1 \leq k \leq M\,,
}
\\ \ \\
\displaystyle{
H_k^{\rm{T}}\Big(\{G\left(\{z\},\{\mu\}\right)\},\{z\} \Big) = V^{k-M}H_M^{\rm{T}}\Big(\{F\left(\{\mu\},\{z\}\right)\},\{\mu\} \Big) \quad \mathrm{for} \quad M< k \leq N\,.
}
\end{array}
\eq
\end{predl}

\noindent\underline{\em{Proof:}}\quad   The proof is similar to the one for (\ref{q43}). It is performed by induction on the number of Bethe roots $M$. The base of the induction $M=0$ is the statement of Lemma \ref{L3} (\ref{q66}). Suppose (\ref{q79}) holds true for $M-1$ Bethe roots.

\underline{1. Consider the case $1 \leq k \leq M$.} Compute the residues of the l.h.s of (\ref{q79}):
\beq \label{q80}
\begin{array}{c}
\displaystyle{
\res\limits_{\mu_1=z_1}H_k^{\rm{T}}\Big(\{G\left(\{z\},\{\mu\}\right)\},\{z\} \Big) =
}
\\ \ \\
\displaystyle{
-Vz_1\prod\limits_{l \neq 1}^M\frac{z_1}{z_1-\mu_l} \prod\limits_{k \neq 1}^N\frac{z_k}{z_k-z_1}\,H_{k-1}^{\rm{T}}\Big(\Big\{G\left(\{z\}/\{z_1\},\{\mu\}/\{\mu_1\}\right)\Big\},\{z\}/\{z_1\} \Big)
}\,,
\end{array}
\eq
while for the r.h.s. of (\ref{q79}) one obtains
\beq \label{q81}
\begin{array}{c}
\displaystyle{
\res\limits_{\mu_1=z_1}H_k^{\rm{T}}\Big(\{F\left(\{\mu\},\{\z\}\right)\},\{\mu\} \Big) =
}
\\ \ \\
\displaystyle{
-V z_1 \prod\limits_{l \neq 1}^M\frac{z_1}{z_1-\mu_l} \prod\limits_{k \neq 1}^N\frac{z_k}{z_k-z_1}\,H_{k-1}^{\rm{T}}\Big(\Big\{F\left(\{\mu\}/\{\mu_1\},\{z\}/\{z_1\}\right)\Big\},\{\mu\}/\{\mu_1\} \Big)\,.
}
\end{array}
\eq
Using the induction hypothesis one gets
\beq \label{q82}
\displaystyle{
\res\limits_{\mu_1=z_1}H_k^{\rm{T}}\Big(\{G\left(\{z\},\{\mu\}\right)\},\{z\} \Big)=\res\limits_{\mu_1=z_1}H_k^{\rm{T}}\Big(\{F\left(\{\mu\},\{z\}\right)\},\{\mu\} \Big)\,.
}
\eq
By comparing the behaviour at  infinity
\beq \label{q83}
\displaystyle{
\lim\limits_{\mu_1 \rightarrow \infty}H_k^{\rm{T}}\Big(\{G\left(\{z\},\{\mu\}\right)\},\{z\} \Big) = 0\,,
}
\eq
\beq \label{q84}
\displaystyle{
\lim\limits_{\mu_1 \rightarrow \infty}H_k^{\rm{T}}\Big(\{F\left(\{\mu\},\{ z\}\right)\},\{\mu\} \Big) = 0\,,
}
\eq
we finish the proof for $1 \leq k \leq M$.

\underline{2. Consider the case when $M < k \leq N$.} Let us perform the same calculations as in the previous case:
\beq \label{q85}
\begin{array}{c}
\displaystyle{
\res\limits_{\mu_1=z_1}H_k^{\rm{T}}\Big(\{G\left(\{z\},\{\mu\}\right)\},\{z\} \Big) =
}
\\ \ \\
\displaystyle{
-Vz_1\prod\limits_{l \neq 1}^M\frac{z_1}{z_1-\mu_l} \prod\limits_{k \neq 1}^N\frac{z_k}{z_k-z_1}\,H_{k-1}^{\rm{T}}\Big(\Big\{G\left(\{z\}/\{z_1\},\{\mu\}/\{\mu_1\}\right)\Big\},\{z\}/\{z_1\} \Big)
}\,,
\end{array}
\eq
exactly as in the (\ref{q80}), while for the r.h.s. of (\ref{q79}) one computes
\beq \label{q86}
\begin{array}{c}
\displaystyle{
\res\limits_{\mu_1=z_1}H_M^{\rm{T}}\Big(\{F\left(\{\mu\},\{ z\}\right)\},\{\mu\} \Big) =
}
\\ \ \\
\displaystyle{
-V z_1 \prod\limits_{l \neq 1}^M\frac{z_1}{z_1-\mu_l} \prod\limits_{k \neq 1}^N\frac{z_k}{z_k-z_1}\,H_{M-1}^{\rm{T}}\Big(\Big\{F\left(\{\mu\}/\{\mu_1\},\{z\}/\{z_1\}\right)\Big\},\{\mu\}/\{\mu_1\} \Big)\,.
}
\end{array}
\eq
The induction hypothesis (\ref{q85})-(\ref{q86}) then yields
\beq \label{q87}
\displaystyle{
\res\limits_{\mu_1=z_1}H_k^{\rm{T}}\Big(\{G\left(\{z\},\{\mu\}\right)\},\{z\} \Big)=V^{k-M}\res\limits_{\mu_1=z_1}H_M^{\rm{T}}\Big(\{F\left(\{\mu\},\{ z\}\right)\},\{\mu\} \Big)\,.
}
\eq
Finally, we compare the behaviour at the infinity using (\ref{q83})-(\ref{q84}), which are true even for the $M < k \leq N$. $\blacksquare$

Now we may prove (\ref{q64}). The main idea is to apply determinant identities (\ref{q79}) together with the Bethe equations (\ref{q30})-(\ref{q32}).

\subsubsection*{\underline{Proof of Theorem \ref{th2} (\ref{q64})}}

 Let us apply the determinant identity (\ref{q79}) to the l.h.s. of (\ref{q64}):
\beq \label{q88}
\begin{array}{c}
\displaystyle{
H_k^{\rm{T}}\Big(\{h\left(\{z\},\{\mu^1\}\right)\},\{z\} \Big) = H_k^{\rm{T}}\Big(\{F^{(1)}\left(\{\mu^1\},\{z\}\right)\},\{\mu^1\}\Big) \quad \mathrm{for} \quad 1 \leq k \leq M_1\,,
}
\\ \ \\
\displaystyle{
H_k^{\rm{T}}\Big(\{h\left(\{z\},\{\mu^1\}\right)\},\{z\} \Big) =
}
\\ \ \\
\displaystyle{
=V^{k-M_1}H_M^{\rm{T}}\Big(\{F^{(1)}\left(\{\mu^1\},\{z\}\right)\},\{\mu^1\} \Big) \quad \mathrm{for} \quad M_1< k \leq N\,,
}
\end{array}
\eq
where
\beq \label{q89}
\displaystyle{
F_i^{(1)}\left(\{\mu^1\},\{z\}\right) = V_1 \prod\limits_{l \neq i}^{M_1}\frac{\mu^1_i}{\mu^1_i-\mu^1_l}\prod\limits_{j=1}^N\frac{z_j}{z_j-\mu^1_i}\,.
}
\eq
By applying the Bethe equations (\ref{q30})-(\ref{q32}) one obtains
\beq \label{q90}
\displaystyle{
F_i^{(1)}\left(\{\mu^1\},\{z\}\right) = V_2\prod\limits_{l \neq i}^{M_1}\frac{\mu^1_l}{\mu^1_l-\mu^1_i}\prod\limits_{j=1}^{M_2}\frac{\mu^1_i}{\mu^1_i-\mu^2_j} = \frac{V_2}{V_1}\,h_i\left(\{\mu^1\},\{\mu^2\} \right)\,.
}
\eq
Thus, one can rewrite (\ref{q88}) using the  Bethe roots $\{\mu^1\}$ and $\{\mu^2\}$  only without inhomogeneities $\{z\}$:
\beq \label{q91}
\begin{array}{c}
\displaystyle{
H_k^{\rm{T}}\Big(\{H\left(\{z\},\{\mu^1\}\right)\},\{z\} \Big) = H_k^{\rm{T}}\Big(\Big\{\frac{V_2}{V_1}\,h\left(\{\mu^1\},\{\mu^2\}\right)\Big\},\{\mu^1\} \Big) \quad \mathrm{for} \quad 1 \leq k \leq M_1\,,
}
\\ \ \\
\displaystyle{
H_k^{\rm{T}}\Big(\{h\left(\{z\},\{\mu^1\}\right)\},\{z\} \Big) =
}
\\ \ \\
\displaystyle{
=V_1^{k-M_1}H_M^{\rm{T}}\left(\Big\{\frac{V_2}{V_1}\,h(\{\mu^1\},\{\mu^2\})\Big\},\{\mu^1\} \right) \quad \mathrm{for} \quad M_1< k \leq N\,.
}
\end{array}
\eq
It follows from (\ref{q91}) that for the case $1 \leq k \leq M_1$ all the twists $V_1$ are replaced by $V_2$, however for the case $M_1<k \leq N$ there is an additional multiplier $V_1^{k-M_1}$. The procedure is repeated until all Bethe roots are eliminated  except $\{ \mu^{n-1}\}$, for which one can apply (\ref{q66}), and obtain the statement of the theorem (\ref{q64}).
 $\blacksquare$
%\newpage

%??? Let us also remark that the special definition of $[M]_k$ from (\ref{q65}) means that for $M>k$ each twist is %replaced by the next one, so it does not appear in  (\ref{q64}).

%
%
%
%
%
%
%
%
%
\section{Quantum-quantum duality for higher rank five-vertex model}
In the previous Section we described the quantum-classical duality between 5-vertex models and
% rational and trigonometric
the classical Calogero's Goldfish models. The quantization of the classical
 systems of particles %leads to the common eigenvalue problem for
means that we deal with the quantized Hamiltonians
\beq \label{w1}
\displaystyle{
\hat{H}^{\rm{R}}_k = \sum\limits_{I_k}\left[\prod\limits_{j \in I_k}\prod\limits_{l \notin I_k}\frac{1}{q_j-q_l}\right] \prod\limits_{n \in I_k}e^{\hbar \partial_{n}}\,,
}
\eq
\beq \label{w2}
\displaystyle{
\hat{H}^{\rm{T}}_k =(-1)^{k(N-k)} \sum\limits_{I_k}\left[\prod\limits_{j \in I_k}\prod\limits_{l \notin I_k}\frac{z_l}{z_j-z_l}\right] \prod\limits_{n \in I_k}\hat{T}^q_{n}\,,
}
\eq
in the rational and trigonometric cases respectively, where
\beq \label{w02}
\displaystyle{
\hat{T}^q_i f(z_1,..,z_i,..,z_N) = f(z_1,..,qz_i,..,z_N)\,.
}
\eq
In the quantum case the Hamiltonians commute as operators
% on the space of functions
\beq \label{w3}
\displaystyle{
\begin{array}{cc}
\left[\hat{H}^{\rm{R}}_k,\hat{H}^{\rm{R}}_j\right]=0, & \left[\hat{H}^{\rm{T}}_k,\hat{H}^{\rm{T}}_j\right]=0.\,
\end{array}
}
\eq

We are interested in eigenfunctions and eigenvalues of a family of operators (\ref{w1}) or (\ref{w2}). In this section we will argue how one can construct eigenfunctions for (\ref{w1}) and (\ref{w2}) with certain eigenvalues from the solutions of quantum Knizhnik-Zamolodchikov equations \cite{FR}. This type of construction is a well-known Matsuo-Cherednik correspondence \cite{M1,Ch1,Ch2},\cite{ZZ1} producing the eigenfunctions for Macdonald-type difference operators.
% Now we have to specify the case we are working with, firstly, we deal with the rational model.

\subsection{Quantum-quantum duality for rational QKZ}
The quantum Knizhnik-Zamolodchikov equations are a system of %holonomic
difference equations \cite{FR}:
\beq \label{w4}
\displaystyle{
e^{\hbar \partial_i}\ket{\Psi} = \hat{K}_i^{\hbar} \ket{\Psi}\,,
}
\eq
where
\beq \label{w5}
\displaystyle{
\hat{K}_i^{\hbar} =\tilde{R}_{i,i-1}\left(q_i-q_{i-1}+\hbar\right)...\tilde{R}_{i,1}\left(q_i-q_1+\hbar \right)V^{(i)}\tilde{R}_{i,n}\left(q_i-q_N\right)...\tilde{R}_{i,i+1}\left(q_i-q_{i+1}\right)\,,
}
\eq
with $V^{(i)}$ being the twist matrix $\mathrm{diag}\left(V_1,...,V_n\right)$ acting in $i$-th tensor component as in (\ref{q16}) and $\ket{\Psi} = \ket{\Psi} \left(q_1,...,q_N\right)$ is a $\left(\mathbb{C}^n\right)^{\otimes N}$ valued function (in what follows we omit the dependence on $q$ for brevity). The quantum $R$-matrix $\tilde{R}(x)$ entering
(\ref{w5}) has different normalization with respect to (\ref{q10}):
\beq \label{w6}
\displaystyle{
\tilde{R}(u)= P+u\sum\limits_{\al > \be}^{n}E_{\al \al}\otimes E_{\be \be}\stackrel{(\ref{q10})}{=}uR(u)\,.
}
\eq
It satisfies the unitary condition
\beq \label{w7}
\displaystyle{
\tilde{R}_{12}(u)\tilde{R}_{21}(-u)= 1_n\otimes 1_n\,.
}
\eq
Introduce the following momenta operators
\beq \label{w8}
\displaystyle{
\hat{M}_k =\sum\limits_{i=1}^{N}E^{(i)}_{kk}\,,\quad k=1,...,n\,.
}
\eq
They commute with each other and with operators (\ref{w4}):
\beq \label{w9}
\displaystyle{
\begin{array}{ccc}
[\hat{K}^{0}_i,\hat{K}^{0}_j]=0\,,  &  [\hat{K}^{0}_i,\hat{M}_j]=0, & [\hat{M}_i,\hat{M}_j]=0\,.
\end{array}
}
\eq
Consider also the operators
\beq \label{w10}
\displaystyle{
\hat{H}_i=\hat{K}^{0}_i \prod\limits_{k \neq i}^N\frac{1}{q_i-q_k}\,.
}
\eq
Notice that operators $\hat{H}_i$ coincide with the non-local Hamiltonians defined in (\ref{q16}). Using (\ref{w9}) one can search for the common eigenfunctions
\beq \label{w11}
\displaystyle{
\left\{ \begin{array}{l}
\hat{H}_i\ket{\psi} = H_i \ket{\psi}\,,\\
\hat{M}_i \ket{\psi} = M_i \ket{\psi}\,.
\end{array}\right.
}
\eq
With the help of the action of momenta operators (\ref{w8}) the Hilbert space ${\cal V}=\left(\mathbb{C}^n \right)^{\otimes N}$ decomposes into eigenspaces for momenta operators
\beq \label{w12}
\displaystyle{
{\cal V}=
\bigoplus_{M_1, \ldots , M_n} \!\! \!\! {\cal V}(\{M_a \})\,,\qquad
 \mbox{dim} {\cal V}(\{M_a \})=\frac{N!}{M_1! \ldots M_n!}\,.
}
\eq
where $M_1+...+M_n=N$, and $V(\{M\})$ is an eigenspace for the momenta operators.
% (with the eigenvalues $M_a$).

%As we described the basics about quantum Knizhnik-Zamolodchikov equations, we turn to the disccusion of %Matsuo-Cherednik correspondence. Let us
Denote the standard basis in $\mathbb{C}^n$ by $\{\ket{e}\}_{i=1}^n$ and the dual basis $\{ \bra{e}\}_{i=1}^n$.  Introduce the following unique covector
\beq \label{w13}
\displaystyle{
\bra{\Omega} = \bra{e_{i_1}} \otimes ... \otimes \bra{e_{i_N}}, \quad i_1 \geq i_2 \geq ... \geq i_N
}
\eq
from a certain weight subspace $V(\{M\})$.
Namely,
\beq \label{w131}
\displaystyle{
\bra{\Omega} = \underbrace{ \bra{e_{n}} \otimes ... \otimes\bra{e_{n}}}_{M_n\ times}\otimes\, ...\,
\otimes \underbrace{ \bra{e_{1}} \otimes ... \otimes\bra{e_{1}}}_{M_1\ times}\,.
}
\eq
Then the following important identity holds true:
\beq \label{w14}
\displaystyle{
\bra{\Omega}\hat{K}_i^{\hbar} = \bra{\Omega}\hat{K}_i^{0}\,.
}
\eq
Indeed, using the explicit form of the $R$-matrix (\ref{w6}) one can verify that
\beq \label{w141}
\displaystyle{
\bra{\Omega}\tilde{R}_{i,i-1}\left(q_i-q_{i-1}+\hbar\right)\,...\,\tilde{R}_{i,1}\left(q_i-q_1+\hbar \right)=
\bra{\Omega}P_{i,i-1}\,...\,P_{i,1}\,,
}
\eq
which is independent of $\hbar$ as well as the rest of $\hat{K}_i^{\hbar}$.

Let $\ket{\Psi}$ be the solution to the quantum Knizhnik-Zamolodchikov equations (\ref{w5}) belonging to a certain weight subspace $V\left(\{M\} \right)$. Then we claim that the function
\beq \label{w15}
\displaystyle{
\Phi = \braket{\Omega | \Psi}
}
\eq
is an eigenfunction for all the quantum Hamiltonians (\ref{w1}). Let us, firstly, illustrate the technique for the first quantum Hamiltonian
\beq \label{w16}
\begin{array}{c}
\displaystyle{
e^{\hbar \partial_i}\braket{\Omega | \Psi} = \braket{\Omega | \hat{K}_i^{\hbar} | \Psi} = \braket{\Omega | \hat{K}_i^{0} | \Psi}.
}
\end{array}
\eq
Then we pass from $\hat{K}^0_i$ to $\hat{H}_i$ by the following rule
\beq \label{w17}
\displaystyle{
\hat{H}_1^{\rm{R}}\braket{\Omega | \Psi} = \sum\limits_{i=1}^N \prod\limits_{k \neq i}^N \frac{1}{q_i-q_k} e^{\hbar \partial_i} \braket{\Omega | \Psi} = \sum\limits_{i=1}^N \braket{\Omega | \hat{H}_i | \Psi}.
}
\eq
Fortunately, if $\ket{\Psi}$ belongs to a certain weight subspace $V\left(\{m\} \right)$ then due to the quantum-classical duality (\ref{q35}) we know that the operator $\sum\limits_{i=1}^N\hat{H}_i$ acts on every Bethe vector from a subspace $V\left(\{m\} \right)$ by a multiplication on a certain monomial of twists and the action depends not on a vector, but only on the choice of a subspace. Thus, if we assume the completeness of Bethe ansatz then we have
\beq \label{w18}
\displaystyle{
\hat{H}_1^{\rm{R}} \braket{\Omega | \Psi} = \la_1 \braket{\Omega | \Psi}.
}
\eq
We do not write here an explicit expression for the eigenvalue as we find it more convenient to write them all later together.

To deal with the higher quantum Hamiltonians we follow the calculations from section 4 of \cite{ZZ1}. We skip the technical calculations for the higher Hamiltonians as they are exactly the same as in the \cite{ZZ1} and claim that one can obtain

\noindent {\bf Proposition 3.} {\em For operators $\hat{K}^0_i$ defined in (\ref{w5}) and the wave function (\ref{w15}) $\braket{\Omega | \Psi}$ satisfy
\beq \label{w19}
\displaystyle{
\prod\limits_{j=1}^{k}e^{\hbar \partial_{i_j}} \braket{\Omega | \Psi} = \braket{\Omega | \hat{K}^{0}_{i_1}...\hat{K}^{0}_{i_k} | \Psi}, \quad \mathrm{for} \quad i_l \neq i_m.
}
\eq
}
Using the proposition (\ref{w19}) one can obtain
\beq \label{w20}
\begin{array}{c}
\displaystyle{
\hat{H}_k^{\rm{R}}\braket{\Omega | \Psi}=\sum\limits_{I_k}\left[\prod\limits_{j \in I_k}\prod\limits_{ l \notin I_k}\frac{1}{q_j-q_l}\right] \prod\limits_{n=1}^k e^{\hbar \partial_{i_n}}=
}
\\ \ \\
\displaystyle{
 =(-1)^{\frac{k(k-1)}{2}} \sum\limits_{I_k} \braket{\Omega | \hat{H}_{i_1} ... \hat{H}_{i_k} | \Psi} \prod\limits_{m < r}^k(q_{i_m} - q_{i_r})^2 =\la_k \braket{\Omega | \Psi},
}
\end{array}
\eq
where the last equality is true due to the quantum-classical duality (\ref{q35}) and completeness assumption. Now we write the formula for the eigenvalues with the help of quantum-classical duality (\ref{q35})
\beq \label{w21}
\displaystyle{
\la_k = \left\{ \begin{array}{l}
V_1^{M_1}V_2^{M_2}\,...\,V_{n-1}^{M_{n-1}}V_{n}^{M_{n}},\;\; k=N,\\\\
V_{i+1}^{M_{i+1}}V_{i+2}^{M_{i+2}}\,...\,V_{n}^{M_{n}}, \;\; k=N -\sum\limits_{l=1}^i M_l,\\\\
0, \;\; \mathrm{otherwise}.
\end{array} \right.
}
\eq

\subsection{Quantum-quantum duality for trigonometric QKZ}
The trigonometric QKZ equation is the following system of holonomic equations
\beq \label{w22}
\displaystyle{
\hat{T}_i^{q} \ket{\Psi} = \hat{K}_i^{q} \ket{\Psi},
}
\eq
where $\ket{Psi} = \ket{\Psi}(z_1,...,z_N)$ is the same as in the previous section and
\beq \label{w23}
\displaystyle{
\hat{T}^{q}_i f(z_1,..,z_i,..,z_N) =f(z_1,..,qz_i,..,z_N),
}
\eq
\beq \label{w24}
\displaystyle{
\hat{K}^q_i = \tilde{R}_{i,i-1}\left(qz_i/z_{i-1}\right)...\tilde{R}_{i,1}\left(qz_i/z_1 \right)V^{(i)}\tilde{R}_{i,n}\left(z_i/ z_N\right)...\tilde{R}_{i,i+1}\left(z_i/z_{i+1}\right)
}
\eq
for the $R$-matrix from (\ref{q24})
\beq \label{w25}
\begin{array}{c}
\displaystyle{
R(x) =\left( \sum\limits_{a=1}^{n}E_{aa}\otimes E_{aa} + (1-z)\sum\limits_{a > b}^{n}E_{aa}\otimes E_{bb} + z\sum\limits_{a<b}^{n}E_{ab}\otimes E_{ba} + \right.
}
\\ \ \\
\displaystyle{
+\left. \sum\limits_{a>b}^{n}E_{ab}\otimes E_{ba} \right),
}
\end{array}
\eq
satisfying unitarity condition
\beq
\displaystyle{
R_{12}(z)R_{21}(z^{-1}) =1_n\otimes 1_n\,.
}
\eq
As in the previous section we introduce the momenta operators and split the space $\left(\mathbb{C}^n\right)^{\otimes N}$ into the direct sum
\beq \label{w26}
\displaystyle{
\left(\mathbb{C}^n\right)^{\otimes N} = \bigoplus\limits_{m_1,...,m_n}V\left(\{m\} \right)
}
\eq
and the operators
\beq \label{w27}
\displaystyle{
\hat{H}_i = \hat{K}^1_i \prod\limits_{k \neq i}^N \frac{z_k}{z_k-z_i}\,,
}
\eq
which coincide with non-local Hamiltonians (\ref{q27}). Let us also consider the following covector
\beq \label{w28}
\displaystyle{
\bra{J} = \sum\limits_{i_1,...,i_N=1}^{n}\bra{e_{i_1}} \otimes ... \otimes \bra{e_{i_N}}
}
\eq
with the following property similar to the (\ref{w14}) from the previous section
\beq \label{w29}
\displaystyle{
\bra{J} \hat{K}^q_i = \bra{J} \hat{K}^1_i.
}
\eq
Then we claim that the following proposition analogous to (\ref{w19})

\noindent{\bf Proposition 4.} {\em
For operators $\hat{K}^1_i$ defined in (\ref{w24}) and let $\ket{\Psi}$ be the solution for the QKZ equations (\ref{w22}) then the wave function  $\braket{J | \Psi}$ satisfy
\beq \label{w30}
\displaystyle{
\prod\limits_{j=1}^{k}\hat{T}^q_j \braket{J | \Psi} = \braket{J | \hat{K}^{1}_{i_1}...\hat{K}^{1}_{i_k} | \Psi}, \quad \mathrm{for} \quad i_l \neq i_m.
}
\eq
The proof is exactly the same as in \cite{ZZ1}.
}

\noindent Now we proceed as for the rational case
\beq \label{w31}
\begin{array}{c}
\displaystyle{
\hat{H}^{\rm{T}}_k \braket{J | \Psi} =(-1)^{k(N-k)} \sum\limits_{I_k}\left[\prod\limits_{j \in I_k}\prod\limits_{l \notin I_k}\frac{z_l}{z_j-z_l}\right] \prod\limits_{n=1}^{k}\hat{T}^q_{i_n} \braket{J | \Psi}=
}
\\ \ \\
\displaystyle{
=(-1)^{k(N-k)}\sum\limits_{I_k} \left[\prod\limits_{j \in I_k}\prod\limits_{l \notin I_k}\frac{z_l}{z_j-z_l}  \prod\limits_{r \in I_k}\prod\limits_{m \neq r} \frac{z_m-z_r}{z_m}\right]  \braket{J |H_{i_1}..H_{i_k} | \Psi}=
}
\\ \ \\
\displaystyle{
=(-1)^{\frac{k(k-1)}{2}} \sum\limits_{I_k}\prod\limits_{l<r}\frac{(z_{i_l} - z_{i_r})^2}{z_{i_l}z_{i_r}}\braket{J | H_{i_1}..H_{i_k} | \Psi} = \la_k \braket{J | \Psi}
}
\end{array}
\eq
where the eigenvalue
\beq \label{w32}
\displaystyle{
\la_k=V_{1}^{k-[N-m_1]_k} V_{2}^{[N-m_1]_{k}-[N-m_1-m_2]_{k}}\,...\,V_{n}^{[N-m_1-..-m_{n-1}]_k}.
}
\eq
Thus, we constructed the eigenfunctions for the quantum trigonometric goldfish model (\ref{w2}) from the solutions of QKZ equations (\ref{w22}) with eigenvalues (\ref{w32}).
%\newpage
%

\section{Degenerate affine nil-Hecke algebra and open Toda chain}
In this section we consider a holonomic system of differential equations which after a projection on a certain subspace produces the open Toda chain quantum problem. We follow the ideas of \cite{Ch1}, \cite{Ch2}, \cite{M1}, for the geometric viewpoint and generalization to arbitrary root systems see \cite{BMO}. It is well-known that Calogero-Moser-Sutherland and Ruijsenaars quantum operators can be constructed with the help of Hecke algebras. Also, one can obtain open Toda chain from the trigonometric Calogero-Moser-Sutherland system via Inozemtsev limit \cite{I}. We claim that on the level of KZ equations Inozemtsev limit could be seen as a limit from Hecke algebra to a corresponding nil-Hecke algebra. As mentioned above trigonometric Calogero-Moser-Sutherland model is a certain projection of KZ equations written for affine degenerate Hecke algebra (dAHA) \cite{Ch1}. To obtain quantum open Toda chain one should write KZ equations for nil-dAHA
\beq \label{ww1}
\begin{array}{c}
\displaystyle{
\partial_i \ket{\Psi} = \left( -\sum\limits_{j >i}^N e^{x_i-x_j}T_{ij}+\sum\limits_{j<i}^N e^{x_j-x_i}T_{ji} +y_i\right) \ket{\Psi},
}
\end{array}
\eq
where $\ket{\Psi}$ lives in some vector space which we will define later. The consistency condition for (\ref{ww1}) is an algebraic restriction on operators $y_i$ and $T_{ij}$. To define these operators we introduce nil-dAHA algebra:

\noindent {\bf Definition \cite{K}:} {\em Degenerate affine nil-Hecke algebra (proper reference for first appearance) is an associative unital algebra $\mathcal{H}^0_N$ (over $\mathbb{C}$) generated by $\left(T_1,..,T_{N-1},y_1,..,y_N\right)$ with relations
\beq \label{ww2}
\begin{array}{c}
\displaystyle{
\begin{array}{cc}
 T_i^2=0, & T_i T_j = T_j T_i \;\; {\rm for} \;\; |i-j|>1,
\end{array}
}
\\ \ \\
\displaystyle{
T_i T_{i+1} T_i = T_{i+1}T_i T_{i+1},
}
\\ \ \\
\displaystyle{
\begin{array}{cc}
y_i y_j = y_j y_i,
\end{array}
}
\\ \ \\
\displaystyle{
y_j T_i = T_i y_j \;\; {\rm for} \;\; j \neq i,i+1,
}
\\ \ \\
\displaystyle{
T_iy_i - y_{i+1}T_i = y_i T_i- T_i y_{i+1} = k \in \mathbb{C}.
}
\end{array}
\eq
}
\noindent{\bf Proposition:} {\em System (\ref{ww1}) is consistent if
\beq \label{ww3}
\begin{array}{c}
\displaystyle{
T_i = T_{i,i+1}, \;\; 1 \leq i \leq N-1,
}
\\ \ \\
\displaystyle{
T_{ij} = (-k)^{i-j+1} T_i T_{i+1}...T_{j-2}T_{j-1}T_{j-2}...T_{i+1}T_i, \;\; j>i
}
\end{array}
\eq
and $\left\{T_1,..,T_{N-1},y_1,..,y_N \right\}$ are generators of nil-dAHA algebra (\ref{ww2}).
}

\noindent{\bf Proof} The proof can be given by direct calculations, thus we omit the details, also see \cite{BMO}. $\blacksquare$

Before moving on to construction of Toda chain from (\ref{ww1}) let us summarize several facts \cite{GL}, \cite{JB} about nil-dAHA algebra (\ref{ww2}):
\begin{itemize}
    \item Center of algebra (\ref{ww2}) is a ring of symmetric polynomial in $\{y_i\}$: $\mathcal{Z}(\mathcal{H}_n^{0})=S(y_1,..,y_n)$.
     \item To study (\ref{ww1}) we need a special type of representation of nil-dAHA. It could be described as "highest weight" type of representation $V_{\la}$, let $\ket{0}$ be the highest weight vector and the module of $\mathcal{H}_n^{0}$ is generated by the action of $T_i$
     \beq \label{ww5}
     \begin{array}{c}
     \displaystyle{
     y_i \ket{0} = \lambda_i \ket{0}.
     }
     \end{array}
     \eq
     As a vector space this module is isomorphic to nil-Coxeter algebra generated by $\left. \left\{T_i \right\} \right|_{i=1}^{N-1}$. The action of $y_i$ should be deduced from the relations. For general $\lambda_i$ this representation is irreducible.
\end{itemize}
Now we specify the representation $V_{\la}$ of nil-dAHA in (\ref{ww1}) and introduce the dual highest weight vector $\bra{0} \in V_{\la}^{*}$. Notice that it has a remarkable property that it is annihilated by the action of $T_i$
\beq \label{ww6}
\begin{array}{c}
\displaystyle{
\bra{0} T_i = 0.
}
\end{array}
\eq
 Let us apply derivative to (\ref{ww1})
\beq \label{ww7}
\begin{array}{c}
\displaystyle{
\partial_i^2 \ket{\Psi} = \left( -\sum\limits_{k >i}^n e^{x_i-x_k}T_{ik}-\sum\limits_{k<i}^n e^{x_k-x_i}T_{ki}\right)\ket{\Psi} +
}
\\ \ \\
\displaystyle{
+\left( -\sum\limits_{k >i}^n e^{x_i-x_k}T_{ik}+\sum\limits_{k<i}^n e^{x_k-x_i}T_{ki} +y_i\right)^2 \ket{\Psi}.
}
\end{array}
\eq
Now we project (\ref{ww7}) on the dual highest weight vector $\bra{0}$ noticing that it is annihilated by  all $T_{ml}$
\beq \label{ww8}
\begin{array}{c}
\displaystyle{
\bra{0} \partial_i^2 \ket{\Psi} =\bra{0} y_i\left( -\sum\limits_{k >i}^n e^{x_i-x_k}T_{ik}+\sum\limits_{k<i}^n e^{x_k-x_i}T_{ki} +y_i\right)\ket{\Psi}.
}
\end{array}
\eq
We denote $\braket{0|\Psi} = \Phi(x_1,..,x_n) $ and summing all the terms we obtain
\beq\label{ww9}
\begin{array}{c}
\displaystyle{
\sum\limits_{i=1}^n \partial_i^2 \Phi =  \bra{0} \left( -\sum\limits_{k >i}^n e^{x_i-x_k}y_iT_{ik}+\sum\limits_{k<i}^n e^{x_k-x_i}y_iT_{ki} +\sum\limits_{i=1}^ny_i^2 \right)\ket{\Psi} =
}
\\ \ \\
\displaystyle{
=  \bra{0} \left( -\sum\limits_{k >i}^n e^{x_i-x_k}y_iT_{ik}+\sum\limits_{k<i}^n e^{x_k-x_i}y_iT_{ki} +\sum\limits_{i=1}^ny_i^2 \right)\ket{\Psi} =
}
\\ \ \\
\displaystyle{
=\bra{0} \left( \sum\limits_{k >i}^n e^{x_i-x_k}\left(y_k-y_i\right)T_{ik}+y_i^2 \right)\ket{\Psi}=
}
\\ \ \\
\displaystyle{
=\bra{0} \left( \sum\limits_{k >i}^n e^{x_i-x_k}\left[y_k-y_i,T_{ik}\right]+\sum\limits_{i=1}^ny_i^2 \right)\ket{\Psi} =
}
\\ \ \\
\displaystyle{
\bra{0} \left( \sum\limits_{k >i}^n e^{x_i-x_k}\left[y_k+y_i,T_{ik}\right] -2\sum\limits_{k>i}^ne^{x_i-x_k}\left[y_i,T_{ik}\right]+\sum\limits_{i=1}^ny_i^2 \right)\ket{\Psi}=
}
\\ \ \\
\displaystyle{
\bra{0}\left(-2\sum\limits_{k>i}^ne^{x_i-x_k}y_iT_{ik}+\sum\limits_{i=1}^ny_i^2 \right)\ket{\Psi}.
}
\end{array}
\eq
We proceed with the following observation
\beq \label{ww10}
\begin{array}{c}
\bra{0} y_i T_{ik} = 0 ,\;\;{\rm for}\; k>i+1,
\end{array}
\eq
because for $k>i$ $T_{ik}$ is given by more than one operator $T_j$, thus one of them would inevitably annihilate the dual vacua $\bra{0}$. Thanks to (\ref{ww10}) we continue the calculation (\ref{ww9})
\beq\label{ww11}
\begin{array}{c}
\displaystyle{
\sum\limits_{i=1}^n \partial_i^2 \Phi = \bra{0}\left(-2\sum\limits_{i=1}^{n-1} y_i T_i e^{x_i-x_{i+1}} +\sum\limits_{i=1}^n y_i^2 \right)\ket{\Psi} = \bra{0}\left(-2k\sum\limits_{i=1}^{n-1}e^{x_i-x_{i+1}} + \sum\limits_{i=1}^ny_i^2 \right)\ket{\Psi}.
}
\end{array}
\eq
Remember that symmetric polynomials of $\left\{y_i\right\}$ are Casimir elements, thus they act identically in the irreducible representation
\beq \label{ww12}
\begin{array}{c}
\displaystyle{
\left(\sum\limits_{i=1}^n \partial_i^2 +2k\sum\limits_{i=1}^{n-1}e^{x_i-x_{i+1}}\right) \Phi =  \left(\sum\limits_{i=1}^{n}\lambda_i^2 \right) \Phi,
}
\end{array}
\eq
which means that function $\Phi\left(x_1,..,x_n\right)$ is an eigenfunction of non-relativistic  open Toda chain problem. We see that linear problem (\ref{ww1}) generates nil-dAHA as it's consistency and reproduces solutions for quantum open Toda chain problem. The parameter $k$ in the Hecke algebra plays the role of coupling constant (which can be eliminated by appropriate shifts of $\left\{x_i\right\}$) and parameters of representation of degenerate affine nil-Hecke algebra account for the eigenvalue of quantum many-body problem.

To obtain the third open Toda Hamiltonian one should follow the same strategy as in previous derivations. Let us denote the operator in the r.h.s of (\ref{ww1})
\beq \label{ww13}
\begin{array}{c}
\displaystyle{
\begin{array}{cc}
H_i =  -\sum\limits_{j >i}^N e^{x_i-x_j}T_{ij}+\sum\limits_{j<i}^N e^{x_j-x_i}T_{ji} +y_i, & \left[H_i,H_j\right] =0.
\end{array}
}
\end{array}
\eq
We calculate the following expression
\beq \label{ww14}
\begin{array}{c}
\displaystyle{
\left(\sum\limits_{i=1}^N \partial_i^3 + 3k \sum\limits_{i=1}^{N-1}e^{x_i-x_{i+1}} \left( \partial_i + \partial_{i+1}\right) \right)\braket{0|\Psi} =
}
\\ \ \\
\displaystyle{
\bra{0} \left(\sum\limits_{i=1}^N \left(H_i \partial_i H_i +2\left(\partial_i H_i \right)H_i + H_i^3 + \partial_i^2 H_i\right) + 3k\sum\limits_{i=1}^{N-1} e^{x_i-x_{i+1}}\left(H_i + H_{i+1} \right) \right) \ket{\Psi} =
}
\\ \ \\
\displaystyle{
= \left(\sum\limits_{i=1}^N \la_i^3 \right) \braket{0| \Psi}.
}
\end{array}
\eq
Let us comment on the structure of (\ref{ww1}) from the point of view of quantum integrable systems. For the case of trigonometric KZ equations which produces trigonometric Calogero-Moser-Sutherland model the operators in the r.h.s of (\ref{ww1}) are exactly commuting Hamiltonians of trigonometric inhomogeneous Gaudin system. Thus, we can think about operators (\ref{ww13}) as integrals of motion of some quantum integrable "spin chain" acting in the representation of nil-dAHA and try to study it's spectra. We conjecture that joint spectra of (\ref{ww13}) correspond to the intersection points of two Lagrangian submanifolds for classical open Toda chain: the first one corresponds to the submanifold of fixed momenta $p_i = \Lambda_i(x)$, where $\Lambda_i(x)$ is the eigenvalue of Hamiltonian $H_i$ from  (\ref{ww13}) and the second one is a submanifold of fixed classical integrals of motion of open Toda chain. We also note that since qKZ equation (\ref{w4}) - (\ref{w5}) provide solutions for rational Goldfish model and the KZ equation (\ref{ww1}) for the open Toda chain we expect the (q)KZ equations (\ref{w4}) and (\ref{ww1}) to be compatible as Toda chain and rational Goldfish models are spectrally dual, however we do not know how to check this conjecture since they are written in totally different languages \footnote{Recently, after the publication of the paper, Tarasov and Varchenko \cite{TVh} considered KZ equations of type (\ref{w5}) and (\ref{ww1}) and proved their compatibility}.

%\section{Discussion}

\section{Conclusion}

In this study we have demonstrated the mapping between the
QC and QQ dualities familiar
for integrable many-body systems and duality
relations known in the realm of integrable probabilities.
It is a real surprise that these relations have not
been noticed so far because of the non-overlapping
communities. The mapping explains the observed
dualities for integrable probabilities as QQ dualities
between the integrable long-range many-body systems
and inhomogeneous spin chains under the proper mapping
of parameters. The results obtained for the observables
averaged over the boundary conditions with the measure defined in the
stochastic spin models yield the interesting multiple integral representations for the spectrum  averaged observables  for the CM and RS model.

In this paper we restrict ourselves by the QQ dualities for
wave functions and measures for
integrable probabilities. In the forthcoming paper we shall discuss
in details the corresponding generalization
of QQ dualities for the processes. The realization of the spectral
duality in the probabilistic setting will be considered as well.
Since inhomogeneous ASEP
is related by gauge transformation with XXZ spin chain
we expect that ASEP model is spectrally self-dual.
It was  argued in  \cite{borodindual} that indeed
the duality relations are exactly the same as expected
from spectral self-duality of XXZ. The useful tKZ
interpretation of duality in stochastic processes
can be found in \cite{chen}.

We have conjectured that the  representation
of the wave functions from the CM-RS family in the QQ dual form as
sum of the colored paths in the discrete stochastic model
can be considered as  manifestation of more general phenomenon. Namely
it is a kind of  "dual Feynman path integral" for the wave function
in terms of open paths in a Hilbert space instead of the conventional
Feynman integral over the open paths in the coordinate space. In this
dual representation the coordinate $x$ serves as a weight
for some characteristic of a path. We noted such interpretation
for the simplest oscillator example. Similar and more complicated
combinatorial representations are known for
Macdonald polynomials as well. This issue certainly deserves further
study.

To clarify the details of the correspondence  we
elaborate new  duality between inhomogeneous periodic multi-species TASEP and
Goldfish models.
First we have formulated the QC duality for rational and
trigonometric classical and quantum Goldfish models. We have found that these
systems are QC dual to the rational 5-vertex model and
multi-species TASEP model. As usual the coordinates in classical Goldfish systems were
identified with jump inhomogeneities of the TASEP and the velocities of
classical systems were identified with the eigenvalues of non-local
quantum Hamiltonians of quantum systems. The values of classical
Goldfish Hamiltonians are fixed by the twist parameters of TASEP. This correspondence
may be seen as a limit $\hbar \rightarrow \infty$ of known correspondence
between XXX/XXZ spin chains and rational and trigonometric
RS systems respectively. We have argued how to extend duality to QQ level and obtain the eigenfunctions for
quantum Goldfish models from the solutions of quantum Knizhnik-Zamolodchikov
equations for the 5-vertex $R$-matrices.

At present the spin Macdonald process   is on top
of hierarchy of the integrable probabilities while the
six-vertex higher spin stochastic model
is considered as the  top of another hierarchy.  It is clear that there is
elliptic generalization of the integrable probabilities
and therefore the dualities at the elliptic level
shall involve the
elliptic RS, double elliptic models and corresponding dual elliptic
stochastic processes generalizing colored 6-vertex models.
Some steps in this direction have been recently
made both from the stochastic side \cite{borodinell}
and integrable elliptic many-body systems
side \cite{shakirov,awata}.

It is clear that the $(q,t)KZ$
equation and representations of the DIM algebra
will play an essential role for  elliptic
stochastic processes. In particular the spectral
duality is the automorphism of DIM algebra while
some ingredients of QQ duality have been observed
in the context of representations of DIM algebra
as well \cite{zenkevich}.

It would be also interesting to utilize some
known results for TASEP and ASEP models at the
dual CM-RS side. In particular it would be interesting
to identify at CM-RS side the explicit
expressions for the vacuum expectation value of current in TASEP,
the TASEP spectral curve \cite{prolhac} and
the jamming phase transitions known in the
TASEP models with different boundary conditions.

Oppositely some known results at the CM-RS side
certainly can be used at the stochastic side. We could
mention the $\nu\rightarrow \frac{1}{\nu}$ and
$\nu\rightarrow 1 -\nu$ dualities in Calogero model
which implies dualities involving $SL(2,Z)$ action
on the Planck constant $\hbar$ in the ASEP and TASEP
models. Another interesting issue concerns the role
of $SL(2,R)$ algebra known in the CM model (see,for
instance \cite{etingof}) which involves the Hamiltonian,
dilatation and boost operator at the ASEP-TASEP side.
It would be also interesting to reformulate the
stochastic models in terms of the 4d-Chern-Simons
theory using the recent derivation of the spin chain Hamiltonian
along this line \cite{costello1,costello2}(see,
also \cite{nikthesis} for the early study). This should
be linked along the logic of QQ duality with the
derivation of trigonometric RS model via perturbed
CS theory \cite{gn}. We plan to discuss these issues
elsewhere.

Another interesting question concerns the large
$N$ limit of the stochastic models. In this limit
the hydrodynamical description is expected for
both sides of the stochastic hierarchies. At the
Macdonald side one expects some version of ILW
equation \cite{koroteevhydro,gkkv,grekov,zzilw}, while
the hydrodynamical version of the six-vertex
stochastic model still has to be formulated. The
duality between two hydrodynamical models,
which is large $N$ generalization of QQ duality for
the many-body systems can be expected.
Another continuation of study may concern the connection between five-vertex models and
integrable Novikov-Veselov equation \cite{NV} in the spirit of \cite{AKLTZ}.

\subsection*{Acknowledgments}
The work of A. Gorsky was supported by Basis Foundation grant 20-1-1-23-1  and by grant RFBR-19-02-00214.
%We are grateful to E. Zenkevich for
%the useful discussions.
The work of M. Vasilyev and A. Zotov was performed at the Steklov International Mathematical Center and supported by the Ministry of Science and Higher Education of the Russian Federation (agreement no. 075-15-2019-1614).

%
%
%
%
%\newpage
\section{Appendix A: classification of five-vertex R-matrices}
 \def\theequation{A.\arabic{equation}}
\setcounter{equation}{0}
Here we classify all possible multiplicative 5-Vertex $R$-matrices\footnote{This part was done together with A. Liashyk}. To be more precise we want to find all the solutions for maps $R:\mathbb{C} \rightarrow \mathrm{End}(\mathbb{C}^2 \otimes \mathbb{C}^2)$ satisfying multiplicative quantum Yang-Baxter equation
\beq \label{A1}
\displaystyle{
R_{12}(\frac{x_1}{x_2})R_{13}(\frac{x_1}{x_3})R_{23}(\frac{x_2}{x_3}) = R_{23}(\frac{x_2}{x_3})R_{13}(\frac{x_1}{x_3})R_{12}(\frac{x_1}{x_2})
}
\eq
for the special forms of the $R$ - matrix
\beq \label{A2}
\displaystyle{
R(x) = \left(
\begin{array}{cccc}
a(x) & 0 & 0& 0\\
0 & 0& b(x)& 0\\
0& c(x) & d(x) & 0\\
0 & 0& 0& 1
\end{array}\right).
}
\eq
Notice that we used the fact that $R$-matrix is defined up to multiplication by the scalar function, thus we set the lower-right element equal to 1. In this respect the form (\ref{A2}) is not generic since we assume the coefficient behind $E_{22}\otimes E_{22}$ to be nonzero.

We search for the solutions such that all functions $a(x)$, $b(x)$, $c(x)$, $d(x)$ are non-zero differentiable functions(for one case $d(x)$ will be equal to zero, but this will be the only case). One can write all the equations for functions explicitly using (\ref{A1}) and find out that all the equations are equivalent to the following set of relations.
\beq \label{A3}
\displaystyle{
\left\{
\begin{array}{lr}
a(\frac{x_1}{x_2})a(\frac{x_2}{x_3}) = a(\frac{x_1}{x_3}), & (1)\\\\
b(\frac{x_1}{x_2})b(\frac{x_2}{x_3}) = b(\frac{x_1}{x_3}), & (2)\\\\
c(\frac{x_1}{x_2})c(\frac{x_2}{x_3}) = c(\frac{x_1}{x_3}), & (3)\\\\
b(\frac{x_2}{x_3})c(\frac{x_1}{x_3})d(\frac{x_1}{x_2}) + a(\frac{x_1}{x_3})c(\frac{x_1}{x_2})d(\frac{x_2}{x_3})-a(\frac{x_2}{x_3})c(\frac{x_1}{x_2})d(\frac{x_1}{x_3}) = 0, & (4)\\\\
b(\frac{x_1}{x_2})a(\frac{x_2}{x_3})d(\frac{x_1}{x_3}) + a(\frac{x_1}{x_3})b(\frac{x_1}{x_2})d(\frac{x_2}{x_3})-b(\frac{x_1}{x_3})c(\frac{x_2}{x_3})d(\frac{x_1}{x_2}) = 0, & (5)\\\\
c(\frac{x_2}{x_3})d(\frac{x_1}{x_3})-c(\frac{x_2}{x_3})d(\frac{x_1}{x_2})-b(\frac{x_1}{x_2})c(\frac{x_1}{x_3})d(\frac{x_2}{x_3})=0, & (6)\\\\
b(\frac{x_2}{x_3})d(\frac{x_1}{x_2})-b(\frac{x_2}{x_3})d(\frac{x_1}{x_3})-b(\frac{x_1}{x_3})c(\frac{x_1}{x_2})d(\frac{x_2}{x_3})=0. & (7)
\end{array}
\right.
}
\eq
Fortunately, equations (1), (2), (3) can be easily solved and we find $a(x) = x^a$, $b(x)=x^b$, $c(x)=x^c$. Therefore, we can rewrite  equations (4)-(7)
\beq \label{A4}
\displaystyle{
\left\{
\begin{array}{lr}
\frac{x_2^b}{x_3^{b+c}}d(\frac{x_1}{x_2})+\frac{x_1^a}{x_3^a x_2^c}d(\frac{x_2}{x_3})-\frac{x_2^{a-c}}{x_3^a}d(\frac{x_1}{x_3}) = 0, & (*)\\\\
\frac{x_2^{a-b}}{x_3^{a}}d(\frac{x_1}{x_3})-\frac{x_1^a}{x_3^a x_2^b}d(\frac{x_2}{x_3})-\frac{x_2^{c}}{x_3^{b+c}}d(\frac{x_1}{x_2}) = 0, & (**)\\\\
x_2^c d(\frac{x_1}{x_3})-x_2^c d(\frac{x_1}{x_2})-\frac{x_1^{b+c}}{x_2^b}d(\frac{x_2}{x_3})=0, & (***)\\\\
x_2^b d(\frac{x_1}{x_2})-x_2^b d(\frac{x_1}{x_3})+\frac{x_1^{b+c}}{x_2^c}d(\frac{x_2}{x_3})=0. & (****)
\end{array}
\right.
}
\eq
Equations (***) and (****) are equivalent. One can easily obtain from (***) setting $x_2=x_3$ that $d(1)=0$. Then, we differentiate (***) with respect to $x_3$ and obtain
\beq \label{A5}
\displaystyle{
x_1 x_2^c d^{'}(\frac{x_1}{x_3}) = \frac{x_1^{b+c}}{x_2^b}d^{'}(\frac{x_2}{x_3}),
}
\eq
then setting $x_2=x_3$ we find
\beq \label{A6}
\displaystyle{
d^{'}(\frac{x_1}{x_2}) = (\frac{x_1}{x_2})^{b+c-1}d^{'}(1).
}
\eq
Now we can find $d(x)$ as we have the first order differential equation and initial condition $d(1)=0$
\beq \label{A7}
\displaystyle{
d(x) = \left\{
\begin{array}{ll}
\ga (x^{b+c}-1),\;\;\; b+c \neq 0,\\
\al \ln{x}, \;\;\; b+c=0,
\end{array}
\right.
}
\eq
where $\ga=\frac{d^{'}(1)}{b+c}$, let us mention that function $d(x)$ is defined by equations (*)-(****) up to multiplicative constant, so $d^{'}(1)$ is an arbitrary number then $\ga$ and $\al$ are arbitrary. Now we have to find out if such solutions solve (*) and (**). Substituting solutions (\ref{A7}) we find 3 series of solutions:
\paragraph{1)} $a=0$.
For this case one obtains no restrictions on $b$ and $c$:
\beq \label{A8}
\displaystyle{
R(x) = \left(
\begin{array}{cccc}
1 & 0 & 0 & 0\\
0 & 0 & x^b & 0\\
0 & x^c & \ga(x^{b+c}-1) & 0\\
0 & 0 & 0 & 1
\end{array}
\right).
}
\eq
\paragraph{2)} $a \neq 0$.
For this case we obtain from (*) and (**) that $a=b+c$:
\beq \label{A9}
\displaystyle{
R(x) = \left(
\begin{array}{cccc}
x^{b+c} & 0 & 0 & 0\\
0 & 0 & x^b & 0\\
0 & x^c & \ga(x^{b+c}-1) & 0\\
0 & 0 & 0 & 1
\end{array}
\right).
}
\eq
\paragraph{3)} $b+c=0$.
For this case we obtain $a=0$:
\beq \label{A10}
\displaystyle{
R(x) = \left(
\begin{array}{cccc}
1 & 0 & 0 & 0\\
0 & 0 & x^b & 0\\
0 & x^{-b} & \al \ln{x} & 0\\
0 & 0 & 0 & 1
\end{array}
\right).
}
\eq
%
%
%
%

%%%%%%%%%%%%%%%%%%%%%%%%%%%%%%%%%%%%%%%%%%%%%%%%%%%%%%%%%%%%%%%%%%%%%%%%%%%%%%%%%%%%%%%%%%%%%%%%%%%%%%%%%%%%%%%%%%%

\section{Appendix B: wave function as correlator
of height functions in KPZ-like models. Polymer picture }
\def\theequation{B.\arabic{equation}}
\setcounter{equation}{0}

Integrable probabilities are interested in life in a realm of
the non-equilibrium statistical physics and belong to the KPZ
universality class involving the different versions of
stochastic growth problems.
The continuum KPZ stochastic equation
for the height function $h(x,t)$  in $(1+1)$ reads as
\beq
\frac{\partial h(x,t)}{\partial t}=\frac{1}{2} \frac{\partial^2 h(x,t)}{\partial^2 x} +
\frac{1}{2} (\frac{\partial h(x,t)}{\partial x})^2 + \eta(x,t)\,,
\eq
where $\eta(x,t)$ is the Gaussian noise.
Upon the Cole-Hopf transform $Z(x,t)=\exp( h(x,t))$ the KPZ equation
gets mapped into stochastic equation of Fokker-Planck type
\beq\label{rr1}
\frac{\partial Z}{\partial t} = \frac{1}{2}\frac{\partial^2 Z}{\partial x^2} + \eta Z\,.
\eq
The function $Z(x,t)$ can be treated
as a partition function of polymer of length $t$ in the random environment.

The evaluation of $Z(x,t)$ can be performed via the replica trick
when instead of the single polymer $n$ polymers are considered, hence
one introduces the joint probability
averaged over disorder:
$$\Psi(x_1,\dots x_n,t)= <Z(x_1,t)\dots Z(x_n,t)>\,.$$
It turns out
(see \cite{spohn1,corwin2} for reviews)  that the
joint probability obeys the non-stationary equation for the
$n$-particle attractive  Lieb-Lineger
many-body system:
\beq\label{ll}
\displaystyle{
\frac{d\Psi}{dt}= H_n \Psi\,, \qquad H_n= -
\frac{1}{2} \sum^n \frac{\partial^2}{\partial x_j^2} -\frac{1}{2} \sum_{i\neq j}^n \delta(x_i -x_j)\,.
%\label{ll}
}
\eq
This example explains one of the roles which integrable many-body system
plays in the context of non-equilibrium dynamics. The Bethe ansatz equations
now came into the game and yield the spectrum of eigenvalues of (\ref{ll}).
Similarly one can consider the polymer in the semi-discrete (1+1) space-time
when a space coordinate is discretized. If one introduces
$N$ directed polymers in semi-discrete space-time in the random medium  starting from one point the joint polymer partition function
obeys the non-stationary Schrodinger equation for N-particle
open Toda chain \cite{ocon}.

Trigonometric Calogero-Moser system appears in the similar polymer setting as follows \cite{cardy}.
Consider the multiple radial SLE stochastic process which describes the growth
of N interacting polymers in the random environment on the disc.
The polymers grow from points $x_i(t)$ at the boundary circle
and obey the process
\beq\label{rr3}
dx_i(t)= \sum_{i\neq j} cot \frac{(x_i(t) -x_j(t))}{2} dt 
- \sqrt{k} \sum_{k\neq i} dB_k\,,
\eq
where $B_k$ are independent  Brownian motions started from the origin.

The joint probability up to conjugation coincides
with the wave function of Calogero-Moser Hamiltonian at energy $E$:
\beq\label{rr4}
H_{CM}\Psi(x_1,\dots x_n)= E \Psi(x_1,\dots x_n)\,,
\eq
\beq\label{rr5}
H_{CM}= -\frac{k}{2}\sum \frac{\partial^2}{\partial x_i^2} + \frac{2-k}{2k} \sum_{i\neq j}
\frac{1}{sin^2(x_i -x_j)/2} - \frac{N(N-1)}{2k}\,.
\eq
where  $E=\frac{k^2 -16}{32}$.

\section{Appendix C: from wave function to measure and process. Toy example}
\def\theequation{C.\arabic{equation}}
\setcounter{equation}{0}

\paragraph{On the dual Feynman integral for the wave function}
 We recall now the quantum mechanics of one degree of freedom. From
the textbook two following equivalent representations for
the partition function are known
\begin{equation}
    Z(\beta)= Tr_{\cal{H}} e^{-\beta H} = \int dx(t)e^{\frac{-S(x(t))}{\hbar}}\,,
\end{equation}
where in the Feynman path integral representation the periodic boundary
condition is imposed for trajectories in the Euclidean time $x(t)=x(t+\beta)$.
The first representation can be considered as the "closed paths in
the Hilbert space $\cal{H}$".

Turn now to the  representation of the wave function $\Psi(x,E)$ considered
as the function of two arguments. What are two analogous representations in this
case? The first one is the conventional Feynman representation for the wave function
\begin{equation}
    \Psi_E(x) = \int dx(t)e^{\frac{-S(x(t),E)}{\hbar}}\,,
\end{equation}
where we consider all open paths which end up at point $x$, while the energy E enters
the weight in the path integral. Hence in this representation one
argument of the wave function fixes the boundary condition for all open paths,
while the second argument enters the measure in the path statistical model.

Is there the dual Feynman integral for a wave function, where we expect
that $E$ provides a kind of boundary condition for the open paths in the Hilbert space,
while $x$ enters the measure for such collection of open paths?
The self-duality property of the Macdonald
polynomial serves as the simple argument in favour that such dual Feynman path integral
in the Hilbert space does exist. We should however answer what are the
paths in the Hilbert space involved and how the coordinate $x$ enters the measure for the summation over the paths.

\paragraph{Statistical model for oscillator}

It turns out that the dual Feynman integral for a wave function defines
the peculiar statistical model. The simplest example concerns the dual
path integral representation for the oscillator wave function given by Hermite polynomials \cite{gnv}.
In this case the statistical model gets identified with the path counting on the particular tree with varying degrees.

\begin{figure}[h!]
    \centering
    \includegraphics[width=0.5\linewidth]{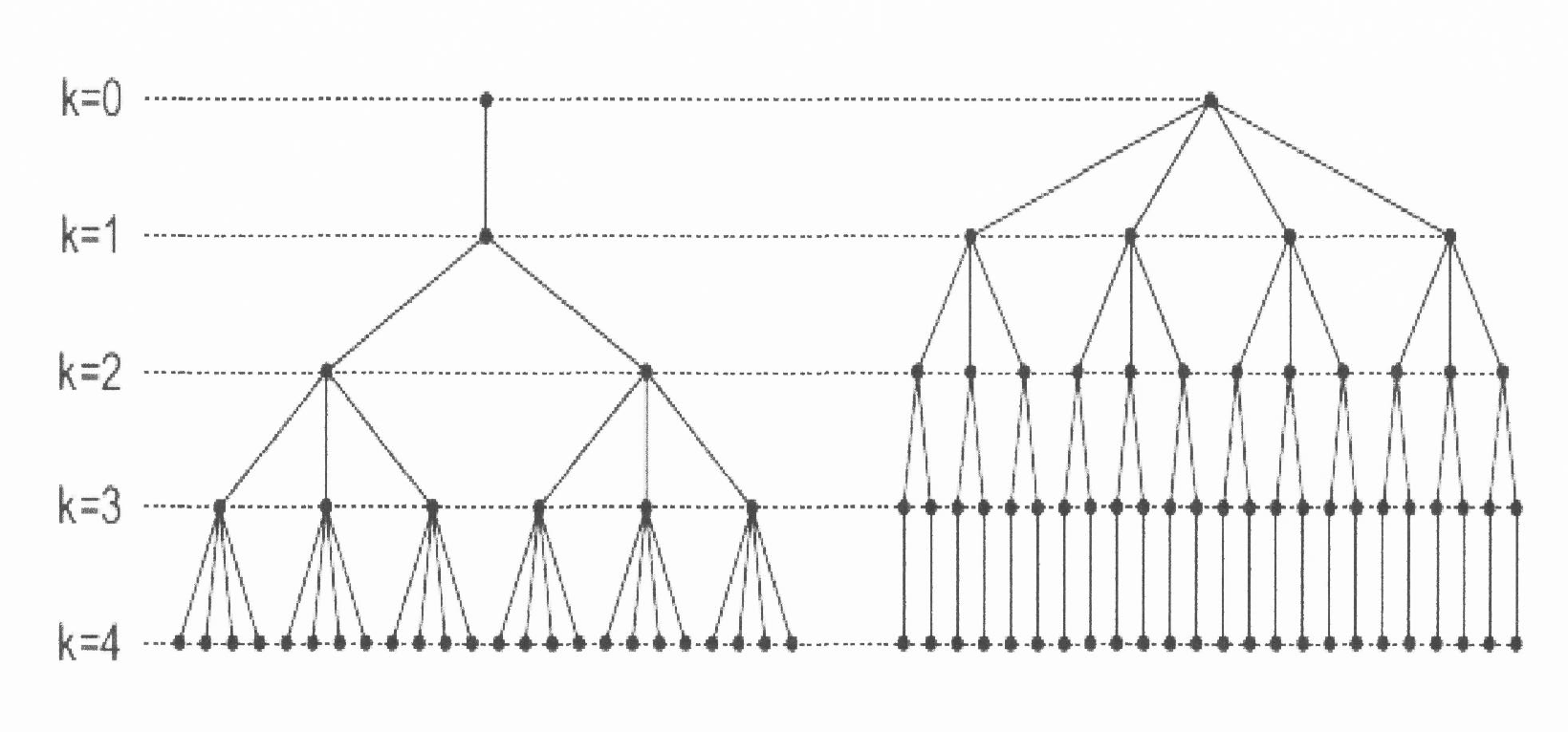}
    \caption{Super trees: (a) growing tree ${\cal T}^+$ with $p_0$ branches at the root point and $a=1$, (b) descending tree ${\cal T}^-$ with $p_0=4$ branches and $a=-1$.}
    \label{fig:01}
\end{figure}

Let us consider the following counting problem: given a regular finite tree ${\cal T}$, compute the partition function $Z_N(k)$, of all $N$-step trajectories starting at the tree root $(k=1)$ and ending at some tree level, $k$ ($k=0,1,...,K-1$). The ${\cal T}$ is the standard Cayley tree with the constant branching $p$ in each vertex at all tree levels.  The uniform $p$-branching Cayley tree, is regarded as a discretization of the target space possessing the hyperbolic geometry -- the Riemann surface of the constant negative curvature.
We consider paths statistics on symmetric finite "super trees", ${\cal T}^+$ and ${\cal T}^-$, of $K$ levels, for which the branching (vertex degree) $p$ is not constant, but linearly depends on the current level, $k$ ($k=0,1,2,...,K-1$), i.e.
\begin{equation}
p_k=\begin{cases} p_0, & \mbox{for $k=0$}\,,  \medskip \\ 2+ak, & \mbox{for $k\ge 1$, $a\ge 0$} \end{cases}
\label{01}
\end{equation}
for "growing trees", ${\cal T}^+$, and
\begin{equation}
p_k=\begin{cases} p_0, & \mbox{for $k=0$}\,, \medskip \\ p_0+ak, & \mbox{for $k\ge 1$, $a\le 0$} \end{cases}
\label{01b}
\end{equation}
for "descending trees", ${\cal T}^-$, where "branching velocity", $a$, is some integer-valued constant, and $p_0$ is the branching at the tree root, which is labelled by the index $k=0$. The extension of $a$ to the set of real numbers will be discussed below.  The growing tree, ${\cal T}^+$, with $p_0=1$ branches at the root point, and $a=1$, is shown on figure \ref{fig:01}a (left part of figure), while the descending tree, ${\cal T}^-$, with $p_0=4$ branches at the root point, and $a=1$, is depicted on figure \ref{fig:01}b (right part of figure).
%
%\begin{figure}
%\epsfig{file=super-fig01.eps,width=15cm}
%\caption{Super trees: (a) growing tree ${\cal T}^+$ with $p_0$ branches at the %root point and $a=1$, (b) descending tree ${\cal T}^-$ with $p_0=4$ branches and %$a=-1$.}
%\label{fig:01}
%\end{figure}
%
%
For a growing tree, the partition function, $Z_N(k)$, defined above, satisfies the recursion ($k=0,1,...,K-1$):
\begin{equation}
\begin{cases}
Z_{N+1}(k)=(p_{k-1}-1)Z_N(k-1)+Z_N(k+1) & \mbox{for $2\le k \le K-1$}\,, \medskip \\
Z_{N+1}(k)=Z_N(k+1), & \mbox{for $k=0$}\,, \medskip \\
Z_{N+1}(k)=p_{k-1} Z_N(k-1)+Z_N(k+1) & \mbox{for $k=1$}\,,\medskip \\
Z_{N+1}(k)=(p_{k-1}-1) Z_N(k-1), & \mbox{for $k=K-1$}\,, \medskip \\
Z_{N=0} = \delta_{k,0}\,.
\end{cases}
\label{02}
\end{equation}
To rewrite \ref{02} in a matrix form, make a shift $k\to k+1$ and construct the $K$-dimensional vector $\mathbf{Z}_N=(Z_N(1),Z_N(2),... Z_N(K))^{\top}$. Then \ref{02} sets the evolution of $\mathbf{Z}_N$ in $N$:
\begin{equation}
\mathbf{Z}_{N+1}=T\mathbf{Z}_N; \quad T=
\left(\begin{array}{cccccc}
0 & 1 & 0 & 0 &   \ldots & 0 \medskip \\
p_0 & 0 & 1 & 0 &  &  \medskip \\
0 & p_1-1 & 0 & 1 &  & \vdots \medskip \\
0 & 0 & p_2-1 & 0 &  & \medskip \\
\vdots &  &  &  & \ddots &  \medskip \\
0 &  & \dots & & p_{K-2}-1 & 0  \end{array}\right); \quad \mathbf{Z}_{N=0}=\left(\begin{array}{c} 1 \medskip \\ 0 \medskip \\ 0 \medskip \\ 0 \medskip \\ \vdots \medskip \\ 0\end{array} \right)\,.
\label{03}
\end{equation}
Now we proceed in a standard way and diagonalize the matrix $T$. The characteristic polynomials, $P_k(\lambda)=\det(T-\lambda \hat{I})$ of the matrix $T$ of size $k\times k$, satisfy the recursion
\begin{equation}
\begin{cases}
P_k(\lambda)=-\lambda P_{k-1}(\lambda)-(p_{k-2}-1)P_{k-2}(\lambda), & \mbox{for $3\le k \le K$}\,,
\medskip \\
P_1(\lambda)=-\lambda\,,\medskip \\
P_2(\lambda)=\lambda^2-p_0\,.
\end{cases}
\label{04}
\end{equation}

Now let us discuss the analytic solution of (\ref{04}) for a growing tree ${\cal T}^+$ for a special choice $p_0=1$ and $a=1$, and analyze the corresponding asymptotics of $P_K$. The characteristic polynomials, $P_k(\lambda)$ of the transfer matrix $T$ satisfy the recursion
\begin{equation}
\begin{cases}
P_k(\lambda)=-\lambda P_{k-1}(\lambda)-(k-1)P_{k-2}(\lambda), & \mbox{for $3\le k \le K$}\,,
\medskip
\\ P_1(\lambda)=\lambda\,, \medskip
\\ P_2(\lambda)=\lambda^2-1\,,
\end{cases}
\label{05}
\end{equation}
which coincide with the recursion for the so-called monic Hermite polynomials, ${\cal H}_k(\lambda)$, also known as the "probabilists Hermite polynomials":
\begin{equation}
P_k(\lambda) \equiv {\cal H}_k(\lambda)=(-1)^k e^{\frac{\lambda^2}{2}}{\frac{d^k} {d\lambda^k}}e^{-{\frac{\lambda^2}{2}}}; \qquad{\cal H}_k(\lambda)=2^{-k/2}H_k(\lambda / \sqrt{2)}\,,
\label{05a}
\end{equation}
where $H_k(\lambda)$ are the standard Hermite polynomials. Hence the eigenvalues of the matrix $T$ of size $K\times K$ (see (\ref{03})) are the roots of the monic Hermite polynomial, ${\cal H}_k(\lambda)$.

\paragraph{Wave function from matrix model}
The determinant representation has a clear counterpart in the
matrix model realization of quantum mechanics \cite{krefl} where
the potential is fixed by the measure in the matrix model.
Consider the integral over the eigenvalues of $N\times N$ matrix $M$
\begin{equation}
Z=\int \prod_{i}^N d\lambda_i\, (\lambda_i -\lambda_j)^{\beta}\, e^{\frac{\beta}{g_s}W(\lambda_i)}\,,
\end{equation}
where the "weight function" $W(x)$ is called  the "superpotential" and $g_s$ is constant.
It can be identified with stringy coupling constant which counts the genus expansion in the stringy
partition function.
It has been argued in  \cite{krefl} that the matrix element of the operator
\begin{equation}
\langle \det(M-x)^{\beta}\rangle = Z^{-1}\int \prod_{i}^N d\lambda_i\, (\lambda_i -\lambda_j)^{\beta} \det(M-x)^{\beta}\,  e^{\frac{\beta}{g_s}W(\lambda_i)}
\end{equation}
in the large $N$ limit plays the role of the wave function of the Shr\"odinger equation. Specifically, at $N\to \infty$, one sets $g_s\beta N = {\rm const}$, $\beta N = {\rm const}$ and the explicit form of the Schr\"odinger equation becomes
\begin{equation}
\hbar^2 \frac{d^2\Psi(x)}{dx^2} = \left[\left(\frac{d W(x)}{dx}\right)^2-f(x)\right]\Psi(x)\,,
\end{equation}
where the function $f(x)$ is the polynomial of degree $(d-2)$ if $W(x)$ has degree $d$. The function $f(x)$ is defined as
\begin{equation}
f(x)=\hbar\big(W'(x) + 2c(x) + d(x)\big)
\end{equation}
with
\begin{equation}
c(x)= \lim_{\beta \rightarrow 0} \left[\beta N \frac{W'(x)-W(0)}{x}\right], \qquad
d(x)= \hbar \lim_{\beta \rightarrow 0}\left[\beta \hat{D} Z\right]\,,
\end{equation}
where $W'(x) = \frac{dW(x)}{dx}$ and the operator $\hat{D}$ acts on the parameters of the polynomial superpotential
$W(x)=\sum_{i=1}^{n}t_i x^i$ as $\hat{D}=\sum_{i=1}^{n}\partial_{t_i} $  \cite{krefl}.   Effectively, all Virasoro constraints are combined into a single equation. If $\beta$ is finite, the non-stationary Schr\"odinger equation with the same potential emerges.

From the geometric viewpoint one considers the refined topological string in the Calabi-Yau geometry parameterized by the superpotential $W(x)$ in the IIB model. The matrix model microscopically describes the refined topological string, for which the $\beta$-parameter of the matrix model is identified as $\beta=-\frac{\epsilon_1}{\epsilon_2}$ where $\epsilon_1,\epsilon_2$ are the standard equivariant parameters of the $\Omega$--deformation. We are interested in the limit $\beta\rightarrow 0$ that is the Nekrasov-Shatashvili limit of the refined topological string. It was recognized long time ago \cite{aga} that the operator $\det(M-z)$ in the matrix model corresponds to the insertion of the Lagrangian brane in the Calabi-Yau geometry hence from the geometrical viewpoint we are considering the wave function of the Lagrangian brane in the particular geometry. In the context of the Liouville theory such operator corresponds to the insertion of FZZT brane.

The immediate question dealing with the Shr\"odinger equation concerns the identification of the particular energy level in the spectrum. For this purpose it is useful to consider the Gaussian potential $W(x)= x^2$ \cite{krefl}. The spectrum of the corresponding oscillator Hamiltonian can be obtained from the matrix model, it reads
\begin{equation}
E=\hbar\left(\frac{1}{2} + \lim_{\beta \rightarrow 0}[\beta N]\right)\,,
\end{equation}
where one immediately recognizes the energy level $k = \lim_{\beta \rightarrow 0} [\beta N]$. Therefore, to derive the energy level and the corresponding determinant representation for the wave function, one has to start from large $N$ matrices and consider the suitable scaling limits.
In both cases $E$ provides the boundary condition -
end point for open paths or size of matrix, while $x$ effectively
defines the measure ("chemical potential for lengths").

This dual Feynman representation in terms of the
statistical model for one-body quantum mechanics
can be generalized to the
many-body systems. The nice example, which in fact underlies
the QQ duality is the path integral representation
for the Macdonald polynomials \cite{BW19}, which
we shall use in the next Section. The corresponding
statistical model gets identified exactly as
the colored inhomogeneous path model on the cylinder.
In that case as expected the coordinates (specializations) $x_i$
of Macdonald polynomials enter the weights of the
statistical model, while the spectral variables $\lambda_i$
enter a boundary condition for  paths on a cylinder.

\paragraph{Measure}

The Schur or Macdonald measures on partitions have
a clear-cut quantum-mecha\-nical interpretation.
Indeed, consider the bilinear in the wave functions
$M_{x,y}(E)= \Psi^{*}_{E}(x)\Psi_{E}(y)$, which has been
discussed in integrable probabilities when
wave function is, say, the Macdonald polynomial.
The probabilistic measure which depends on $(x,y)$ as parameters
is the measure on the
energy space, or in more general situation
on the space of integrals of motion.
In the context of integrable probabilities one
evaluates with this measure  the expectation values of observables
which are the functions of partitions.

To fit the place of these  objects in the conventional
quantum mechanics let us consider the transition amplitude for
oscillator between points $x, y$ during time T
\begin{equation}
<x|e^{iHT}|y>=\sum_N\Psi^{*}_{N}(x)\Psi_{N}(y)e^{iTE_N}\,,
\end{equation}
which has exactly the structure of the expectation
value of the function of the energy evaluated
with the probabilistic measure depending on the initial and
final points of evolution. Due to the discreteness
of the spectrum we have the sum over the spectrum
instead of the integration over the continuum.

As is well known that the transition amplitude can be evaluated explicitly
\begin{equation}
 <x|e^{iHT}|y>= (\frac{m\omega}{2\pi i\hbar sin \omega T})^{1/2}
 exp(\frac{im\omega}{2\hbar sin\omega T}[(x^2+y^2) cos \omega T -2xy])\,.
\end{equation}
Notice that the transition amplitude is non-analytic in $T$.
According to the standard analogy with the statistical mechanics
such transition amplitude can be related upon the Wick rotation
with the partition function of the polymer whose length is fixed
by T. In Appendix B we recall how this interplay between the transition
amplitudes and polymer partition functions can be extended
from the case of one degree of freedom to the
interacting integrable many-body systems.

It is instructive to compare this probabilistic
measure with another conventional quantum mechanical
objects. The spectral density defined as
\begin{equation}
    \rho(E)=\int dx \Psi^{*}_{E}(x)\Psi_{E}(x)
\end{equation}
can be considered as the Laplace transform of the partition
function
\begin{equation}
    \rho(E)=\int d\beta e^{\beta E} Z(\beta)\,.
\end{equation}
The spectral correlator $<\rho(E)\rho(E')>$ is more important
characteristics of the theory and its behavior defines  stochastic
or integrable properties of a model.

Remark that from the quantum mechanical
perspective the Wigner
function is another object
defined on the product of wave functions
evaluated at different points:
\beq
W_{E_1,E_2}(x,p)= \int dy e^{ipy}\Psi^*_{E_1}(x+\hbar y)
\Psi_{E_2}(x-\hbar y)\,.
\eq
However contrary to the probabilistic measure
yielding the measure on the spectrum the Wigner function
defines measure on the phase space
which allows to evaluate the matrix elements
\beq
<E_1|\hat{O}|E_2>=\int dxdp O(x,p)W_{E_1,E_2}(x,p)\,.
\eq
which can be considered as the "transition amplitude"
in the Hilbert space.
The Wigner function is projector  in terms of
Moyal product
\begin{equation}
W_{nm}* W_{kl} = \hbar^{-1} \delta_{ml} W_{kn}\,,
\end{equation}
where the Moyal product is defined as
\begin{equation}
    *= e^{i\hbar(\partial_{x,\leftarrow}\partial_{p,\rightarrow}
    - \partial_{p,\leftarrow}\partial_{x,\rightarrow})}\,.
\end{equation}
where in our notations  $\partial_{x,\leftarrow}$ denotes the action on
$x$ variable  at the left.
For instance, the Wigner function for oscillator is expressed
in terms of the generalized Laguerre polynomials
\begin{equation}
W_{nm}(z,\theta))= \frac{(-1)^m}{\pi} \sqrt{\frac{m!}{n!}}z^{\frac{n-m}{2}}
e^{i(n-m)\theta}L_{m}^{n-m}(z)\,,
\end{equation}
where $ z=2(x^2+p^2)$ and $\theta$ is the angular variable
on the phase plane.

Summarizing comments above, we can claim that the measure considered
in the probability is the measure on the spectrum from the
quantum mechanical viewpoint and a pair of coordinates correspond
to the specializations in the probabilistic measure.

\paragraph{Probabilistic process in quantum mechanics}

The processes have been introduced in
the context of probability theory and  they define the measure
on the plane partitions or 3D Young diagrams. Let
us comment on their meaning in the one-body quantum
mechanics. As we have argued above for the case
of oscillator the wave function $\Psi(x,E)$ allows
determinantal dual Feynman representation as
the sum of weighted paths. Consider a bit artificial
representation for each term in the expansion
of determinant
\begin{equation}
  <0|\hat{X}^N|E>= \sum_{E_1,E_2\dots E_N}
  <0|\hat{X}|E_1> <E_1|\hat{X}|E_2> <E_2|\hat{X}|E_3> \dots  <E_{N}|\hat{X}|E>\,,
\end{equation}
where we just insert multiple units in the matrix element.

If we consider a single path in the representation above
we get the analog of the measure on the probabilistic process
in quantum mechanics:
\begin{equation}
    P(E_1,E_2,\dots E,x)\propto  <0|\hat{X}|E_1> <E_1|\hat{X}|E_2> <E_2|\hat{X}|E_3> \dots  <E_{N}|\hat{X}|E>\,.
\end{equation}
To keep more relation with integrable probability
we could impose the ascending condition $E_1<E_2< \dots <E_N<E$.
If this condition is imposed we have the link to GUE process \cite{or03}
in terms of the matrix model representation of quantum mechanics \cite{krefl}.
Indeed, the size of the matrix corresponds to the energy in this representation
and the process can be considered as the growth of the matrix in the spirit
of \cite{or03}.

Of course, the one-body problem is  oversimplified.
For instance, for Macdonald process we have to consider the product of matrix
elements $<0|A(x_i)|E_i>$ taken at different points $x_i$ and the operators
$A$ obey the Faddeev-Zamolodchikov algebra.
However, the key interpretation of the process as the  particular
path in the dual Feynman representation in the Hilbert space is the same.

\section{Appendix D: relation to SYM dualities}
\def\theequation{D.\arabic{equation}}
\setcounter{equation}{0}
\subsection{Goldfish limit in SYM}
In this Section we shall briefly discuss the specific limit of QC duality
we found above in the framework of brane realization of  SUSY YM theories.
Let us start with duality between the XXZ spin chains(ASEP) and trigonometric
RS model. It was   identified in \cite{GaK} as the correspondence between the quiver $3d$  $\mathcal N=2^*$ gauge theory
on $ \mathbb R^2\times S^1$   and $\mathcal N=2^*$  $4d$ gauge theory on the interval \cite{GaK}. The corresponding elements of the correspondence
are encoded in the parameters and superpotential  of quiver in the $3d$ theory and
in  the boundary conditions for the $4d$ gauge theory with $\mathcal N=2$ SUSY
with $\mathbb R^2\times S^1 \times L$ geometry.

The gauge theory interpretation of QC duality between the XXZ-trigonometric RS duality
has been elaborated in \cite{GaK} and involves the following brane configuration.
We have $n$ parallel  NS5 branes  extended in the
$(x_0,x_1,x_2, x_7,x_8,x_9)$ directions, $N_i$ D3 branes extended
in $(x_0,x_1,x_2,x_3)$ between $i$-th and $(i+1)$-th NS5 branes, and
$K_i$ D5 branes extended in $(x_0,x_1,x_2,$ $x_3,x_4,x_5,x_6)$
directions  between the $i$-th and $(i+1)$-th NS5 branes. From this brane
configuration we obtain the $\prod_{i}^{n} U(N_i)$ gauge group on
the D3 brane worldvolume with additional $K_i$ fundamentals for the
$i$-th gauge group. The distance between the $i$-th and $(i+1)$-th
NS5 branes
yields the gauge coupling for the $U(N_i)$ gauge group while
coordinates of the D5 branes in the $x_7,x_8$ plane correspond to
the masses of fundamentals. The positions of D3 branes in the $x_7,x_8$
plane correspond to the coordinates on the Coulomb branch in the
quiver theory. The additional $\Omega$-deformation reduces the
theory with $N=4$ SUSY to the $N=2^{*}$ theory. At the energy scale
below the scale dictated by the lengths of the intervals the theory
on D3 branes is identified as $N=2^{*}$ 3d quiver gauge theory. In
what follows we assume that one coordinate is compact that is the 3d
theory lives on $R^2\times S^1$.

\begin{figure}[h]
\centering
\begin{tabular}{|c|c|c|c|c|c|c|c|c|c|c|}
%    \label{table:branes}
    \hline
    &0&1&2&3&4&5&6&7&8&9\\
    \hline
    D3&$\times$&$\times$&$\times$&$\times$&&&&&&\\
    \hline
    NS5&$\times$&$\times$&$\times$&&$\times$&$\times$&$\times$&&&\\
    \hline
    D5&$\times$&$\times$&$\times$&&&&&$\times$&$\times$&$\times$\\
    \hline
\end{tabular}
\caption{Brane construction of the $3d$ quiver theory.}
\label{table:branes}
\end{figure}

The QC duality corresponds to the particular Hanany-Witten move of the
brane configuration when we place all the $D5$ branes to the left of the $NS5$ branes.
Hence now we have a $Un(Q)$ four-dimensional gauge theory placed between Neumann boundary conditions provided by $M$ $NS5$ branes and Dirichlet boundary conditions provided by $N=\sum_j  M_i$ D5 branes
\begin{equation}
    Q= \sum_{j=1}^{p} jM_j.
    \label{eq:Q}
\end{equation}
The information about the $3d$ quiver is now encoded in the boundary conditions in the $4d$ theory via embedding $SU(2)\rightarrow U(Q)$ at the left and right boundaries \cite{gw1,gw2,nw}

The mapping of the gauge theory data into the integrability framework goes as follows. In the NS limit of the $\Omega$-deformation the twisted superpotential in $3d$ gauge theory
on the $D3$ branes gets mapped into the Yang-Yang function for the $XXZ$ chain \cite{ns1,ns2}.
The minimization of the superpotential yields the equations
describing the supersymmetric vacua and in the same time they are the Bethe ansatz equations for the $XXZ$ spin chain. That is $D3$ branes
are identified with the Bethe roots which are distributed according to the ranks of the
gauge groups at each of the $p$ steps of nesting $\prod_{i}^p U(N_i)$. Generically the number
of the Bethe roots at the different levels of nesting is different. The distances between
the $NS5$ branes define the twists at the different levels of nesting while the
positions of the D5 branes in the $(45)$ plane correspond to the inhomogeneities
in the $XXZ$ spin chain. To complete the dictionary recall that the anisotropy of the $XXZ$ chain
is defined by the radius of the compact dimensions, while the parameter of the $\Omega$
deformation plays the role of the Planck constant in the $XXZ$ spin chain.

The limit of large Planck constant in XXZ we consider is
identified  as the NS limit with $\epsilon_1=0, \epsilon_2\rightarrow \infty$.
Therefore we consider the strong parameter deformation in the $\Omega$ background
which at the RS side corresponds to large coupling constant $\nu\rightarrow \infty$
which yields the trigonometric Goldfish model.  The twisted superpotential for the TASEP
 which yields the BA equations is the $\hbar\rightarrow \infty $ of the YY function
for XXX model.
To get the gauge counterpart of the QQ duality we have to add the second parameter $\epsilon_1\neq 0$.
The corresponding Planck constant at the Goldfish side gets identified with
$\hbar_{gf}=\frac{\epsilon_1}{\epsilon_2}$
hence to have finite $\hbar_{gf}$ we should assume $\epsilon_1\rightarrow \infty$.
The parameter $\hbar_{gf}$ enters the qKZ equation at TASEP side.

One more point deserves the comments. The trigonometric RS model has clear-cut gauge theory
origin \cite{gn} and corresponds the Chern-Simons theory on $T^2\times R$ with
two inserted space and temporal Wilson lines where the coupling of the model corresponds to
the representation of a temporal Wilson line. When the degeneration of the model
is considered two ways down are possible. First we can consider the plain $\nu\rightarrow \infty $
limit when the trigonometric RS reduces to the trigonometric  Goldfish model. On the other hand another
limit is possible with the additional rescaling of degrees of freedom when it
gets reduced down to the relativistic Toda model \cite{Ruij}. It is the relativistic version
of the Inozemtsev limit. Note that the trigonometric Goldfish and relativistic Toda
are spectrally dual. At further degeneration the trigonometric Goldfish gets down to the rational Goldfish  while the relativistic Toda reduces to the non-relativistic Toda. These models are spectrally dual as well.

It is easy to see that  the Inozemtsev limit and degeneration of the initial
torus are the commutative operations. Indeed we could first consider the two degenerations
of the torus when trigonometric RS gets reduced to the rational RS or to the trigonometric Calogero
model correspondingly. At the next step we can consider $\nu\rightarrow \infty $,
which yields rational Goldfish and non-relativistic Toda once again.

Let us turn to the question about counterpart of two ways of degeneration of trigonometric RS on the spin chain side
via QC duality. In our study we have considered the plain $\nu\rightarrow \infty $ limit
from XXZ model which yields TASEP. On the other hand the different limit briefly mentioned
in Section 5 in \cite{koroteev17} concerns QC dual of  the Inozemtsev limit from the trigonometric RS model to relativistic Toda.
It was claimed there that at the spin chain side one gets 5-vertex model from the XXZ model.
This seems to be doubtful interpretation since we have shown in the previous sections that
5-vertex model is degeneration of TASEP hence it has nothing to do with  the Inozemtsev limit
at the spin chain side. That is limit considered in  \cite{koroteev17} presumably
corresponds to the model spectrally dual to the TASEP. To get this model one has
apply the Inozemtsev limit to the XXZ spin chain, that is similar to trigonometric RS model in addition to  $\nu\rightarrow \infty $ the rescaling of Bethe roots and inhomogeneities has to be done.

\subsection{Measures and multiple surface operators}

We have argued above how the Goldfish-TASEP duality
at the level of wave functions and solutions to KZ is embedded into the
dualities in the perturbative SYM supplemented with the single surface operator.
Let us briefly comment now on the interpretation of the SYM counterpart of the  duality between
expectation values of  particular observables evaluated  with the  Macdonald and
6-vertex model measures. The RS model
wave function involved into the Macdonald measure is the wave function
of the surface operator embedded into the 5d SYM theory which counts
the 3d instantons on the defect worldvolume but neglects the interaction with 5d instantons. The wave functions of defects obey the BPZ and KZ equations
both for theories with adjoint and fundamental matter \cite{nek21}.

The Macdonald measure involves two Macdonald polynomials
with different specializations hence two surface
operators are involved and the matrix element at the Macdonald side
reads as
\begin{equation}
\label{wilson}
\int d\vec{a} \Psi_{\vec{a}}^{surface}(x)e^{\vec{t}\vec{a}} \Psi_{\vec{a}}^{surface}(y)\,.
\end{equation}
It can be presumably interpreted as the vacuum expectation value of
composite observable involving the Wilson loop
 and two surface operators inserted at points $x$ and $y$
on the auxiliary surface. The integral over $\vec{a}$ corresponds
to the integration over Coulomb branch of 5d SYM theory. We shall
turn to the validity of such interpretation elsewhere.

It is instructive to compare this expression with the
vacuum expectation value of circular  Wilson loop in 5d theory
in representation $R_j$ without surface operators \cite{pestun}
when the instantons are taken into account
\begin{equation}
\label{pestun}
<W_{R_j}(a_k)>= \int d\vec{a} |Z_{5d}(\vec{a},\tau)|^2  Tr_{R_j}
(e^{4\pi i a_kt_3})\,,
\end{equation}
where parameter b is expressed via equivariant parameters
in Nekrasov partition function.
In (\ref{pestun}) the Nekrasov instanton partition function
plays the role of the wave function (in the Whitham hierarchy framework),
which depends on the
complexified coupling constant $\tau$.

The (\ref{wilson}) involves the integration over the Coulomb branch
which resembles the situation with AGT correspondence when
the similar integrals of Nekrasov partition functions get mapped
into the correlators in Liouville or $W_N$ theories and the
integral over the Coulomb branch corresponds to the integral over
intermediate
weights in the conformal blocks \cite{agt}.
The insertion of surface operator at point $x$ on the surface
where the Liouville or $W_N$-gravity is defined on corresponds
to the insertion of operator  $\Psi_{2,1}(x)$ \cite{alday}.
Hence the natural guess for
the interpretation of the matrix element of the
operator  evaluated with 6-vertex stochastic model measure is the
correlator in Liouville theory or $W_N$-gravity in the perturbative limit
with insertions of
two $\Psi_{1,2}$ operators at points $x$ and $y$ and
non-local operator which corresponds to a Wilson loop
at the gauge theory side.

The meaning of the Wilson loop operator
at the Liouville side has been clarified in \cite{alday}.
It was argued there that it corresponds to the monodromy
of $\Psi_{1,2}$ operator along the cycle on the Riemann surface
and acts on a surface operator shifting its argument. The
monodromy is defined via a combination of action of  braiding and fusion
operators on the representation ring of $SU(N)$ algebra.
At the stochastic duality side the Wilson loop corresponds
to the height function, or their products $\prod e^{x_N}$
and the natural question concerns if height function can be related to the
monodromy representation of Wilson loop  at the Liouville side.

We restrict ourselves
by short remark postponing the analysis for the separate study.
It was argued in \cite{bprev,orr} that the following
identification holds true
$$\lambda_{NN}=x_N+N$$
between the heights in the ASEP and the boundary weights for the
GT scheme $\lambda_{NN}$. Since the fusion in a representation
ring which defines the monodromy of  $\Psi_{1,2}$
is expressed in terms of the
GT scheme as well it is natural to conjecture that the stochastic
duality for the observables fits with the AGT duality
flavored with two surface operators. It would be interesting
to recognize and utilize the explicit multiple integrals formulae
derived for expectation values of  observables
evaluated with the Macdonald or 6-vertex model measures in the AGT framework.

%\section{Discussion}

%%%%%%%%%%%%%%%%%%%%%%%%%%%%%%%%%%%%%%%%%%%%%%%%%%%%%%%%%%%%%%%%%%%%%%%%%%%%%%%%%%%%%%%%%%%%%%%%%%%%%%
%%%%%%%%%%%%%%%%%%%%%%%%%%%%%%%%%%%%%%%%%%%%%%%%%%%%%%%%%%%%%%%%%%%%%%%%%%%%%%%%%%%%%%%%%%%%%%%%%%%%%%
%%%%%%%%%%%%%%%%%%%%%%%%%%%%%%%%%%%%%%%%%%%%%%%%%%%%%%%%%%%%%%%%%%%%%%%%%%%%%%%%%%%%%%%%%%%%%%%%%%%%%%

%%%%%%%%%%%%%%%%%%%%%%%%%%%%%%%%%%%%%%%%%%%%%%%%%%%%%%%%%%%%%%%%%%%%%%%%%%%%%%%%%%%%%%%%%%%%%%%%%%%%%%
%%%%%%%%%%%%%%%%%%%%%%%%%%%%%%%%%%%%%%%%%%%%%%%%%%%%%%%%%%%%%%%%%%%%%%%%%%%%%%%%%%%%%%%%%%%%%%%%%%%%%%

\begin{small}
 
\end{small}

\end{document}